\documentclass[final,3p,times]{elsarticle}

\usepackage{graphicx}
\usepackage{epsfig}
\usepackage{epstopdf}
\usepackage{amssymb}
\usepackage{amsthm}
\usepackage{bm}
\makeatletter
\def\els@aparagraph[#1]#2{\elsparagraph[#1]{#2}}
\def\els@bparagraph#1{\elsparagraph*{#1}}
\makeatother

\usepackage[colorlinks=true,citecolor=blue,linkcolor=black]{hyperref}
\usepackage{color}
\usepackage{amsmath}
\usepackage{multirow}
\usepackage[justification=centering]{caption}
\usepackage{subcaption}
\usepackage{color}
\usepackage{lineno}

\journal{Engineering Fracture Mechanics}

\begin{document}

\begin{frontmatter}



\title{Numerical Modelling of Crack Initiation, Propagation and Branching under Dynamic Loading}

\author[essteyr,iitkgp]{Md Rushdie Ibne Islam\corref{iitkgp1}}
\ead{rushdie.islam@gmail.com}
\author[iitkgp]{Amit Shaw}

\cortext[iitkgp1]{Corresponding Author}
\address[essteyr]{ESS Engineering Software Steyr GmbH, Berggasse 35, 4400 Steyr, Austria}
\address[iitkgp]{Department of Civil Engineering, Indian Institute of Technology, Kharagpur 721302, West Bengal, India}
 
\begin{abstract}
In this paper crack initiation, propagation and branching phenomena are simulated using the Pseudo-Spring Smooth Particle Hydrodynamics (SPH) in two and three-dimensional domains. The pseudo-spring analogy is used to model material damage. Here, the interaction of particles is limited to its initial immediate neighbours. The particles are connected via springs. These springs do not provide any extra stiffness in the system but only define the level of interaction between the connecting pairs. It is assumed that a crack has passed through a spring connecting a particle pair if the damage indicator of that spring becomes more than a predefined value. The crack branching of a pre-notched plate under dynamic loading and the effect of loading amplitude are studied here. The computed crack speeds, crack paths and surfaces are compared with experimental and numerical results available in the literature and are found to be in good agreement. The effect of notch location for a plate with a circular hole is studied here. The ability of the framework to model arbitrary crack paths and surfaces are demonstrated via three-dimensional simulations of chalk under torsion, Kalthoff-Winkler experiment and Taylor bullet impact.
\end{abstract}


\end{frontmatter}


\section{Introduction}\label{sec-1}
Numerical modelling of crack initiation, propagation and branching under dynamic loading is still an open problem. The tracking of arbitrary paths and their propagation are modelled by different grid-based or mesh-free numerical techniques. The grid-based method such as Finite Element Method (FEM) or mesh-free methods such as Smooth Particle Hydrodynamics (SPH) cannot model crack discontinuity in their standard form. Each of these methods needs some special treatment to model the crack discontinuity.

FEM is used to model crack path and its propagation by element deletion or kill method \cite{beissel1998element}, \cite{borvik2001numerical} or by adaptive re-meshing \cite{de1999elastoplastic}, \cite{martha1993arbitrary}, \cite{marusich1995modelling}, \cite{schollmann2003development}. In the element deletion or kill technique \cite{beissel1998element} \cite{borvik2001numerical}, an element is assumed to fail once the stress, strain or damage values exceed a predefined value to model arbitrary cracks. The edges of the failed meshes act as the new crack surfaces. The main problem in element deletion or kill algorithms is the numerical instabilities arise due to excessive mesh distortion. This may lead to unreal element volume and error in the numerical computation \cite{borvik2002perforation}. Though FEM coupled with automatic and adaptive re-meshing is used for modelling of arbitrary crack propagation \cite{de1999elastoplastic}, \cite{martha1993arbitrary}, \cite{marusich1995modelling}, \cite{schollmann2003development}, this technique is not quite useful for tracking multiple cracks. The method of re-meshing is also complicated as multiple field variables require internal mapping between different meshes during re-meshing. 

The crack growth is also simulated through the element edges using the inter-element separation techniques \cite{camacho1996computational}, \cite{xu1994numerical}, \cite{ortiz1999finite}, \cite{zhou2004dynamic}. The crack path always propagates through the element edges or boundaries. Hence, the element size and shape are important parameters for tracking the crack path correctly. The effect of mesh size in computation is reduced in Embedded Finite Element Method (EFEM) \cite{dvorkin1990finite} \cite{oliver2000discrete}. The EFEM is capable of tracking arbitrary crack propagation without any re-meshing. In this method an enrichment is introduced at the element level \cite{mosler2004embedded} \cite{linder2007finite}. Belytschko and his group \cite{belytschko1999elastic}, \cite{moes2002extended}, \cite{belytschko2003dynamic} developed the extended finite element method (XFEM) for modelling of arbitrary crack propagation. They utilise the partition of unity \cite{melenk1996partition} for local enrichment. In XFEM, only the surrounding elements of a crack and the crack propagation path (known a priori) are enriched and the crack propagated through these enriched elements (not only through the element boundaries). Though these element based methods can model evolving cracks and track their paths, the requirement of re-meshing or local enrichments makes these numerically and computationally intensive. Moreover, these methods may not be so effective for multiple cracks, crack interaction or excessive deformations. In this context, mesh-free methods are more suitable.  

Different types of mesh-free/particle methods are introduced over the years. Belytschko and his colleagues \cite{belytschko1996dynamic}, \cite{lu1995element}, \cite{belytschko1995element}, \cite{belytschko1995crack} developed the Element-Free Galerkin (FFG) method for static and dynamic crack propagation. As the method is mesh-free, there is no re-meshing. In their works, they used moving least square interpolants as test and trial functions. Rabczuk and Belytschko \cite{rabczuk2004cracking} \cite{rabczuk2007three} proposed another cracking particle method for modelling of crack path and its evolution. They introduce discontinuous enrichments, and the crack is modelled by particle splitting. As the cohesive traction at a particle reduces to zero, the particle is assumed to be cracked and split into two particles. Later, another version of cracking particle is proposed \cite{rabczuk2010simple} without any enrichment. However, this still requires particle splitting. The need for discontinuous enrichments or particle splitting or visibility criteria \cite{organ1996continuous} for crack path tracking makes these methods computationally intensive. 

Another particle method, SPH is truly a mesh-free method \cite{MonaghanGingold83}, introduced for astronomical problems. This is extended to fluid problems and later on solids \cite{Libersky93}. In SPH, the computational domain is discretised into a set of particles. At any time instant, a particle interacts with its neighbour particles within its influence domain (known as the neighbourhood) through a kernel function. The bell-shaped kernel function ensures that the level of interaction with the nearest neighbour particles is higher than the particles close to the neighbourhood boundary. The particles have no interaction with particles outside its influence domain which is fixed by the smoothing length of the kernel function. If a particle moves out of the neighbourhood of a particle, their interaction ceases. The kernel function is independent of material properties. Naturally, it is unable to compute the damage state of different particles. But, this does not pose any problem in astronomical or fluid problems as there is no concept of damage. However, for solid continua, it poses significant difficulty as large deformation may often lead to damage/failure. If a particle which moves out of the influence domain of another particle is considered to be damaged, the damage variable will be a function of the smoothing length/radius of the kernel function. As the smoothing length/radius of the kernel function increases, the influence domain of a particle increases. A particle has to move further to get out of the influence domain of another particle. Hence, the damage state of a particle will change if the kernel radius increases or decreases for material under the same loading conditions. This may lead to non-physical behaviour in the numerical computation.

In this context, a `pseudo-spring' strategy in the SPH framework is developed \cite{chakraborty2013pseudo} for crack modelling. Springs connect all the particles in the computational domain. The springs are termed as `pseudo-spring' as these do not provide any extra stiffness to the system but only define the level of interaction between particles based on the damage state. The interaction of particles is not the same as the standard SPH. Here, particles only interact through standard kernel function with its immediate neighbours (from initial configuration). This framework does not require any crack enrichment, particle splitting or tracking of crack paths. When a certain parameter of a pseudo-spring is more than a predefined critical value, the spring is considered to be damaged, and the interaction ceases between the connected particles. Moreover, it is assumed that the crack has passed through that particle pair. For partially damage pseudo-springs, the level of interaction is decreased accordingly. In this framework, the crack path can be tracked automatically by the damaged pseudo-springs. The simplicity of the SPH is preserved in the computation. This method is successfully used to simulated impact induced damages \cite{shaw2015beyond}, \cite{chakraborty2014crack}, \cite{chakraborty2015prognosis}, \cite{chakraborty2017computational}, \cite{islam2017computational}.

In this work, this pseudo-spring analogy in SPH framework is used to simulate the crack initiation, propagation and branching in two and three dimensions. The propagation of cracks, the deviation of crack paths under different conditions and generation of multiple crack branches are simulated. The effects of tensile loading amplitude in crack branching are discussed. The generation of arbitrary crack surfaces under dynamic loading in three dimensions is modelled. The propagation of cracks under impact loading in three dimensions is also demonstrated. The ability of the pseudo-spring SPH to simulate the crack initiation, propagation and branching under dynamic loading in two and three-dimensional space is demonstrated via numerical examples. The computed crack speeds, their paths and crack surfaces are compared with experimental and numerical results available in the literature.

The paper is organised as follows: The necessary steps of pseudo-spring SPH framework and the boundary condition are discussed in Section 2. Numerical examples of two and three-dimensional crack initiation, propagation and branching under dynamic loading are discussed in Section 3. Some conclusions are drawn in Section 4.

\section{Computational framework : \textit{Pseudo-Spring} SPH}\label{sec-2}
A set of particles represents the computational domain in SPH. Here, the basics of SPH and the `pseudo-spring' analogy incorporated in the framework are discussed briefly. A more detailed description of standard SPH and the `pseudo-spring' framework can be found in \cite{Libersky93}, \cite{liu2003smoothed}, \cite{liu2006restoring}, \cite{liu2010smoothed}, \cite{chakraborty2013pseudo}, \cite{chakraborty2017computational}, \cite{islam2017computational}, and references therein. In SPH, a particle interacts with its surrounding neighbourhood particles through a kernel function. The mass, momentum and energy equations and their  approximated particle forms are 

\begin{equation}\label{eq1}
    \frac{d\rho}{dt}=-\rho\frac{\partial u^\beta}{\partial x^\beta}
\end{equation}
\begin{equation}\label{eq5}
    \frac{d\rho_i}{dt}=\sum_{j \in \bar{N}^i} m_{j}u^\beta_{ij} \bar{W}_{ij,\beta}
\end{equation}

\begin{equation}\label{eq2}
    \frac{du^{\alpha}}{dt}=\frac{1}{\rho}\frac{\partial \sigma^{\alpha\beta}}{\partial x^\beta}
\end{equation}
\begin{equation}\label{eq6}
    \frac{du^\alpha_i}{dt}=\sum_{j \in \bar{N}^i} m_{j}\left(\frac{\sigma^{\alpha\beta}_i}{\rho^2_i}+\frac{\sigma^{\alpha\beta}_j}{\rho^2_j}-\pi_{ij}\delta^{\alpha\beta}-P^{a}_{ij}\delta^{\alpha\beta} \right) \bar{W}_{ij,\beta}
\end{equation}

\begin{equation}\label{eq3}
    \frac{de}{dt}=\frac{\sigma^{\alpha\beta}}{\rho}\frac{\partial u^{\alpha}}{\partial x^\beta}
\end{equation}
\begin{equation}\label{eq7}
    \frac{de_i}{dt}=-\frac{1}{2}\sum_{j \in \bar{N}^i} m_{j}u^\alpha_{ij}\left(\frac{\sigma^{\alpha\beta}_i}{\rho^2_i}+\frac{\sigma^{\alpha\beta}_j}{\rho^2_j}-\pi_{ij}\delta^{\alpha\beta} -P^{a}_{ij}\delta^{\alpha\beta} \right) \bar{W}_{ij,\beta}
\end{equation}
where, $i$ denotes the particle of interest, $j$ denotes its neighbour particles within its influence radius and $\mathbb{\bar{N}}^i\in\mathbb{Z^+}$ is the set of total number of neighbour particles within its influence domain. The density of the material is represented by $\rho$, $u^\alpha$ is the velocity component. $\sigma^{\alpha\beta}$ represents the Cauchy stress components. $e$ is the specific internal energy. $\alpha$, $\beta$ are the indices for coordinate axes. The time derivative is in moving Lagrangian frame and $u^\alpha_{ij}=u^\alpha_i-u^\alpha_j$. $\bar{W}_{ij,\beta}$ is the symmetric form of the kernel gradient correction computed as $\bar{W}_{ij,\beta} = 0.5\left(\hat{W}_{ij,\beta}+\hat{W}_{ji,\beta}\right)$ \cite{islam2018consistency}. The kernel gradient correction $\hat{W}_{ij,\beta}$ is calculated as \cite{ChenBeraunCarney99}

\begin{equation}\label{csph_app_1}
    \hat{W}_{ij,\beta}=B^{\beta\alpha}W_{ij,\alpha}
\end{equation}
where, 
\begin{equation}\label{csph_app_2}
   \mathbf{B=A^{-1}},~~~and~~~A^{\beta\alpha}=-\sum_j \frac{m_j}{\rho_j}x^{\beta}_{ij}W_{ij,\alpha}
\end{equation}

The kernel gradient is represented as $W_{ij,\beta}=\frac{\partial W(\bm{x}_i-\bm{x}_j,h)}{\partial x^\beta_i}$. $h$ is the smoothing length of the kernel function. Here, the following cubic B-spline kernel is used.

\begin{equation}\label{eq9}
    W(q, h)=\alpha_d 
\begin{cases}
    1-\frac{3}{2} q^2 +\frac{3}{4} q^3,& \text{if } 0\le q\le 1\\
    \frac{1}{4}(2-q)^3,              & \text{if } 1\le q\le 2\\
    0                                 & \text{otherwise}
\end{cases}
\end{equation}
where, $\alpha_d=\frac{10}{7\pi h^2}$ and $\alpha_d=\frac{1}{\pi h^3}$ in 2D and 3D respectively. $q$ represents the normalised position vector as $q=\frac{|\bm{x}_i-\bm{x}_j|}{h}$.

The artificial viscosity $\pi_{ij}$ shown in Equation \ref{eq8} is utilised in the current computation to remove any unphysical oscillation present near any jump or shock \cite{MonaghanGingold83}.  

\begin{equation}\label{eq8}
    \pi_{ij}= 
\begin{cases}
    \frac{-\beta_1 \bar{C}_{ij}\mu_{ij} + \beta_2 \mu^2_{ij}}{\bar{\rho}_{ij}},& \text{if } \bm{U}_{ij}.\bm{X}_{ij}\le 0\\
    0,              & \text{otherwise}
\end{cases}
\end{equation}
The intensity of the artificial viscosity is determined using $\beta_1$, $\beta_2$. $\mu_{ij}=\frac{h(\bm{U}_{ij}.\bm{X}_{ij})}{\bm{X}^2_{ij}+\epsilon h^2}$ and $\epsilon=0.01$. The sound speed through the medium is computed as $C=\sqrt{\frac{E}{\rho}}$, $E$ denotes the Young's modulus, $\bm{X}_{ij}=\bm{X}_i-\bm{X}_j$, $\bm{U}_{ij}=\bm{U}_{i}-\bm{U}_{j}$, $\bar{C}_{ij}=0.5(C_i+C_j)$ and $\bar{\rho_{ij}}=0.5(\rho_i+\rho_j)$. The artificial/Monaghan pressure \cite{monaghan2000sph} $P^{a}_{ij}$ as shown in Equation \ref{Art_P} is used to remove the \textit{tensile instability}.

\begin{equation}\label{Art_P}
   P^{a}_{ij}=\gamma \left(\frac{|P_i|}{\rho^2_i}+\frac{|P_j|}{\rho^2_j}\right) \left[\frac{W(\Delta x,h)}{W(\Delta p,h)}\right]^l
\end{equation}
where, $\gamma$ is a controlling parameter taken as $0.3$ for tension and $0.01$ for compression. The hydrostatic pressure is denoted by $P$. $\Delta x$ and $\Delta p$ are the current particle spacing and the average particle spacing in the reference configuration. $l$ is 4 \cite{monaghan2000sph}.

The pressure $P$ is calculated from the linear equation of compressibility (Equation of State).

\begin{equation}\label{eq10}
    P=K\left(\frac{\rho}{\rho_0}-1\right)
\end{equation}
where, $K$, $\nu$ and $\rho_0$ are the bulk modulus, Poisson's ratio and initial mass density of the material respectively. The Cauchy stress components $\sigma^{\alpha\beta}$ are calculated from the deviatoric stress components $S^{\alpha\beta}$ and hydrostatic pressure $P$ as $\sigma^{\alpha\beta}=S^{\alpha\beta}-P\delta^{\alpha\beta}$. The rate of change in deviatoric stress components is calculated by Jaumann rate as,

\begin{equation}\label{eq11}
    \dot{S}_{\alpha\beta}=2\mu \left(\dot{\epsilon}_{\alpha\beta}-\frac{1}{3}\delta^{\alpha\beta}\dot{\epsilon}^{\gamma\gamma}\right)+S^{\alpha\gamma}\omega^{\beta \gamma} +S^{\gamma\beta}\omega^{\alpha\gamma}    
\end{equation}
where, $\mu$ represents the shear modulus and the components of strain rate ($\dot{\bm{\epsilon}}$) and spin tensor ($\bm{\omega}$) are computed as, 
\begin{equation}\label{eq12}
    \dot{\epsilon}^{\alpha\beta}=\frac{1}{2} \left(\frac{\partial u^\alpha}{\partial x^\beta}+\frac{\partial u^\beta}{\partial x^\alpha}\right)
\end{equation} 

\begin{equation}\label{eq13}
    \omega^{\alpha\beta}=\frac{1}{2} \left(\frac{\partial u^\alpha}{\partial x^\beta}-\frac{\partial u^\beta}{\partial x^\alpha}\right)
\end{equation} 

Von-Mises yield function $y_f=\sqrt{J_2}-\frac{\sigma_y}{\sqrt{3}}$ is used to account for the effects of metal plasticity. 
$\sigma_y$ and $J_2=\frac{1}{2}S^{\alpha \beta}S^{\alpha \beta}$ are the yield stress of the material and the second invariant of deviatoric stress tensor respectively. The Wilkins criterion $S_n^{\alpha\beta}=c_f S^{\alpha\beta}$ is used for return mapping of the yield surfaces and $c_f=\min \left(\frac{\sigma_y}{\sqrt{3J_2}},1\right)$. The plastic strain increment, effective plastic strain increment and accumulation of plastic work density are estimated as

\begin{equation}
   \Delta \epsilon_{pl}^{\alpha\beta} = \frac{1-c_f}{2\mu}S^{\alpha\beta}    
\end{equation}
\begin{equation}
   \Delta \epsilon_{pl} = \sqrt{\frac{2}{3}  \Delta \epsilon_{pl}^{\alpha\beta}  \Delta \epsilon_{pl}^{\alpha\beta}} = \frac{1-c_f}{3\mu} \sqrt{\frac{3}{2} S^{\alpha \beta} S^{\alpha \beta}} 
\end{equation}
\begin{equation}
   \Delta W_p =  \Delta \epsilon_{pl}^{\alpha\beta} S_{n}^{\alpha\beta}                
\end{equation}   

The standard predictor-corrector method \cite{Monaghan89} is used to update the discretised Equations \ref{eq5}, \ref{eq6}, \ref{eq7} and the stable time step for integration is determined from CFL (Courant-Fredrich-Levy) condition as, $\bigtriangleup t=\min_{i} \left(c_s \frac{h_i}{C_i+|v_i|}\right)$ where, CFL number $c_s$ is 0.3, $C_i$ is elastic wave velocity, $h_i$ is smoothing length and $v_i$ is velocity of the respective particles.

\subsection{\textit{Pseudo-Spring} analogy for crack}\label{sec_spring}
The \textit{pseudo-spring} model for crack propagation is discussed in details in \cite{chakraborty2013pseudo}, \cite{chakraborty2017computational}, \cite{islam2017computational}. Here, a quick review of the algorithm is provided. In SPH, the particles interact with each other through the kernel function. However, the kernel function does not have any material property as shown in Equation \ref{eq8}. Hence, the initiation of crack and its growth cannot be modelled through it directly. A particle influences all the particles in its neighbourhood and the level of interaction between any particle pair decreases as their intermediate distance increases. As the material deforms, a particle may move out of another particle's influence/neighbour domain, and the interaction between the particles stops. However, this cannot be considered as material damage/failure as different values of smoothing length would lead to different types of material response as discussed in \cite{chakraborty2017computational}. Therefore, the \textit{`pseudo-spring'} analogy is used to incorporate the effects of material damage and fracture in the kernel function. 

\begin{figure}[hbtp!]
\centering
\includegraphics[width=0.8\textwidth]{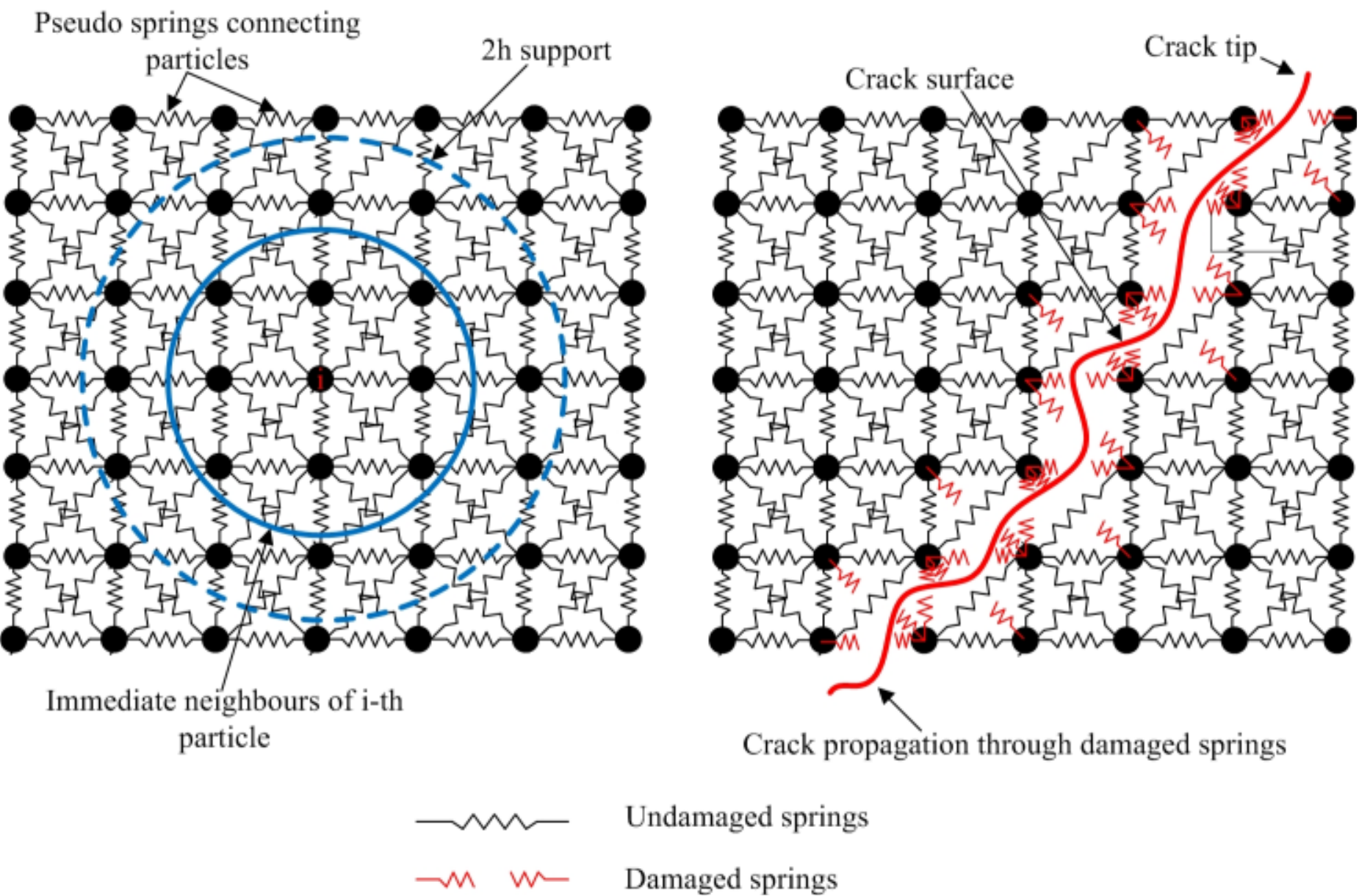}
\caption{Pseudo-springs network on immediate neighbours and crack propagation through damaged springs}\label{imnb}
\end{figure}

\begin{figure}[hbtp!]
\centering
\includegraphics[width=0.8\textwidth]{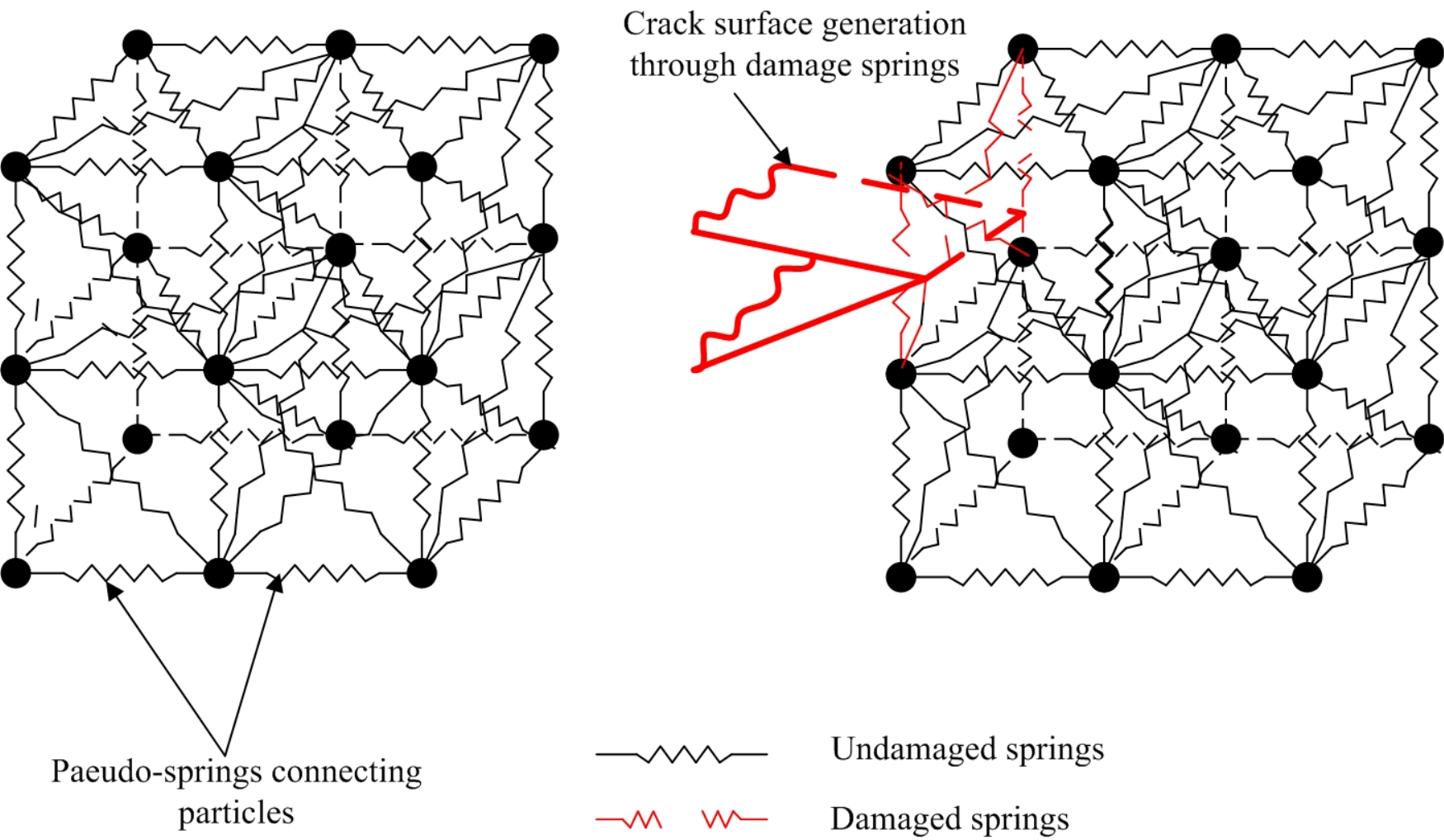}
\caption{Crack propagation through Pseudo-spring networks in three dimension}\label{imnb3d}
\end{figure}

Here, a particle does not interact with all the particles found in its neighbourhood. The interaction of a particle is restricted to its immediate neighbourhood only as shown in Figure \ref{imnb}. All the particles are connected to its modified neighbourhood through a set of springs. These springs define the level of interaction between particles and most importantly these do not provide any extra stiffness to the system. Hence, the term \textit{`pseudo'} is used. At the start of the computation, the material is undamaged and the springs ensure complete interaction between connecting particles. However, as the material undergoes deformations, the damage accumulates in the springs. When the accumulated damage is more than a predefined value, it is assumed that the springs break. This leads to no interaction between connecting particles and propagation of crack path. This analogy does not require any particle splitting \cite{ren2010meshfree}. The concept is illustrated in Figure \ref{imnb} and \ref{imnb3d}. 

The analogy is incorporated in the framework in the following way. A \textit{interaction factor} $f_{ij}$ between a connecting pair of particles is defined as $f_{ij}=1-D_{ij}$ and $D_{ij}$ is the accumulated damage in the spring between the $i-j$ particle pair. At the start of the computation, there is no material damage. Hence, the damage index $D_{ij}=0$ and this leads to the \textit{interaction factor} $f_{ij}=1$ implying complete interaction between particle pairs. However, as the damage accumulates in the springs, the \textit{interaction factor} $f_{ij}$ decreases i.e. $D_{ij}\in(0,1)\Rightarrow f_{ij}\in(0,1)$ or partial interaction. When the damage index $D_{ij}=1$ i.e. the \textit{interaction factor} $f_{ij}=0$ or no interaction (leading to formation of crack surface). $D_{ij}$ is computed as $D_{ij}=0.5(D_i+D_j)$ , where $D_i$ and $D_j$ are the damage parameters at the $i-th$ and $j-th$ particles respectively.

\begin{figure}[hbtp!]
\centering
\begin{subfigure}[t]{0.3\textwidth}    %trim={<left> <lower> <right> <upper>}
\includegraphics[width=\textwidth]{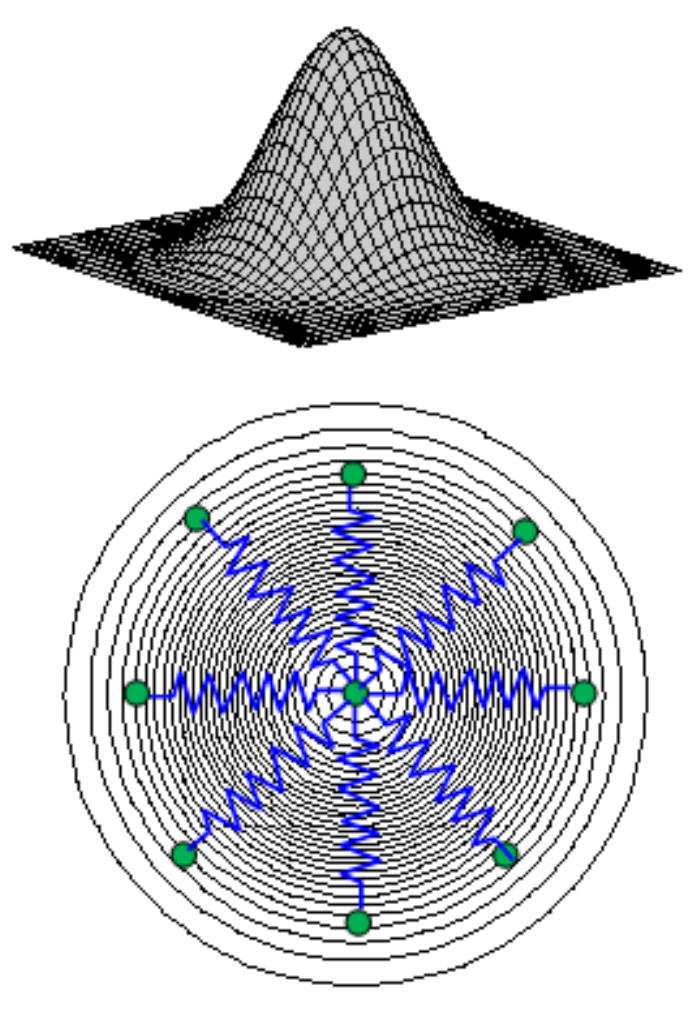}
\caption{Kernel: $f_{ij} = 1$} 
\end{subfigure}
\begin{subfigure}[t]{0.3\textwidth}
\includegraphics[width=\textwidth]{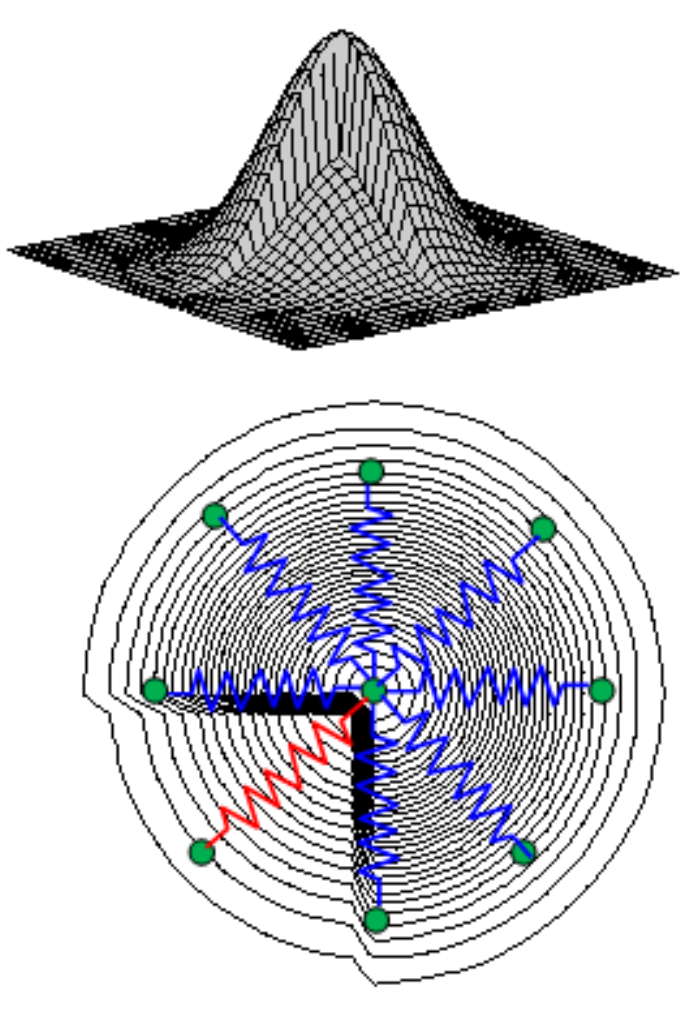}
\caption{Kernel: $0<f_{ij}<1$} 
\end{subfigure}
\begin{subfigure}[t]{0.3\textwidth}    %trim={<left> <lower> <right> <upper>}
\includegraphics[width=\textwidth]{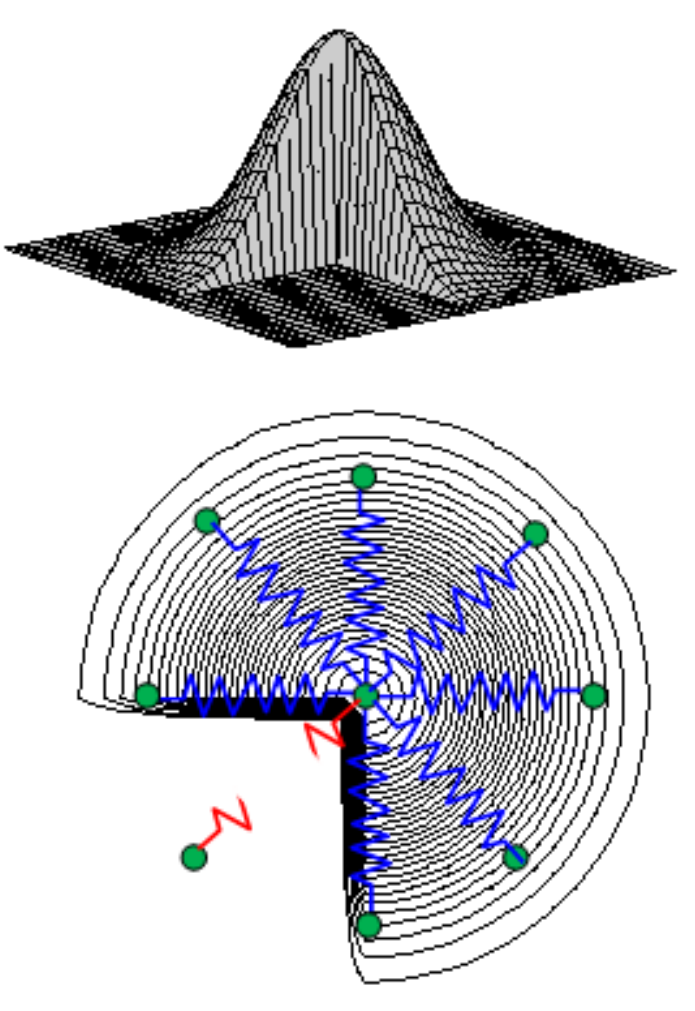}
\caption{Kernel: $f_{ij} = 0$} 
\end{subfigure}
\begin{subfigure}[t]{0.3\textwidth}
\includegraphics[width=\textwidth]{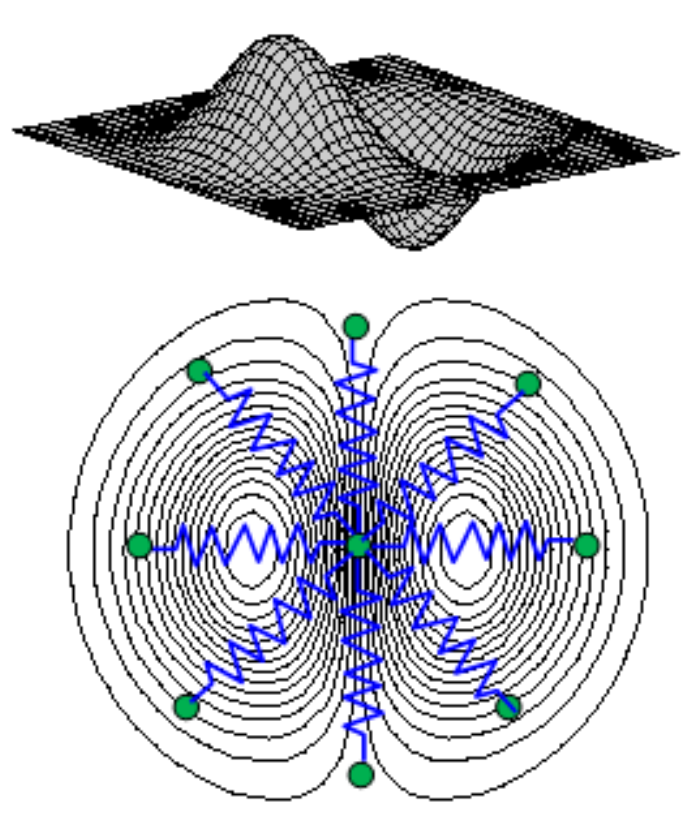}
\caption{Derivative: $f_{ij} = 1$} 
\end{subfigure}
\begin{subfigure}[t]{0.3\textwidth}    %trim={<left> <lower> <right> <upper>}
\includegraphics[width=\textwidth]{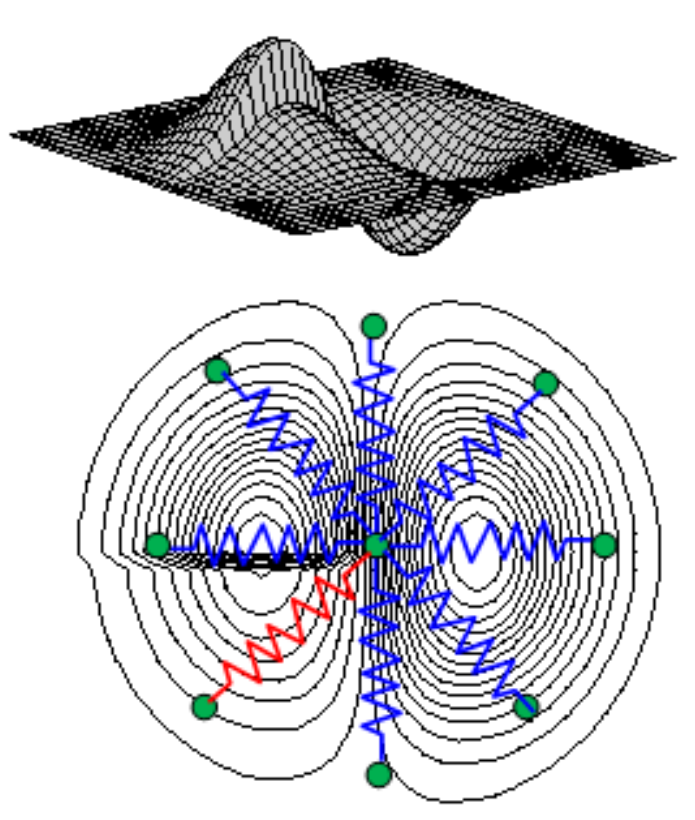}
\caption{Derivative: $0<f_{ij}<1$} 
\end{subfigure}
\begin{subfigure}[t]{0.3\textwidth}
\includegraphics[width=\textwidth]{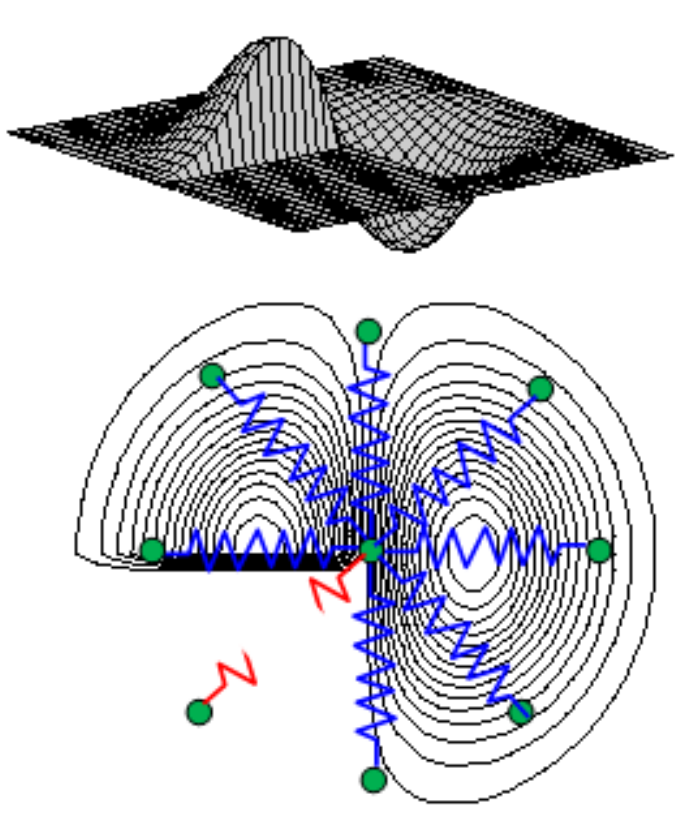}
\caption{Derivative: $f_{ij} = 0$} 
\end{subfigure}
\caption{Modification of kernel in 2D based on damage state \cite{chakraborty2017computational}, \cite{islam2017computational}}\label{fig_kernel2D}
\end{figure}

$\mathbb{N}^i\in\mathbb{Z^+}$, $\mathbb{N}_D^i =\{j\in\ {\mathbb{N}^i}\mid 0\leq f_{ij}<1\} \subseteq \mathbb{N}^i$ and $\mathbb{N}_U^i =\mathbb{N}^i \setminus \mathbb{N}_D^i $ are the set of immediate neighbouring particles that interact with the $i-th$ particle, the set of particles which are connected to the $i-th$ particle through damaged and undamaged springs at a given time instant respectively. The interaction between particles $i$ and $j \in \mathbb{N}^i$ in a damaged configuration is reduced by modifying the corresponding derivative of the kernel function as $\bar{W}_{ij,\beta}=f_{ij}\bar{W}_{ij,\beta} \forall j\in \mathbb{N}_D^i \subseteq \mathbb{N}^i$. Figure \ref{fig_kernel2D} shows the typical modified kernel in 2D for different damage states i.e.,- $f_{ij}=1$ (no damage), $0<f_{ij}<1$ (partial damage) and $f_{ij}=0$ (open crack).

Finally the modified forms of the discrete conservation equations \ref{eq5}, \ref{eq6} and \ref{eq7} become,

\begin{equation}\label{eq5df}
    \frac{d\rho_i}{dt}=\sum_{j \in N^i_U} m_{j}u^\beta_{ij}\bar{W}_{ij,\beta} + \sum_{j \in N^i_D} m_{j}u^\beta_{ij} \left(f_{ij} \bar{W}_{ij,\beta}\right)
\end{equation}

\begin{equation}\label{eq6df}
\begin{split}
    \frac{du^{\alpha}_i}{dt}=\sum_{j \in N^i_U} m_j \left(\frac{\sigma^{\alpha \beta}_i}{\rho^2_i}+\frac{\sigma^{\alpha \beta}_j}{\rho^2_j}-\pi_{ij}\delta^{r\beta}-P^{a}_{ij}\delta^{r\beta}\right)\bar{W}_{ij,\beta}\\
     + \sum_{j \in N^i_D} m_j \left(\frac{\sigma^{\alpha \beta}_i}{\rho^2_i}+\frac{\sigma^{\alpha \beta}_j}{\rho^2_j}-\pi_{ij}\delta^{r\beta}-P^{a}_{ij}\delta^{r\beta}\right) \left(f_{ij} \bar{W}_{ij,\beta}\right)
\end{split}      
\end{equation}

\begin{equation}\label{eq7df}
\begin{split}
    \frac{de_i}{dt}=-\frac{1}{2}\sum_{j \in N^i_U} m_{j}u^\alpha_{ij}\left(\frac{\sigma^{\alpha\beta}_i}{\rho^2_i}+\frac{\sigma^{\alpha\beta}_j}{\rho^2_j}-\pi_{ij}\delta^{\alpha\beta}-P^{a}_{ij}\delta^{\alpha\beta}\right)\bar{W}_{ij,\beta}\\ -\frac{1}{2}\sum_{j \in N^i_D} m_{j}u^\alpha_{ij}\left(\frac{\sigma^{\alpha\beta}_i}{\rho^2_i}+\frac{\sigma^{\alpha\beta}_j}{\rho^2_j}-\pi_{ij}\delta^{\alpha\beta}-P^{a}_{ij}\delta^{\alpha\beta}\right) \left(f_{ij} \bar{W}_{ij,\beta}\right)
\end{split}
\end{equation}

\subsection{Stress boundary condition}\label{sbc}
A stress boundary condition for SPH is developed in the context of solid continua in \cite{douillet2016development}. The boundary condition can be applied with or without ghost/dummy particles. The modified form of the momentum equation to account for the applied stress is shown in Equation \ref{bc_eq}. 

\begin{equation}\label{bc_eq}  
    \frac{du^\alpha_i}{dt}=\left(\frac{du^\alpha_i}{dt}\right)_{Equation~\ref{eq6df}} - \sum_{j \in N^i} m_{j}\left(\frac{\sigma^{\alpha\beta}_0}{\rho_i \rho_j} \right) W_{ij,\beta}
\end{equation}

Here, $\sigma^{\alpha\beta}_0$ is the stress component applied to the boundary particles.  

\section{Numerical simulation}

\subsection{Two dimensional numerical examples}\label{sec-31}
In this section, the crack initiation, propagation and branching in solids under dynamic loading are simulated for two-dimensional geometry. These phenomena are simulated using the \textit{Pseudo-Spring} SPH. First, the crack branching of a pre-notched plate under dynamic axial loading is studied, and the effect of the loading amplitude is also investigated here. The computed crack speeds, their paths, and crack surfaces are compared with experimental and numerical results available in the literature. The crack path from a notch in a plate with a circular hole is also modelled.    

\subsubsection{Crack branching}
\begin{figure}[hbtp!]
\centering
\includegraphics[width=0.5\textwidth]{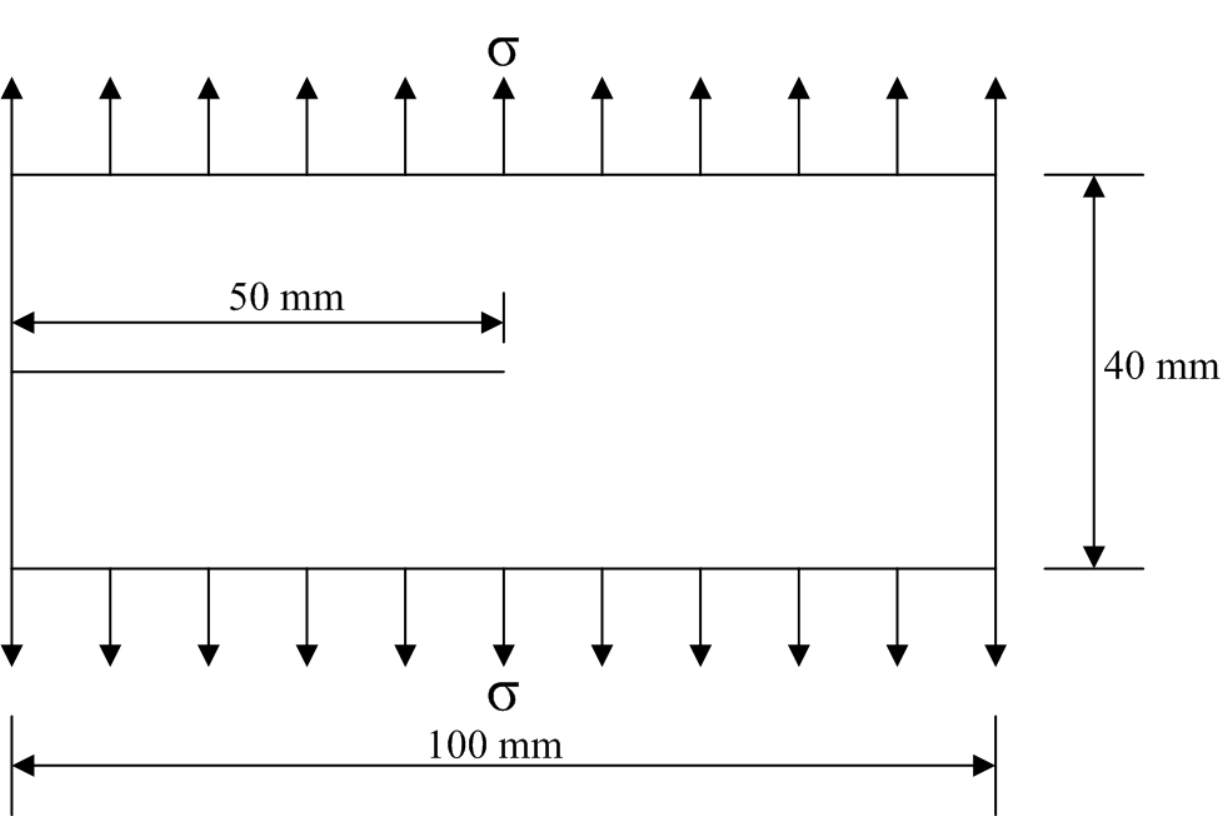}
\caption{Setup of pre-notched glass plate under tensile loading}\label{prenotch}
\end{figure}

A major advantage of the \textit{pseudo spring} analogy compared to other crack tracking approaches is that it can also capture complex crack path evolutions without requiring any special treatment. In this section, the crack branching phenomenon in a brittle material is modelled. Towards this, a rectangular glass plate of 100 mm length and 40 mm width is considered. A pre-existing notch of 50 mm length is located symmetrically as shown in Figure \ref{prenotch}. The plate is under tensile loading ($\sigma$ = 1 MPa) at the boundaries along the length of the plate. The material parameters used for the present computation are from \cite{song2008comparative, agwai2011predicting}. The material and SPH parameters used for the current simulations are given in Table \ref{notch_tab}. The damage parameter $D$ in the pseudo-springs is obtained as below. 
 
\begin{equation}\label{dam_b}
    D=\begin{cases}
    1,          & \epsilon \ge \epsilon_{max}\\
    0,          &  Otherwise
\end{cases}\\
\end{equation} 

\begin{table}[hbtp!]
\centering
\caption{Parameters for pre-notched plate under tensile loading}\label{notch_tab}
\begin{tabular}{ccccccccc}
\hline
                           & \multicolumn{4}{c}{Material Proerties}                 & \multicolumn{2}{c}{Discretization} & \multicolumn{2}{c}{Artificial Viscosity}                \\
\multirow{2}{*}{Parameter} & $\rho$     & $E$ & \multirow{2}{*}{$\nu$} & $\epsilon_{max}$ & $\Delta p$          & $h$          & \multirow{2}{*}{$\beta_1$} & \multirow{2}{*}{$\beta_2$} \\
                           & ($kg/m^3$) & (GPa) &                        &     & (mm)                & (mm)         &                            &                            \\ 
                           \hline
Value                      & 2450       & 32 & 0.2                    & 0.000509     & 0.125                & 0.250         & 1.0                        & 1.0 \\  
\hline                    
\end{tabular}
\end{table}

\begin{figure}[hbtp!]
\centering
\includegraphics[width=0.5\textwidth]{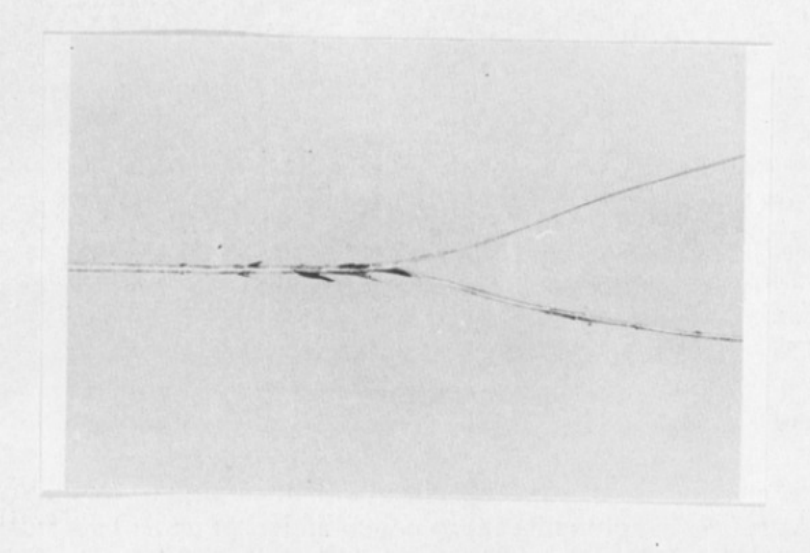}
\caption{Crack propagation and branching : Experimental observatuion \citep{ramulu1985mechanics}}\label{branch_exp}
\end{figure}

\begin{figure}[hbtp!]
\centering
\begin{subfigure}[t]{0.49\textwidth}    %trim={<left> <lower> <right> <upper>}
\includegraphics[width=\textwidth,trim={100 175 100 175}, clip]{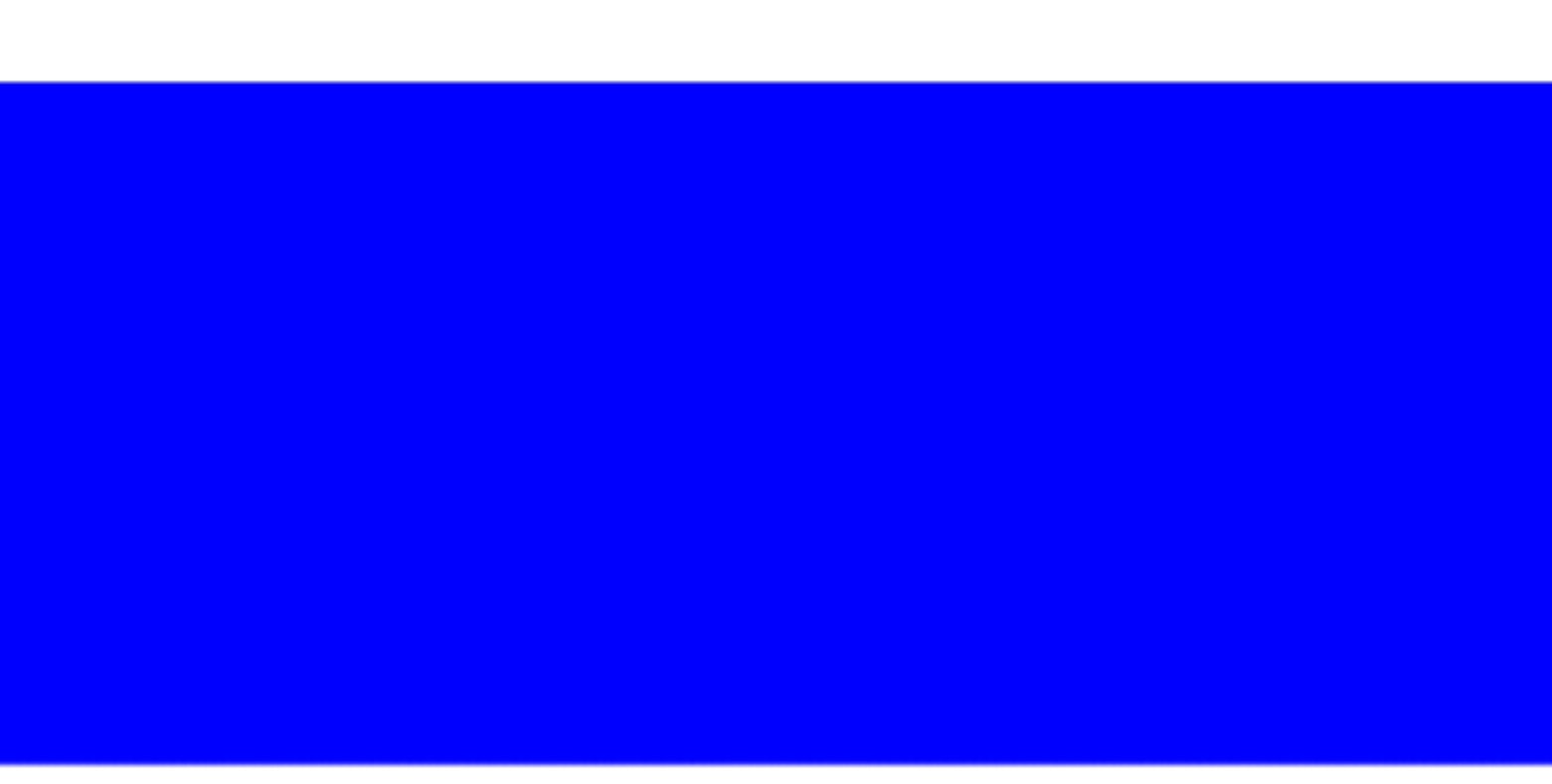}
\caption{Time = 0 $\mu$s} 
\end{subfigure}
\begin{subfigure}[t]{0.49\textwidth}
\includegraphics[width=\textwidth,trim={100 175 100 175}, clip]{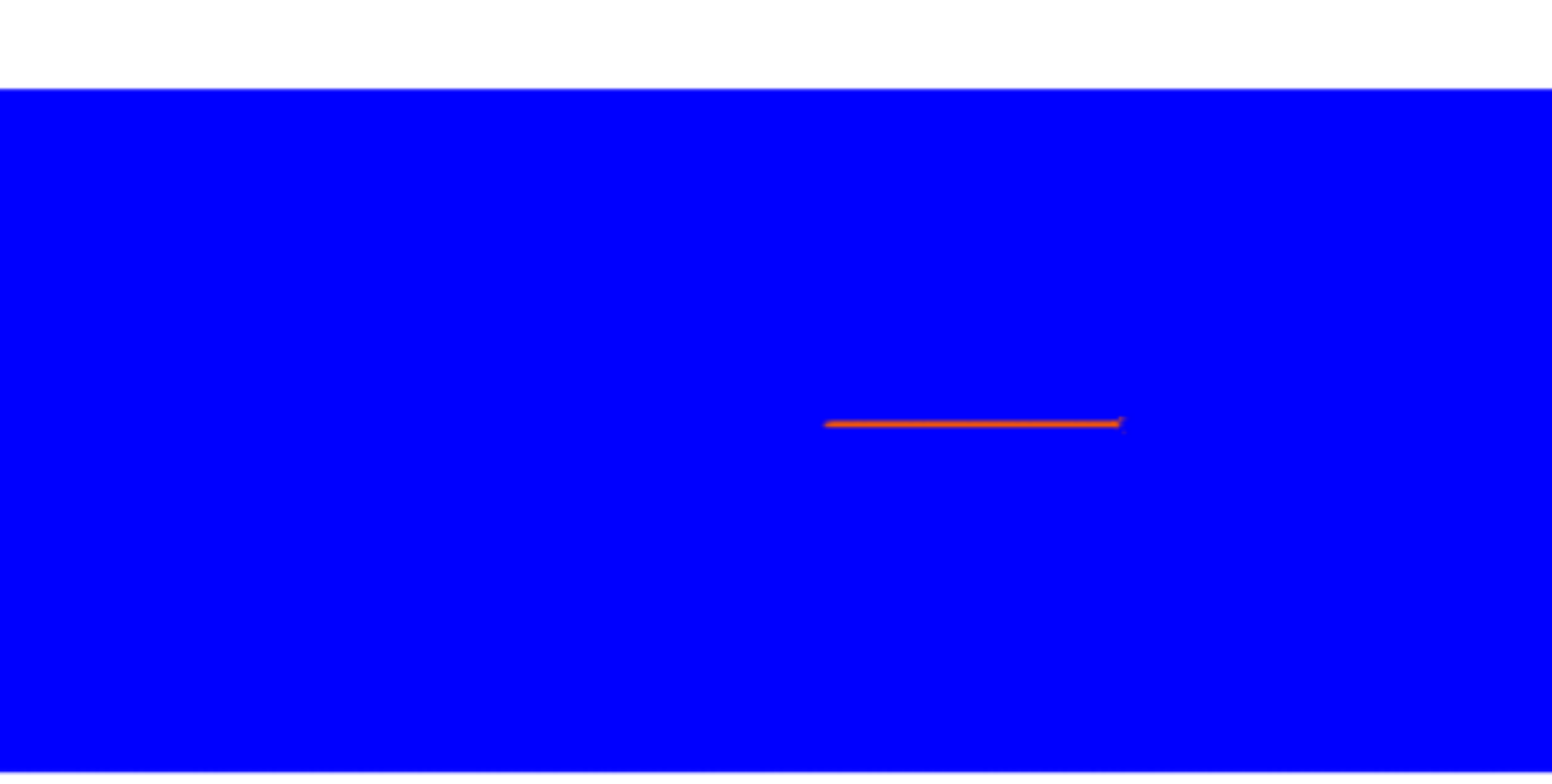}
\caption{Time = 28 $\mu$s} 
\end{subfigure}
\begin{subfigure}[t]{0.49\textwidth}    %trim={<left> <lower> <right> <upper>}
\includegraphics[width=\textwidth,trim={100 175 100 175}, clip]{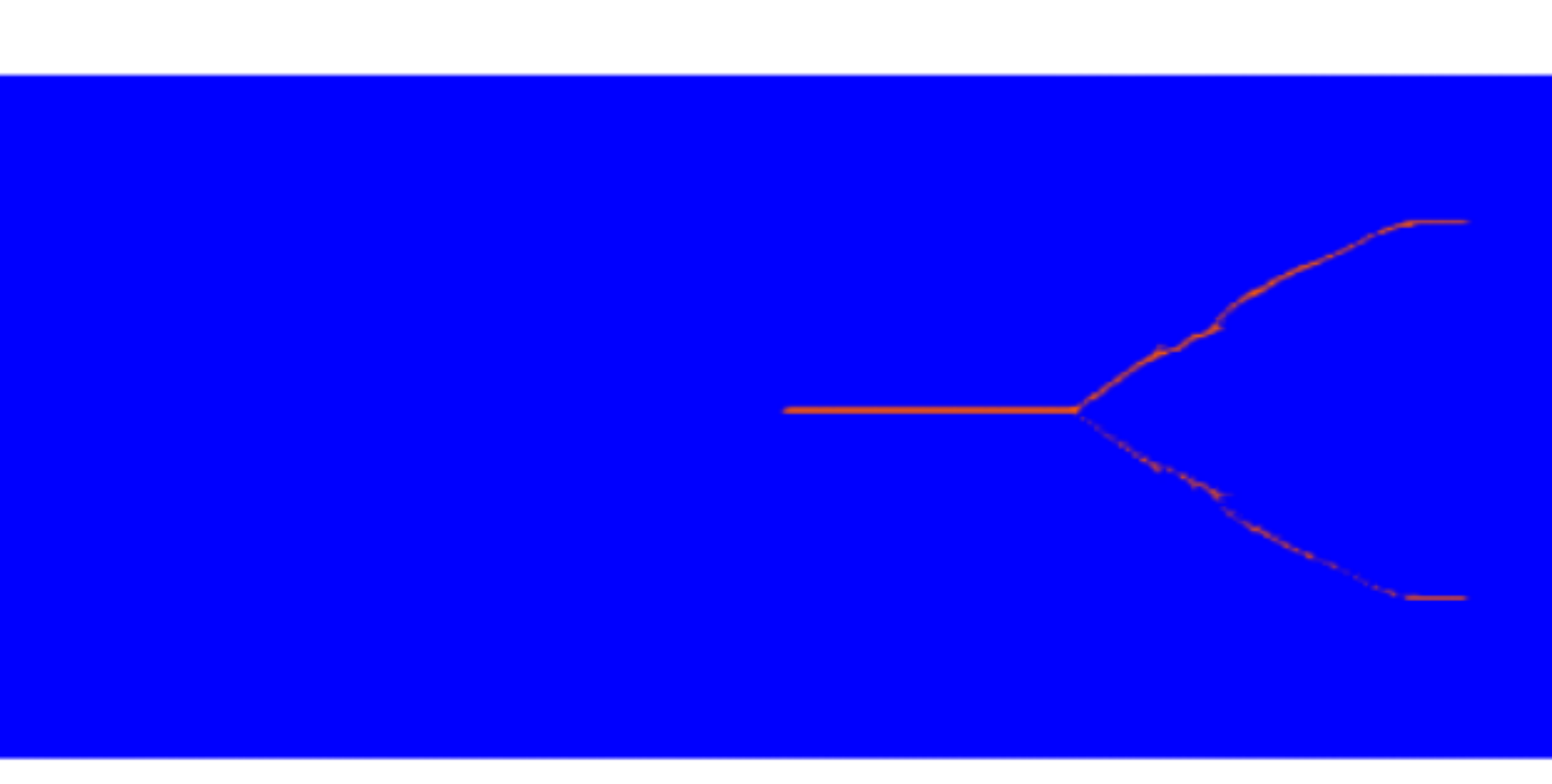}
\caption{Time = 50 $\mu$s} 
\end{subfigure}
\begin{subfigure}[t]{0.49\textwidth}
\includegraphics[width=\textwidth,trim={100 175 100 175}, clip]{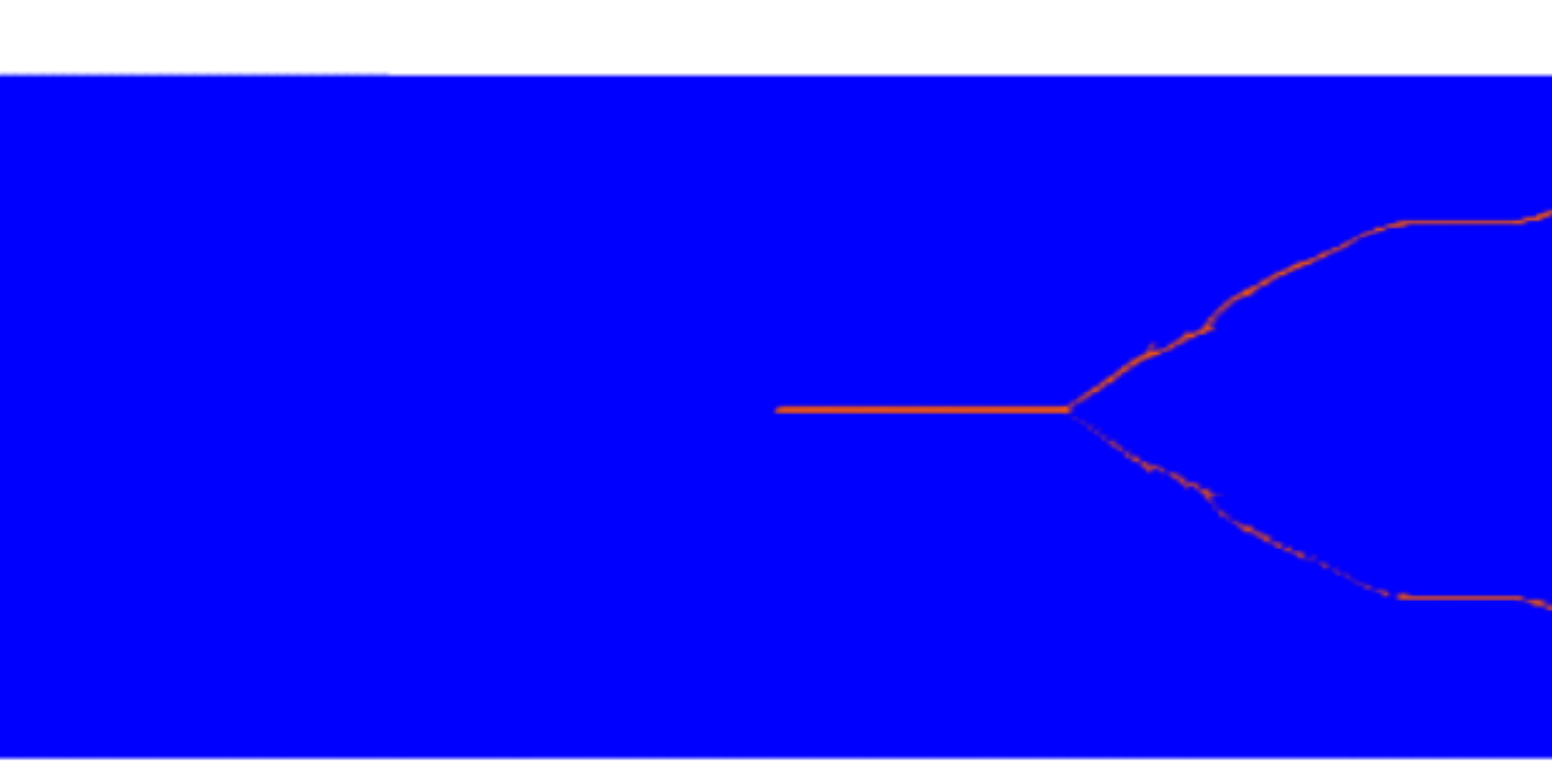}
\caption{Time = 62 $\mu$s} 
\end{subfigure}
\caption{Crack path evoluation over time in pre-notched plate under tensile loading of $\sigma$ = 1 MPa}\label{prenotch_1}
\end{figure}

In the experiment \cite{ramulu1985mechanics}, it is observed that initially the crack propagates in a self-propagating mode and after few microseconds, the branching takes place as shown in Figure \ref{branch_exp}. The predicted crack paths (crack initiation, propagation and branching) as obtained via the present simulations are shown in Figure \ref{prenotch_1}. The crack propagation initiates around 10 $\mu$s and propagates in a straight line up to 27 $\mu$s. The initiation of branching of crack is observed at 28 $\mu$s. The crack propagates in two symmetrical branches and reaches the plate boundary at 62 $\mu$s. 

\begin{figure}[hbtp!]
\centering
\begin{subfigure}[t]{0.49\textwidth}    %trim={<left> <lower> <right> <upper>}
\includegraphics[width=\textwidth]{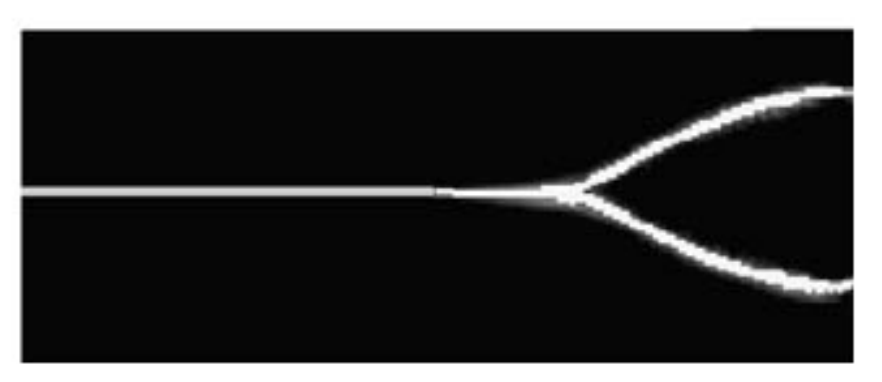}
\caption{} 
\end{subfigure}
\begin{subfigure}[t]{0.49\textwidth}
\includegraphics[width=\textwidth]{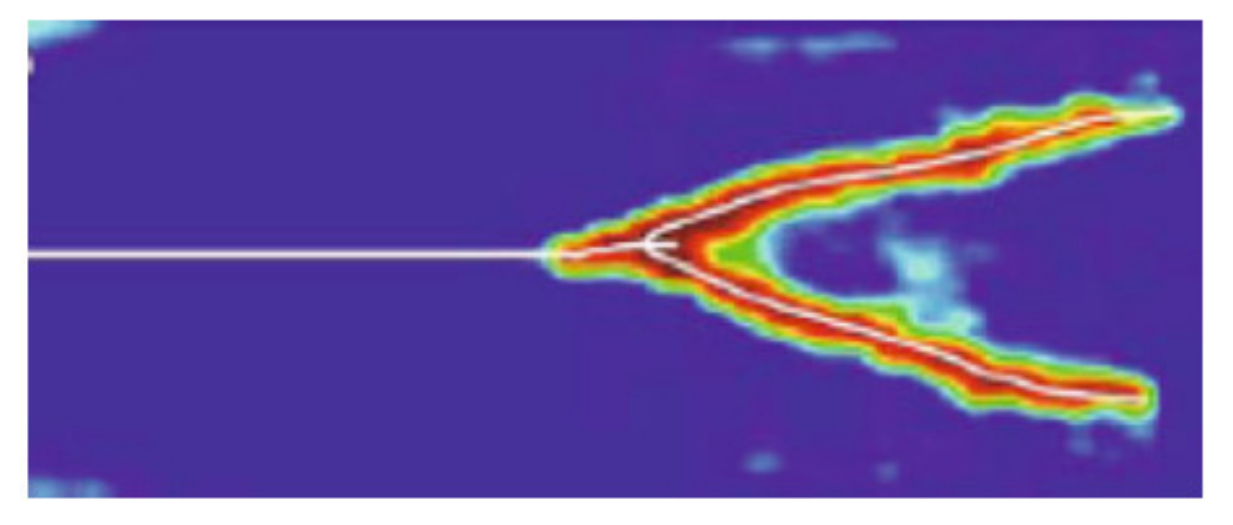}
\caption{} 
\end{subfigure}
\begin{subfigure}[t]{0.49\textwidth}    %trim={<left> <lower> <right> <upper>}
\includegraphics[width=\textwidth]{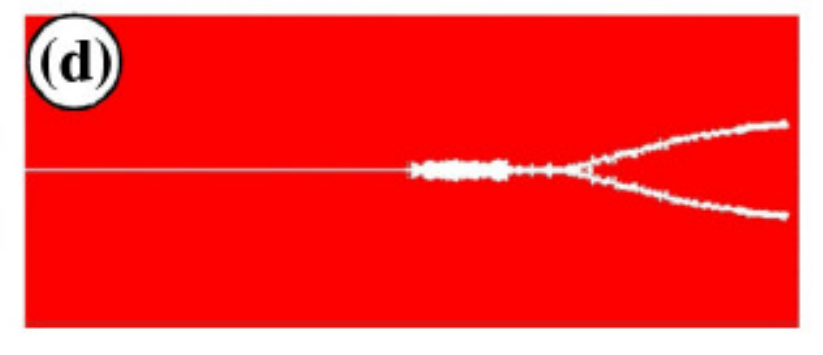}
\caption{} 
\end{subfigure}
\begin{subfigure}[t]{0.49\textwidth}
\includegraphics[width=\textwidth,trim={100 175 100 175}, clip]{prenotch_62micros_1MPa.pdf}
\caption{} 
\end{subfigure}
\caption{Comparison of final crack path under tensile loading of $\sigma$ = 1 MPa: (a) \cite{rabczuk2007meshfree} (b) \cite{song2008comparative} (c) \cite{braun2014new} and (d) Present}\label{prenotch_com}
\end{figure}
 
The crack branching problem is studied in \cite{xu1994numerical, belytschko2003dynamic, rabczuk2004cracking, agwai2011predicting, braun2014new}. The final crack path obtained from the present simulation is compared with other numerical predictions available in the literature as shown in Figure \ref{prenotch_com}. The crack propagation speed over time is compared with different techniques. The present results compare reasonably well with the results obtained in \cite{agwai2011predicting} using peridynamics, \cite{song2008comparative} using XFEM-CZM, and \cite{song2009cracking} using XFEM-CNM. The crack propagation speed increases after the crack initiation and decreases after the crack branching. All the predicted phenomena of the crack branching problem can be captured using the present pseudo-spring SPH framework, and the present predictions are comparable with the experimental and other numerical observations.   

\begin{figure}[hbtp!]
\centering
\includegraphics[width=0.7\textwidth]{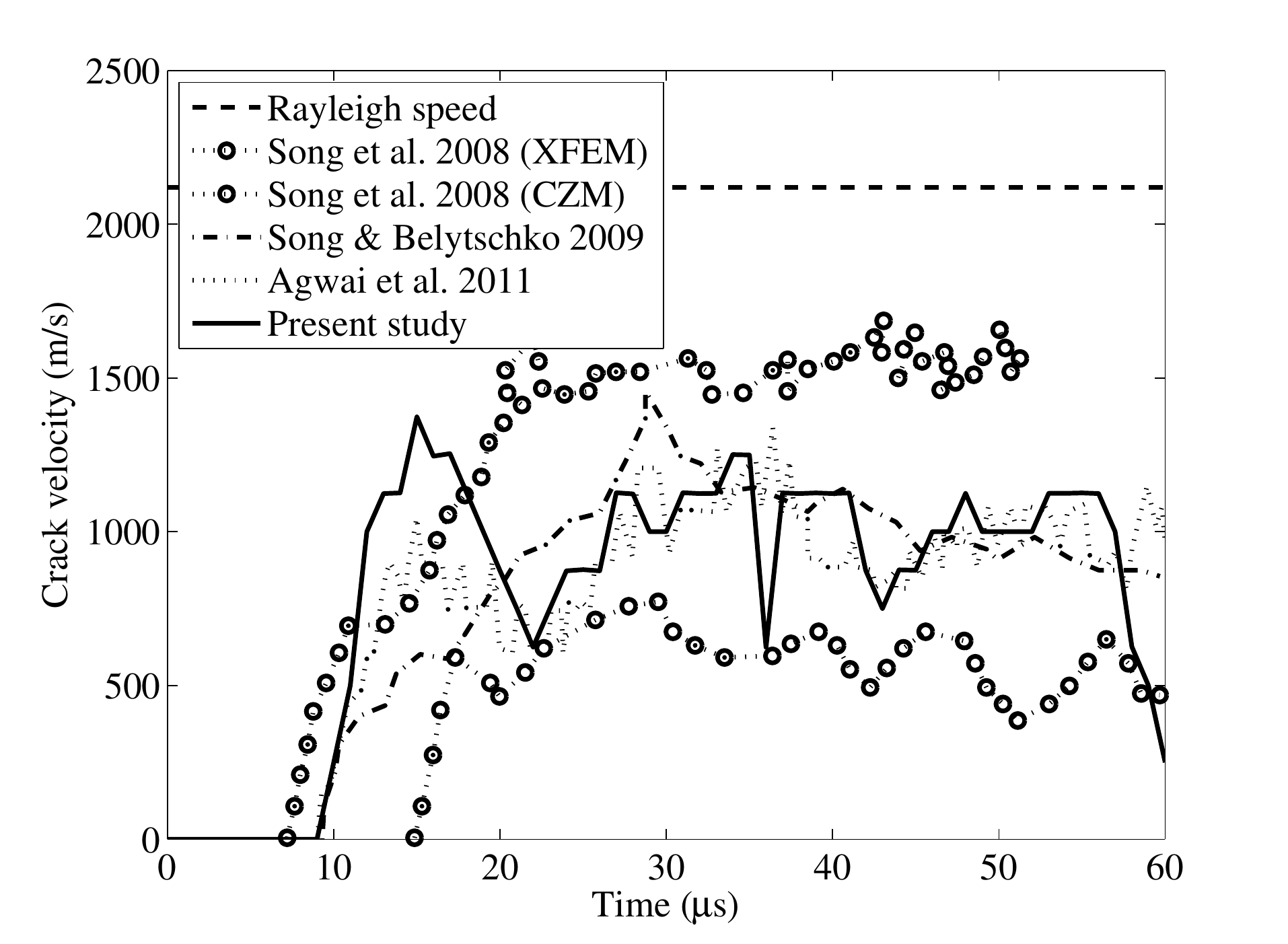}
\caption{Comparison of crack speed over time}\label{prenotch_speed_comp}
\end{figure}

\begin{figure}[hbtp!]
\centering
\begin{subfigure}[t]{0.49\textwidth}    %trim={<left> <lower> <right> <upper>}
\includegraphics[width=\textwidth,trim={100 175 100 175}, clip]{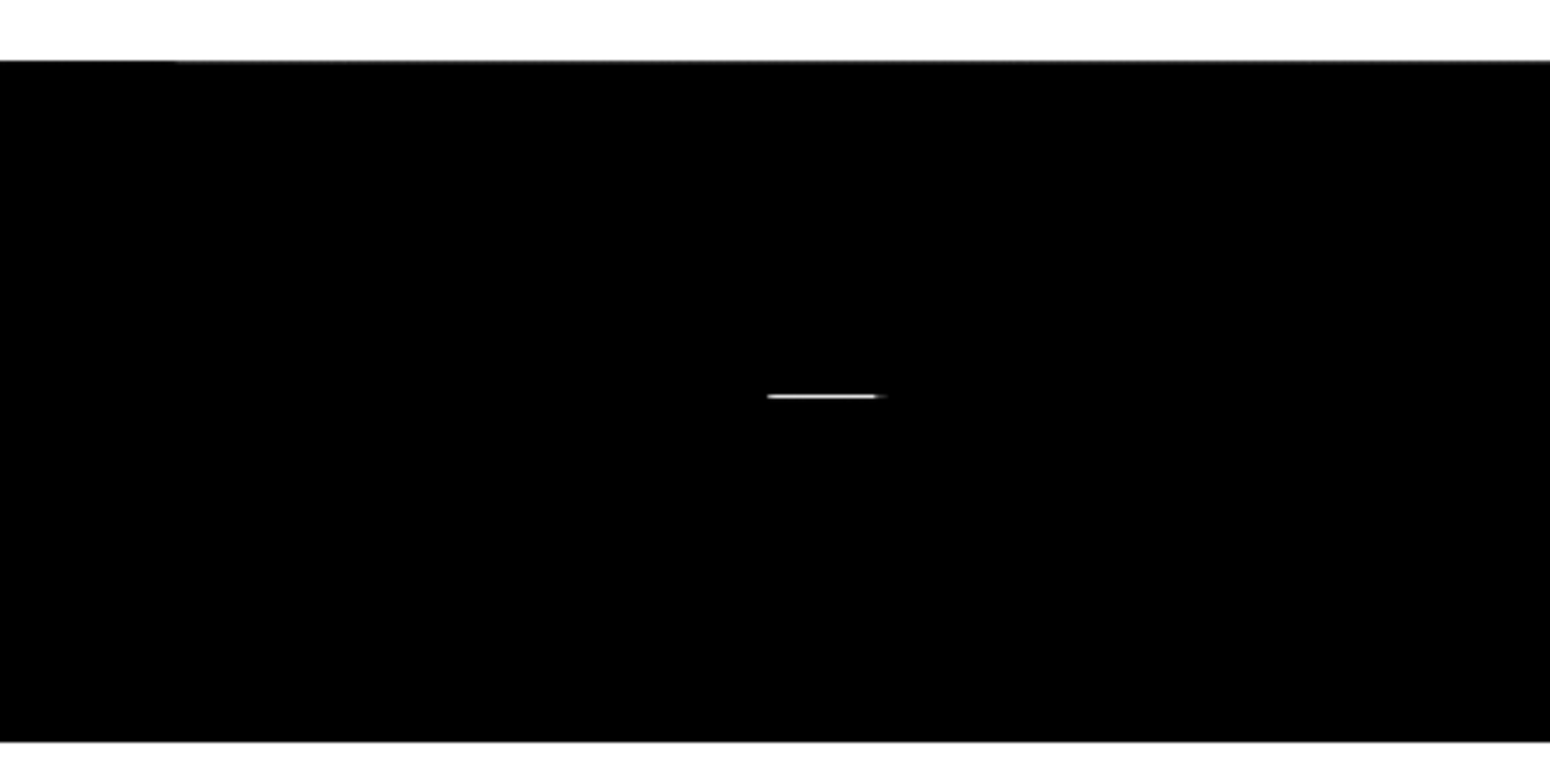}
\caption{Time = 60 $\mu$s} 
\end{subfigure}
\begin{subfigure}[t]{0.49\textwidth}    %trim={<left> <lower> <right> <upper>}
\includegraphics[width=\textwidth,trim={100 175 100 175}, clip]{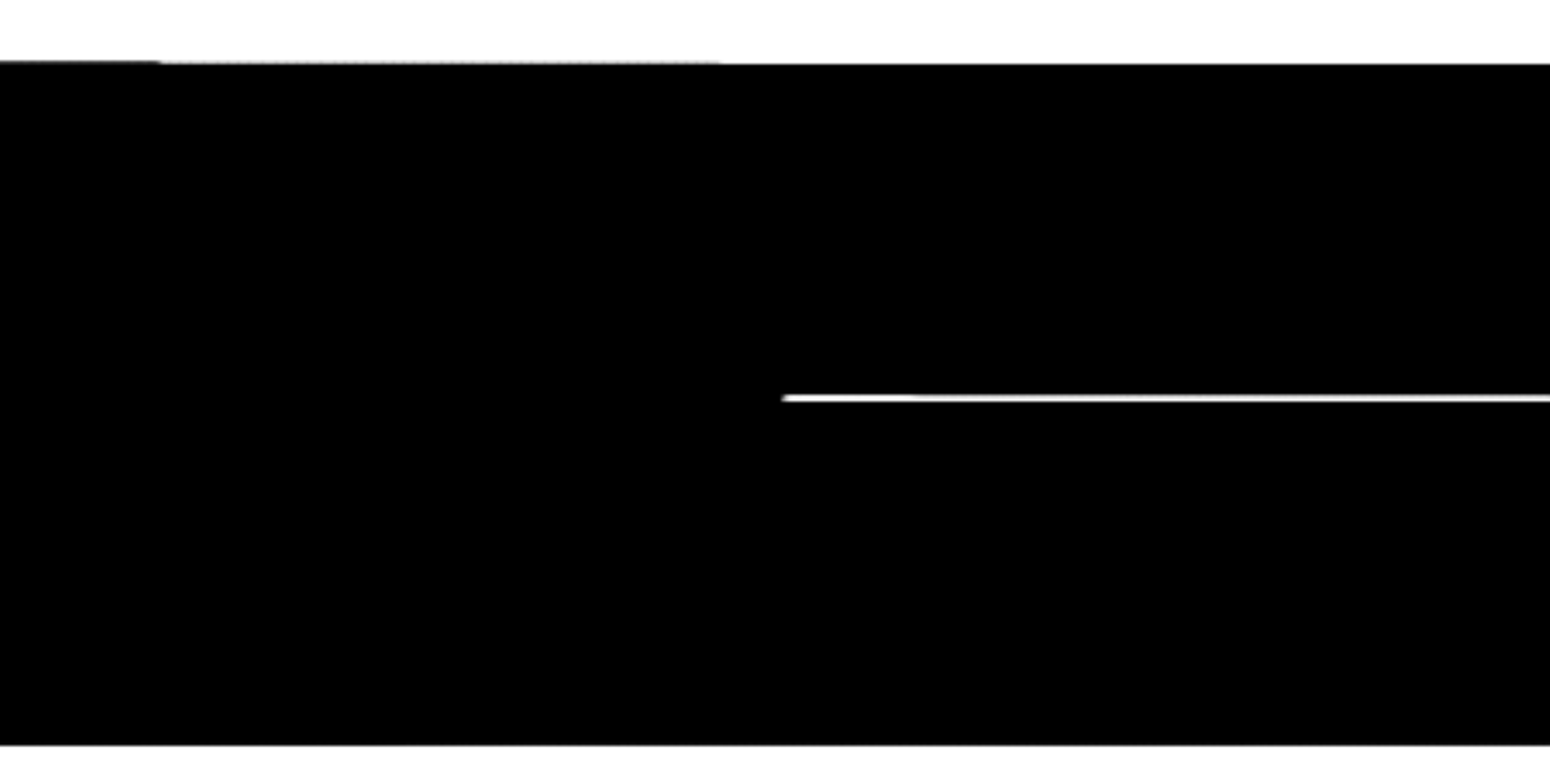}
\caption{Time = 135 $\mu$s} 
\end{subfigure}
\caption{Crack path evoluation over time in pre-notched plate under tensile loading of $\sigma$ = 0.3 MPa}\label{prenotch_03}
\end{figure}

\paragraph{Effect of load intensity}\mbox{} \\
Next, the effect of different tensile loading ($\sigma$) intensities on branching behaviour is explored. $\sigma$ values of 0.3, 1.1, 1.2, 2.0 and 4.0 MPa are considered for the simulations. All the other parameters are kept unchanged. It can be observed that the branching phenomenon depends on the loading ($\sigma$) intensity. For, $\sigma$ = 0.3 MPa loading, no branching is observed, and the crack propagates in a straight line as shown in Figure \ref{prenotch_03}.

\begin{figure}[hbtp!]
\centering
\begin{subfigure}[t]{0.49\textwidth}    %trim={<left> <lower> <right> <upper>}
\includegraphics[width=\textwidth,trim={100 140 100 175}, clip]{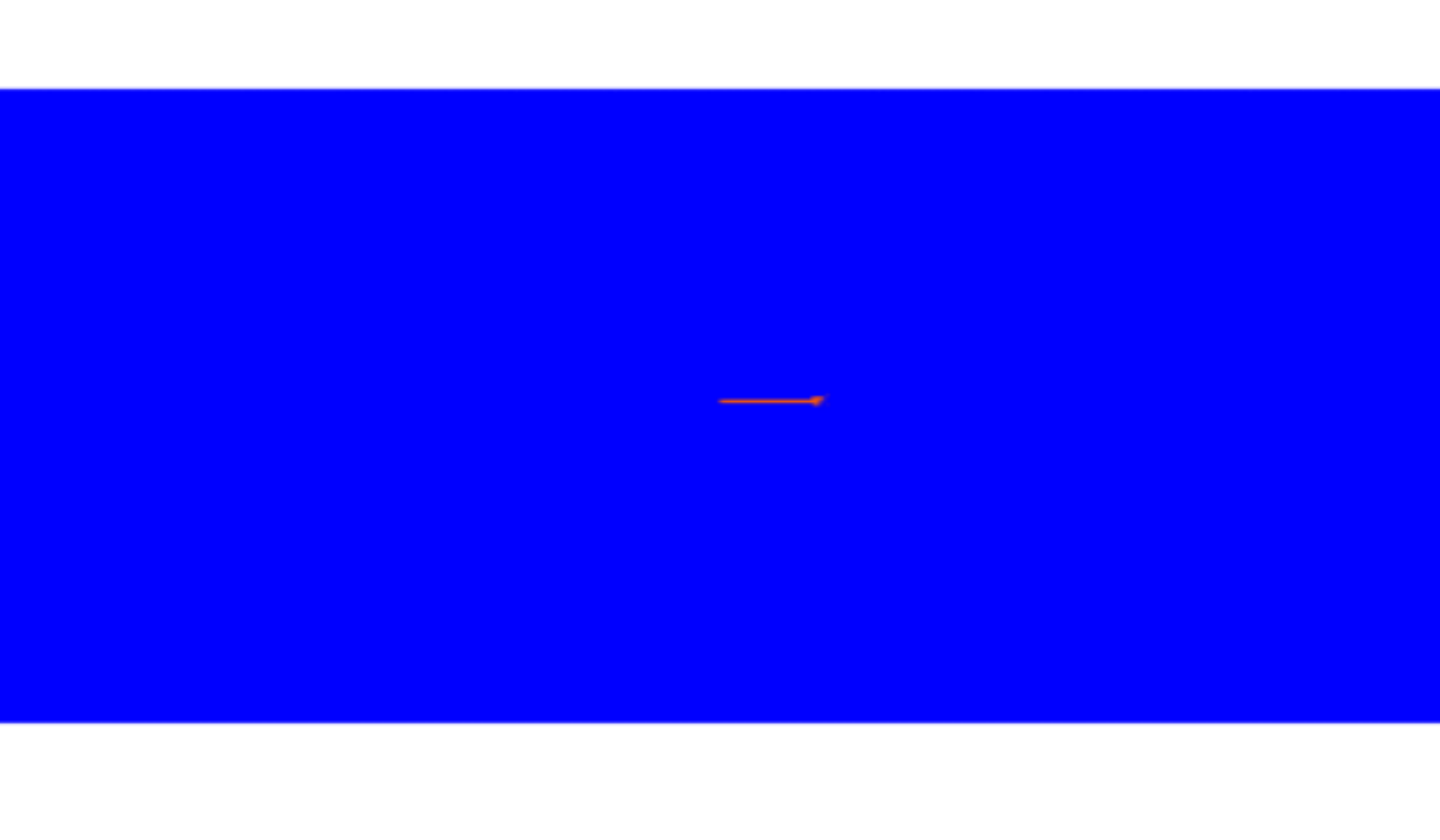}
\caption{Time = 15 $\mu$s} 
\end{subfigure}
\begin{subfigure}[t]{0.49\textwidth}    %trim={<left> <lower> <right> <upper>}
\includegraphics[width=\textwidth,trim={100 25 100 125}, clip]{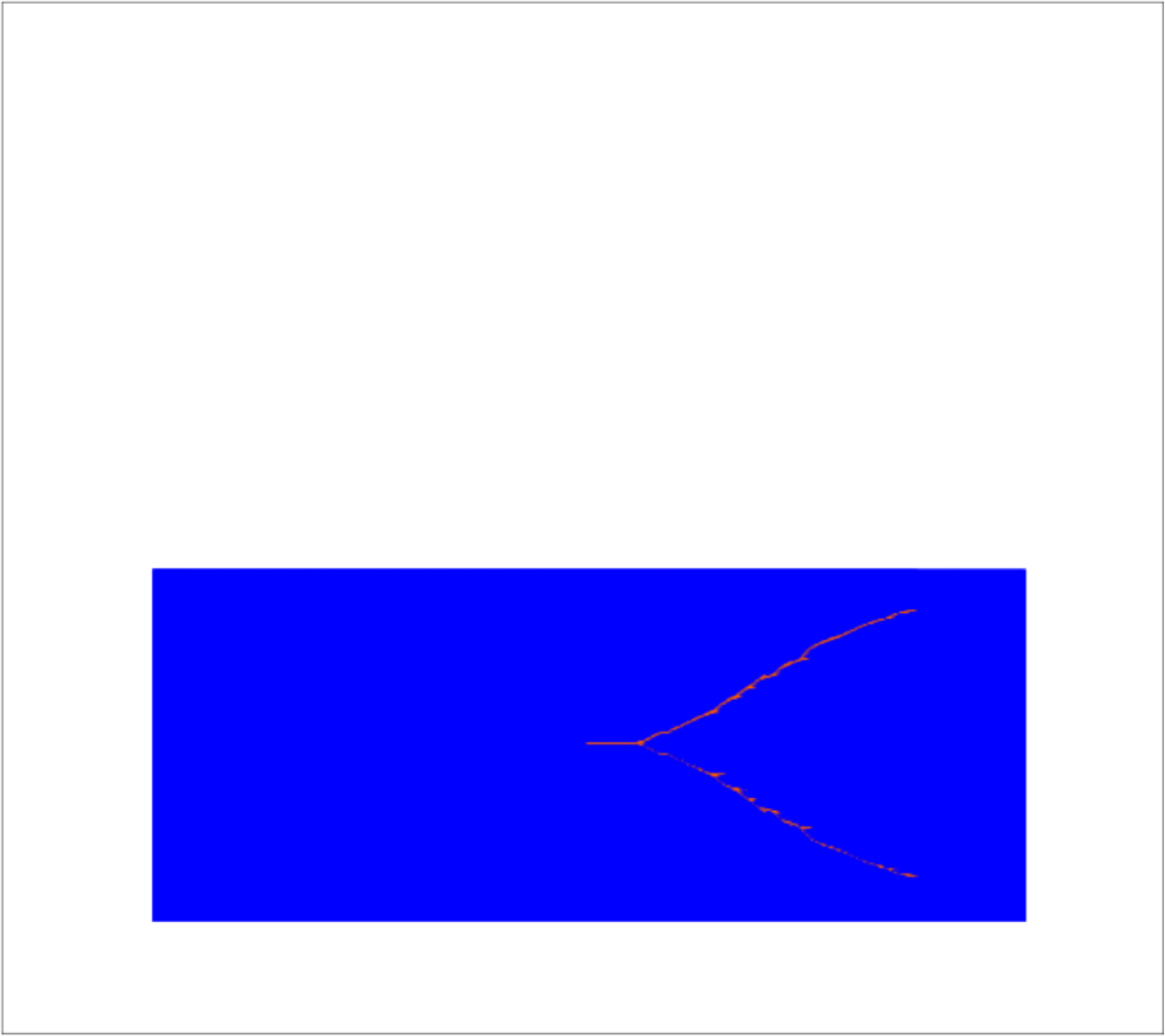}
\caption{Time = 50 $\mu$s} 
\end{subfigure}
\caption{Crack path evoluation over time in pre-notched plate under tensile loading of $\sigma$ = 1.1 MPa}\label{prenotch_11}
\end{figure}

\begin{figure}[hbtp!]
\centering
\begin{subfigure}[t]{0.49\textwidth}    %trim={<left> <lower> <right> <upper>}
\includegraphics[width=\textwidth,trim={100 125 100 175}, clip]{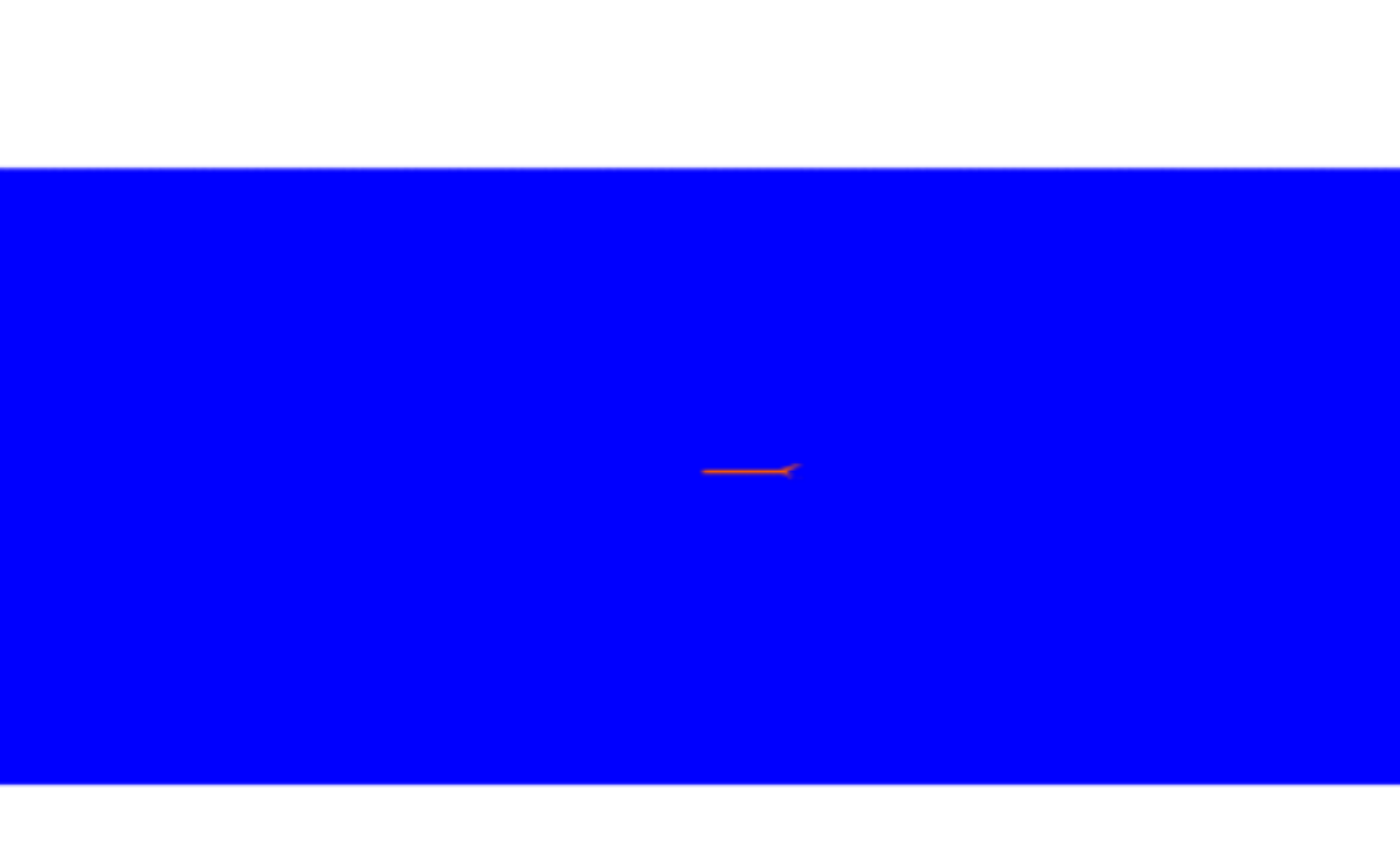}
\caption{Time = 14 $\mu$s} 
\end{subfigure}
\begin{subfigure}[t]{0.49\textwidth}    %trim={<left> <lower> <right> <upper>}
\includegraphics[width=\textwidth,trim={100 125 100 125}, clip]{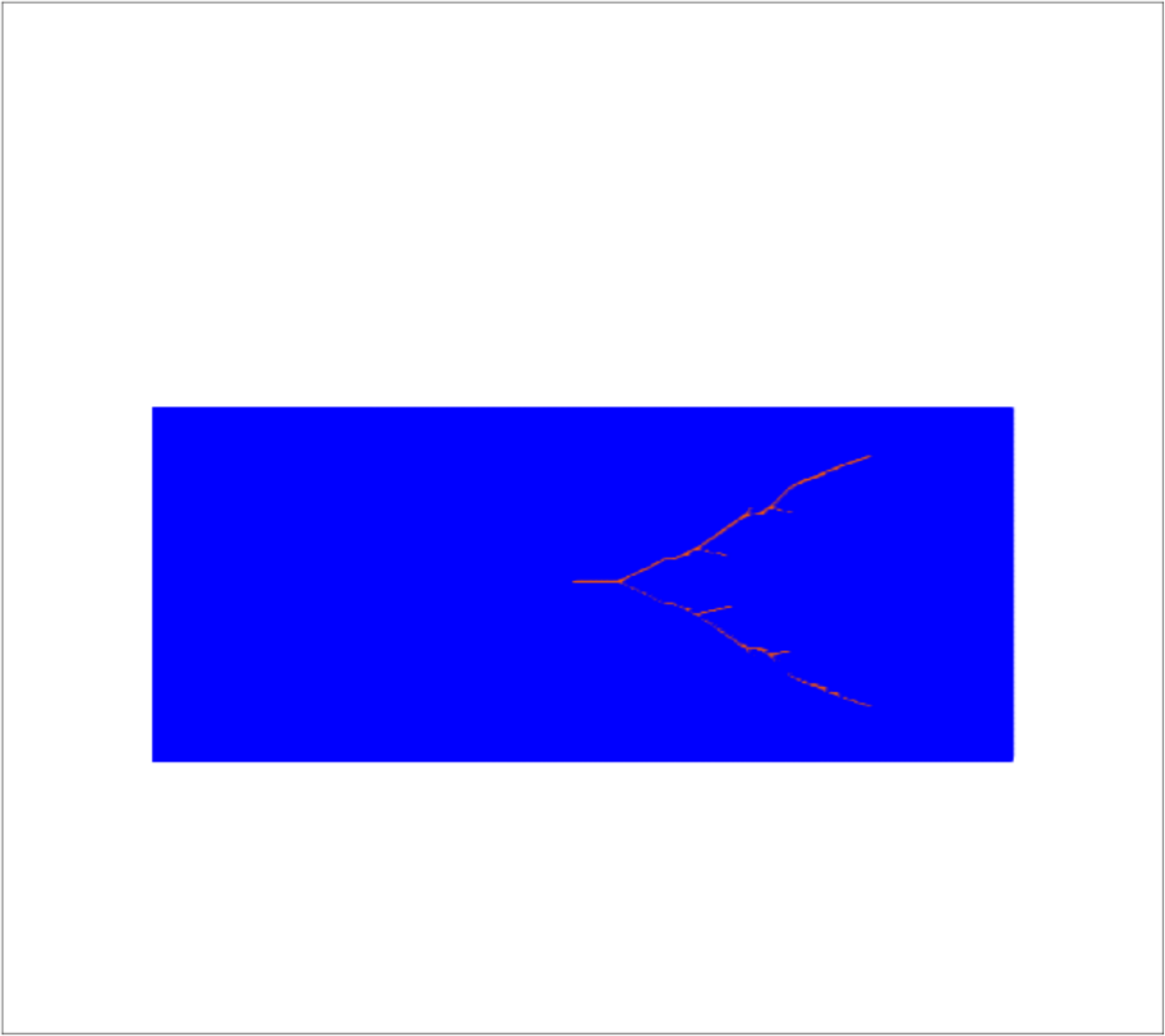}
\caption{Time = 50 $\mu$s} 
\end{subfigure}
\caption{Crack path evoluation over time in pre-notched plate under tensile loading of $\sigma$ = 1.2 MPa}\label{prenotch_12}
\end{figure}

\begin{figure}[hbtp!]
\centering
\begin{subfigure}[t]{0.49\textwidth}    %trim={<left> <lower> <right> <upper>}
\includegraphics[width=\textwidth,trim={100 175 100 175}, clip]{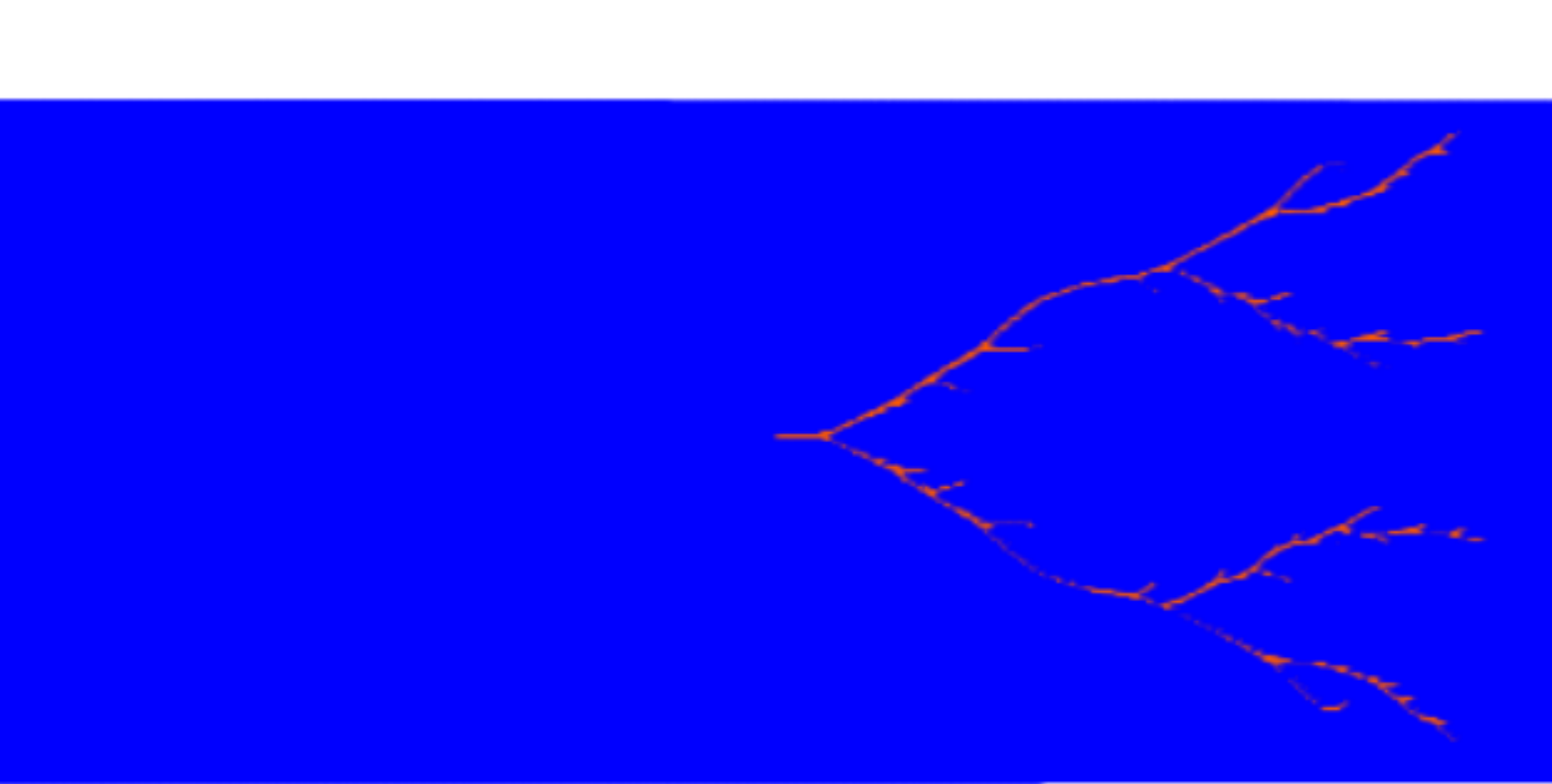}
\caption{$\sigma$ = 2.0 MPa at time = 40 $\mu$s} 
\end{subfigure}
\begin{subfigure}[t]{0.49\textwidth}
\includegraphics[width=\textwidth,trim={100 175 100 175}, clip]{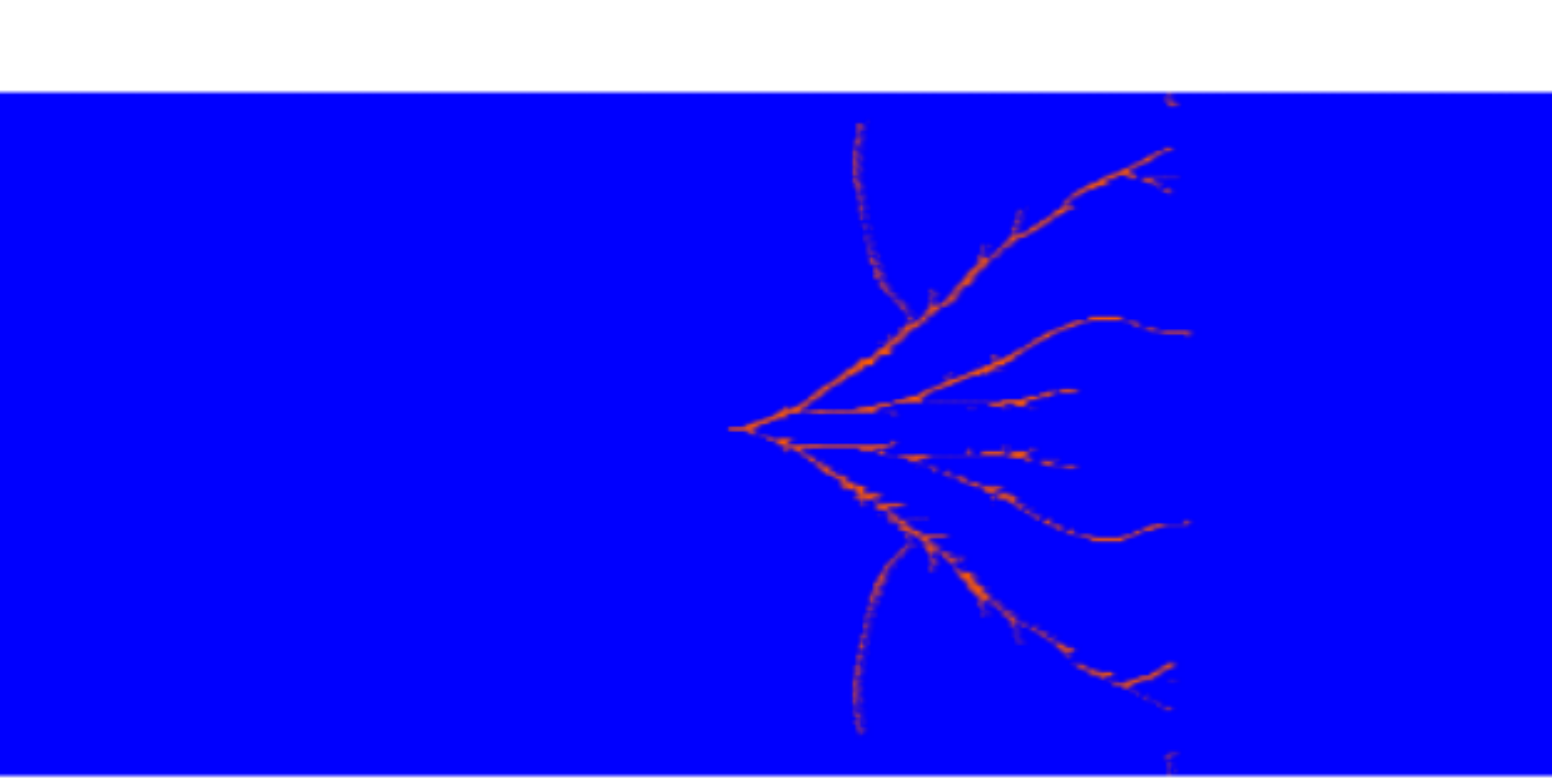}
\caption{$\sigma$ = 4.0 MPa at time = 25 $\mu$s} 
\end{subfigure}
\caption{Crack path evoluation over time in pre-notched plate under different magnitude of tensile loading}\label{prenotch_24}
\end{figure}

But the branching can be observed for all the other load intensities. The crack branching starts around 15 $\mu$s for $\sigma$ = 1.1 MPa (Figure \ref{prenotch_11}), 14 $\mu$s for $\sigma$ = 1.2 MPa (Figure \ref{prenotch_12}), 9 $\mu$s for $\sigma$ = 2.0 MPa and 6 $\mu$s for $\sigma$ = 4.0 MPa (Figure \ref{prenotch_24}). As the loading amplitude ($\sigma$) is increased, the point where crack branching takes place, moves closer to the initial pre-notch position. The increase in loading amplitudes also results in secondary crack branches in the glass plates. No secondary crack branch is observed for $\sigma$ = 1.1 MPa but the secondary branches are found for $\sigma$ = 1.2, 2.0 and 4.0 MPa. The number of secondary branches also increases with the increase in loading amplitudes. These observations are consistent with \cite{zhou2016numerical}.  

Crack branching or splitting of cracks into multiple paths are one of the features of brittle fracture. The brittle materials undergoing fracture have a process zone \citep{ravi1984experimental} and the microcracks in this zone undergo nucleation, growth, and coalescence. These lead to crack path instabilities \citep{ravi1997role}. The crack branching occurs possibly due to the interaction between the reflected wave from the boundaries and the crack tip (waves generated due to crack propagation) or due to the pile-up of waves in the process zone \citep{bobaru2015cracks}. This leads to the multiple branches of crack paths for the higher amplitude of boundary stress. 

\subsubsection{Pre-notched plate with a circular hole}
\begin{figure}[hbtp!]
\centering
\includegraphics[width=0.4\textwidth]{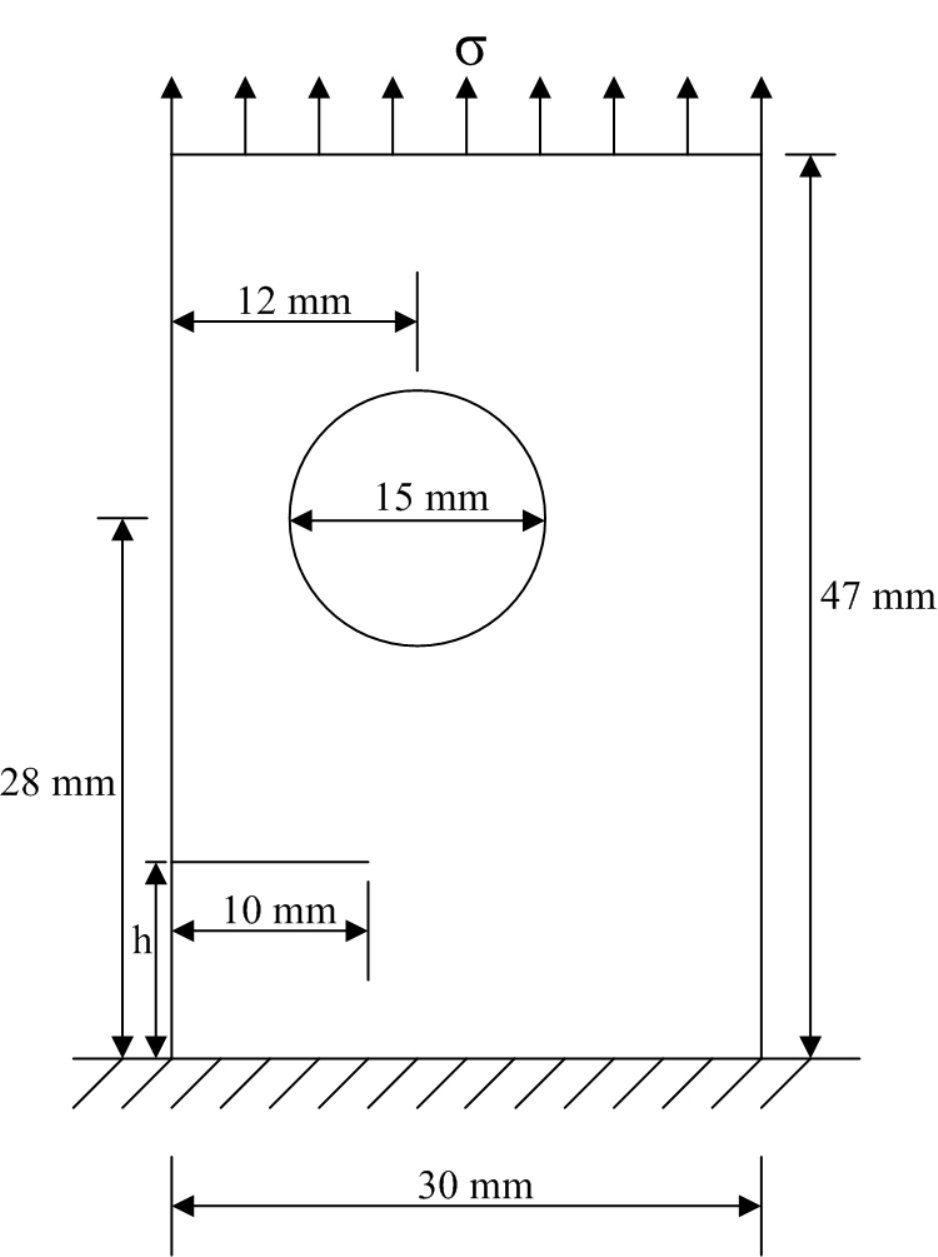}
\caption{Setup of pre-notched plate with a cirular hole under tensile loading}\label{circular_hole}
\end{figure}

In this section, the present framework is applied to model the crack path for a plate with a $10$ mm notch and an off-centre circular hole under tensile loading as shown in Figure \ref{circular_hole}. One end of the plate is clamped, and another end is under uniform tensile stress $\sigma$. This numerical study aims to simulate the crack propagation originating from the pre-existing notch and to track the changes in crack path due to the presence of the off-centre circular hole. The problem is studied in \cite{rashid1998arbitrary} by the arbitrary local mesh replacement method (a finite element based strategy). Tabiei and Wu \cite{tabiei2003development} used DYNA3D and Kosteski et al. \cite{kosteski2012crack} used truss-like discrete element method to model the crack paths. Dipasquale et al. \cite{dipasquale2014crack} used adaptive grid refinement in 2D peridynamics to track the crack paths. Here, the present framework is used to track the crack paths. The material and SPH parameters for the simulation are given in Table \ref{prenotch_tab}. The damage parameter $D$ is obtained at any particle from Equation \ref{dam_b}.

\begin{table}[hbtp!]
\centering
\caption{Parameters for pre-notched plate with a circular hole under tensile stress}\label{prenotch_tab}
\begin{tabular}{ccccccccc}
\hline
                           & \multicolumn{4}{c}{Material Proerties}                 & \multicolumn{2}{c}{Discretization} & \multicolumn{2}{c}{Artificial Viscosity}                \\
\multirow{2}{*}{Parameter} & $\rho$     & $E$ & \multirow{2}{*}{$\nu$} & $\epsilon_{max}$ & $\Delta p$          & $h$          & \multirow{2}{*}{$\beta_1$} & \multirow{2}{*}{$\beta_2$} \\
                           & ($kg/m^3$) & (GPa) &                        &     & (mm)                & (mm)         &                            &                            \\ 
                           \hline
Value                      & 2700       & 71.4 & 0.25                    & 0.00483     & 0.1                & 0.15         & 1.0                        & 1.0 \\  
\hline                    
\end{tabular}
\end{table}

\begin{figure}[hbtp!]
\centering
\begin{subfigure}[t]{0.77\textwidth}    %trim={<left> <lower> <right> <upper>}
\includegraphics[width=\textwidth]{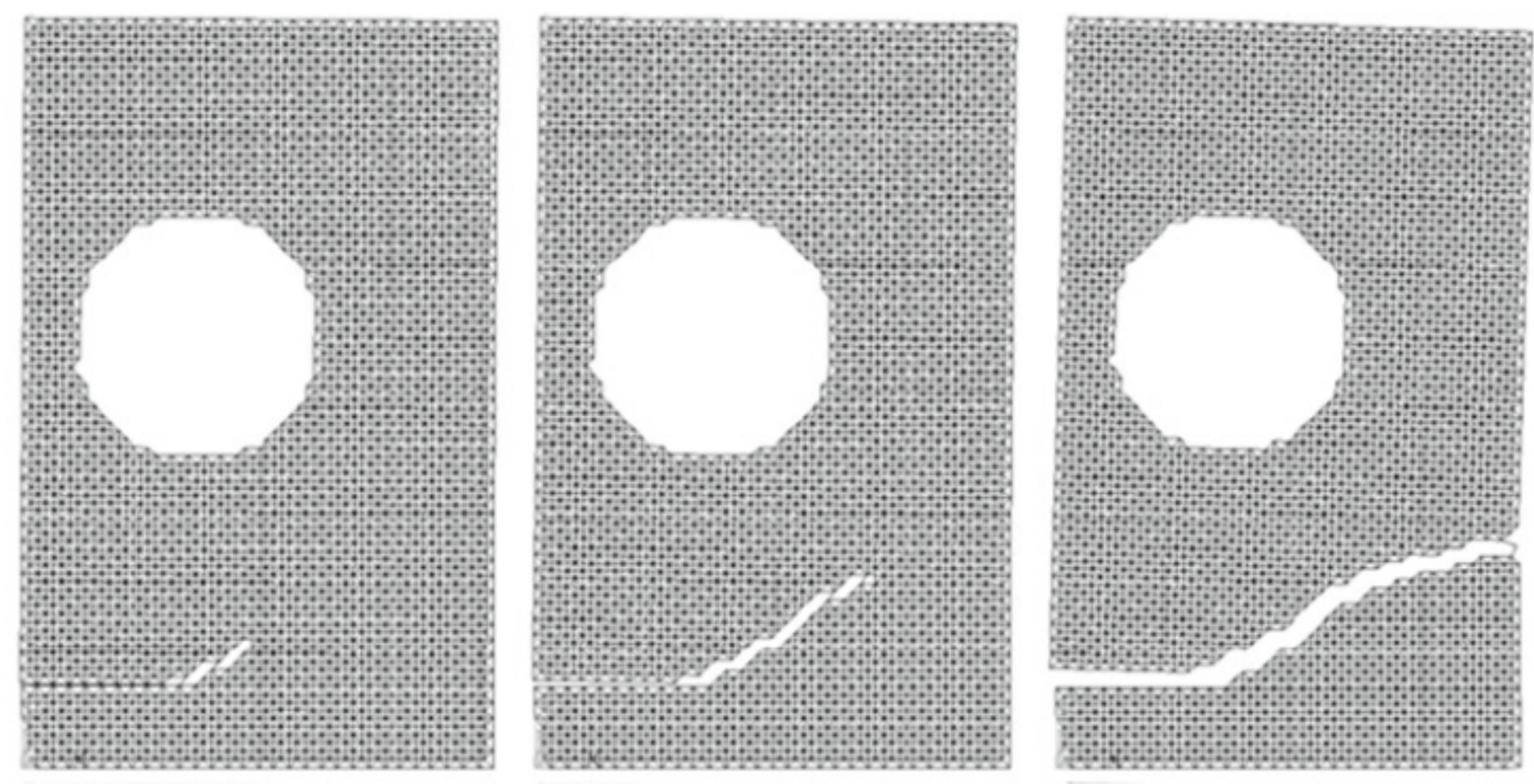}
\caption{Predicted crack path by \cite{kosteski2012crack}} 
\end{subfigure}
\begin{subfigure}[t]{0.8\textwidth}
\includegraphics[width=\textwidth]{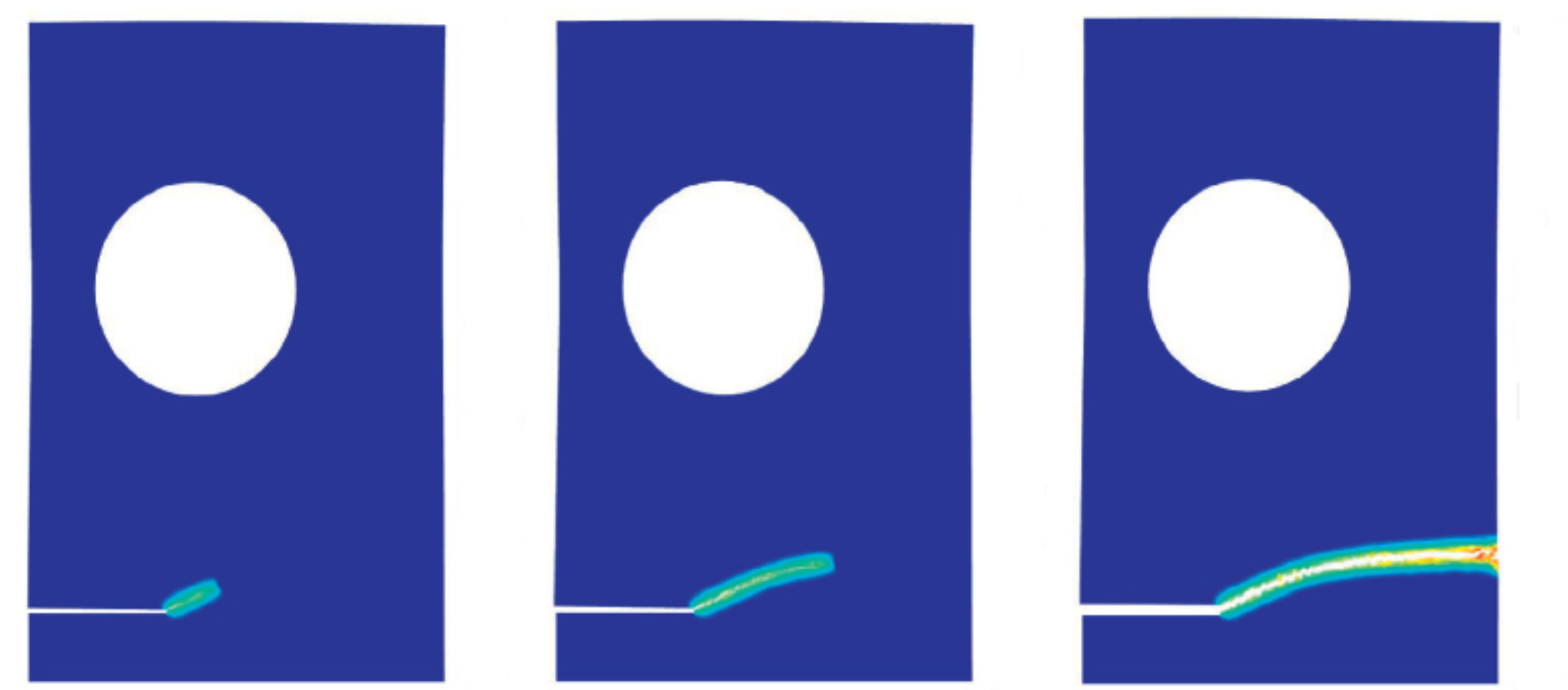}
\caption{Predicted crack path by \cite{ni2017peridynamic}} 
\end{subfigure}
\begin{subfigure}[t]{0.28\textwidth}    %trim={<left> <lower> <right> <upper>}
\includegraphics[width=\textwidth,trim={150 70 150 50}, clip]{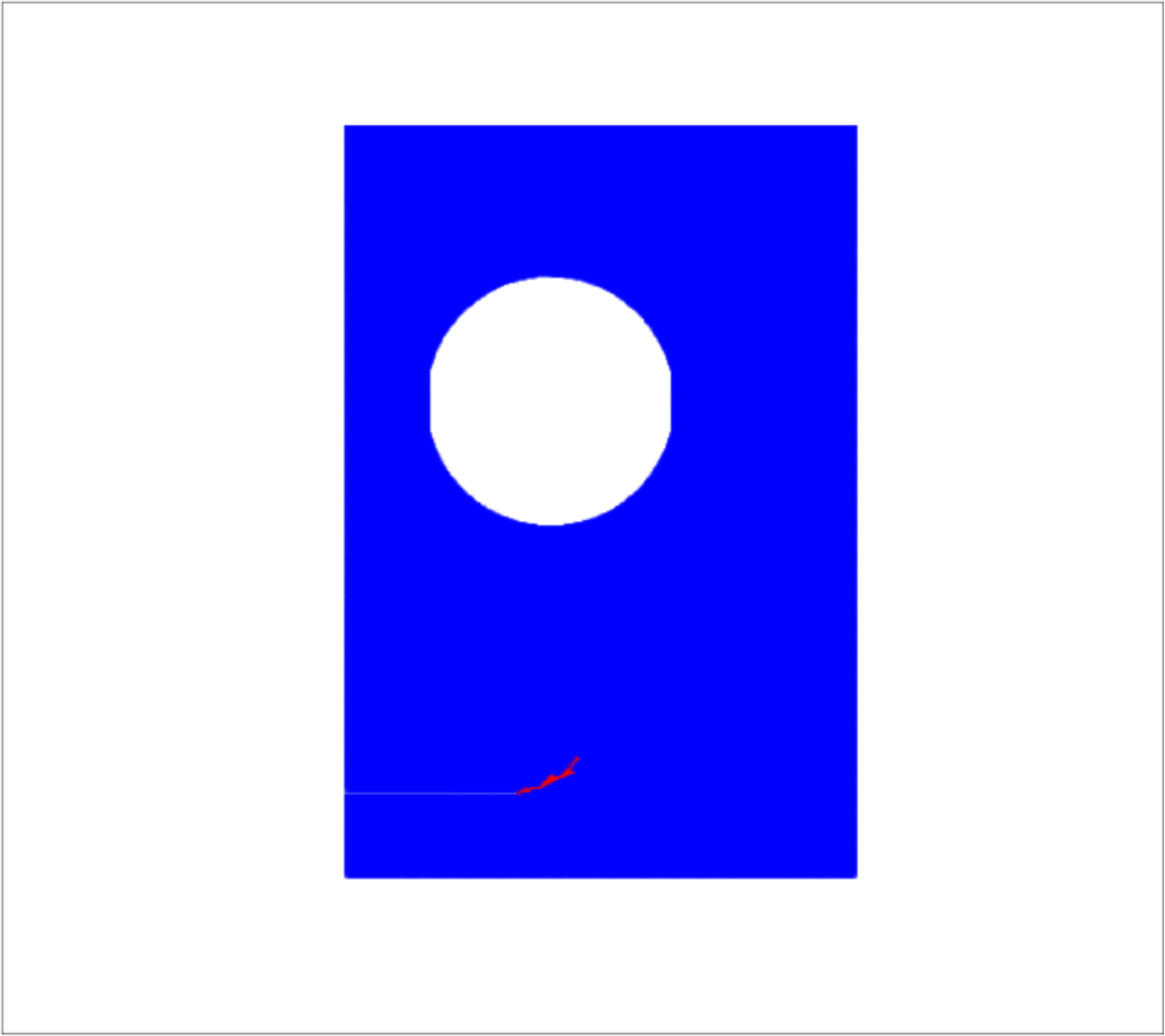}
\end{subfigure}
\begin{subfigure}[t]{0.28\textwidth}    %trim={<left> <lower> <right> <upper>}
\includegraphics[width=\textwidth,trim={35 50 250 50}, clip]{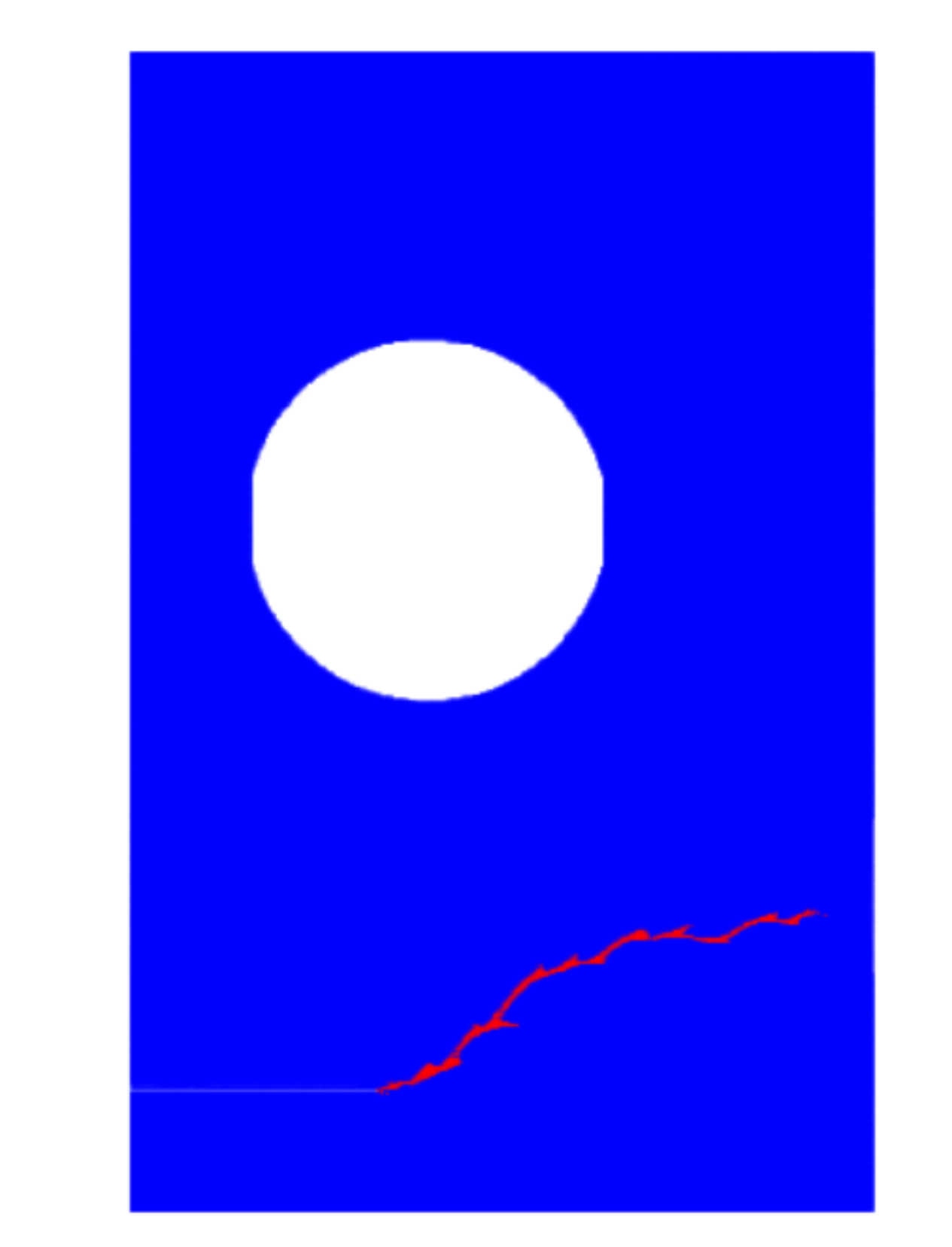}
\caption{Predicted crack path by present framework} 
\end{subfigure}
\begin{subfigure}[t]{0.28\textwidth}    %trim={<left> <lower> <right> <upper>}
\includegraphics[width=0.95\textwidth,trim={150 50 150 50}, clip]{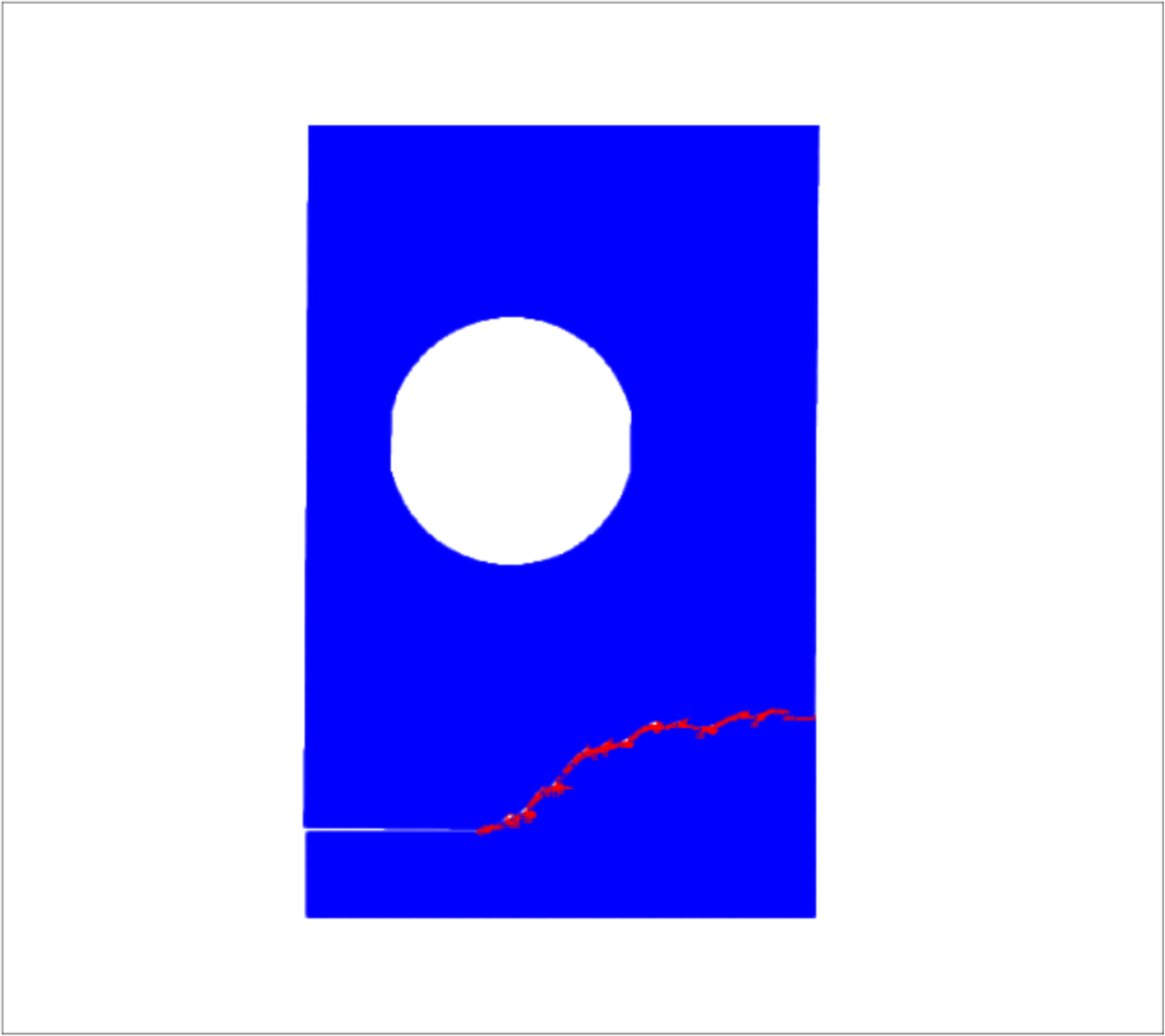}
\end{subfigure}
\caption{Comparision of crack path for notch distance 5 $mm$ from support (case I)}\label{cir_hole_5mm}
\end{figure}

Three different notch locations from the fixed support, named as case I ($h=5$ mm), case II ($h=10$ mm) and case III ($h=15$ mm) \citep{ni2017peridynamic} are considered. Figure \ref{cir_hole_5mm} shows the crack evolution for case I. The crack path does not interact with the circular hole, and the crack propagates from the notch to the other end without going near the circular hole.  

\begin{figure}[hbtp!]
\centering
\begin{subfigure}[t]{0.75\textwidth}    %trim={<left> <lower> <right> <upper>}
\includegraphics[width=\textwidth]{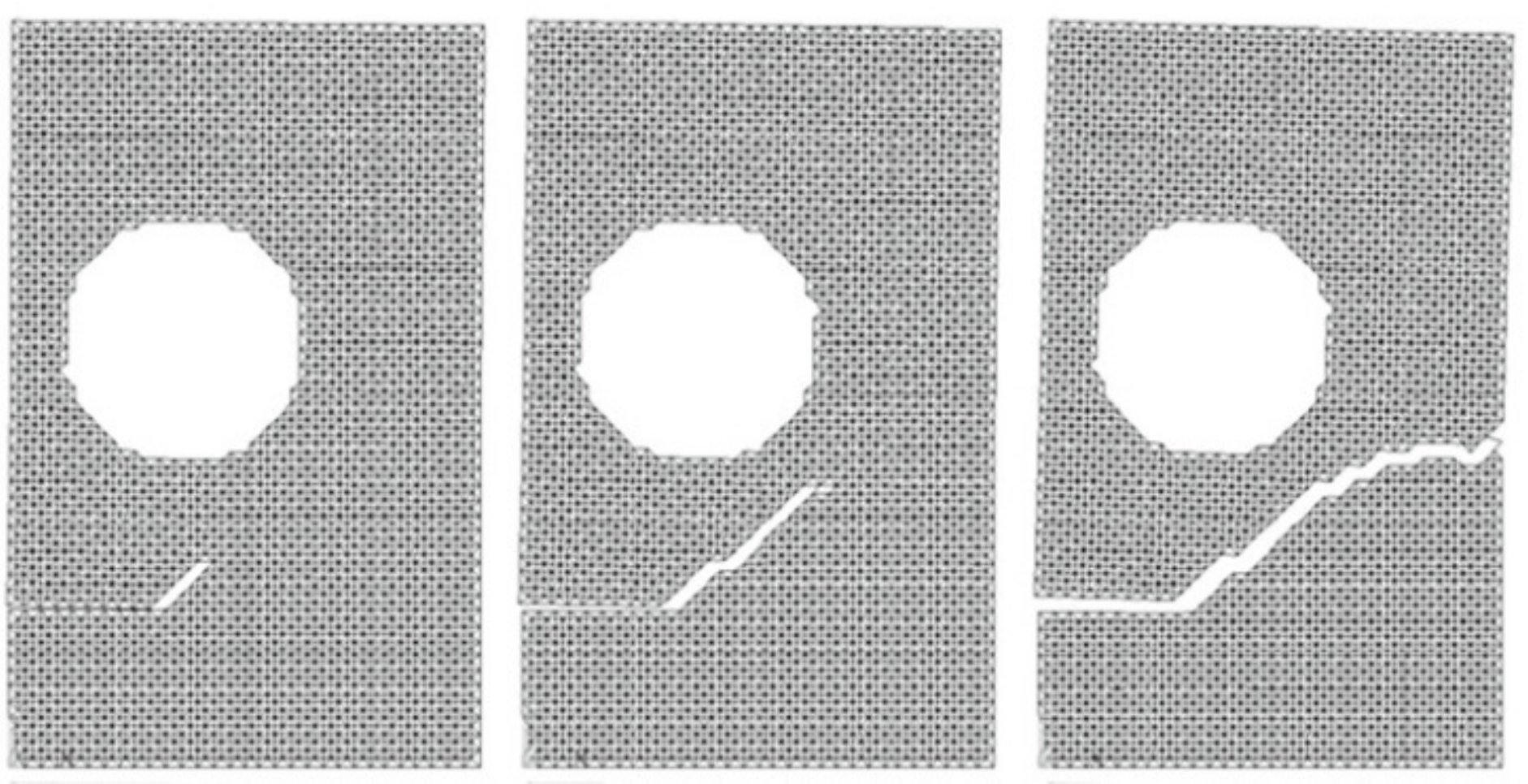}
\caption{Predicted crack path by \cite{kosteski2012crack}} 
\end{subfigure}
\begin{subfigure}[t]{0.82\textwidth}
\includegraphics[width=\textwidth]{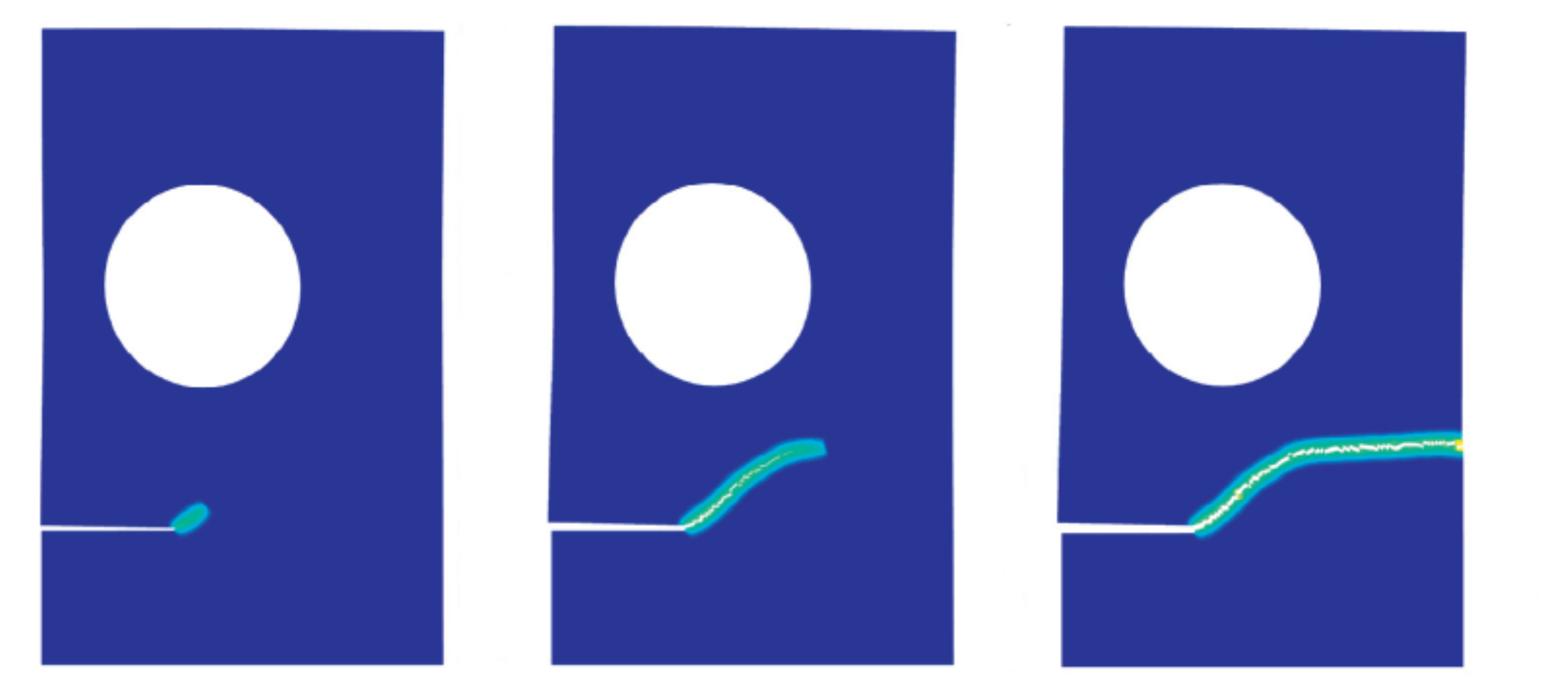}
\caption{Predicted crack path by \cite{ni2017peridynamic}} 
\end{subfigure}
\begin{subfigure}[t]{0.28\textwidth}    %trim={<left> <lower> <right> <upper>}
\includegraphics[width=\textwidth,trim={150 70 150 50}, clip]{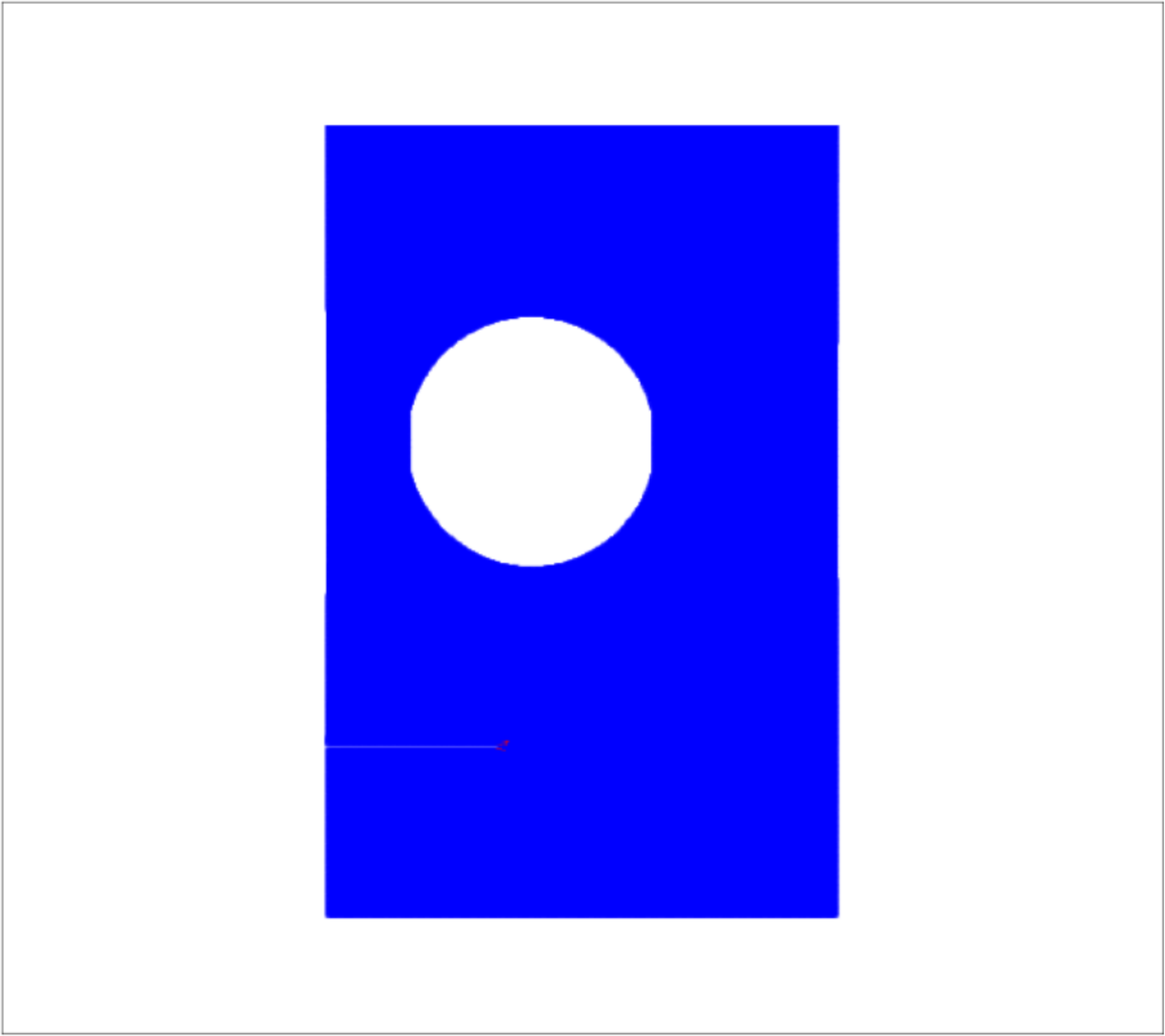}
\end{subfigure}
\begin{subfigure}[t]{0.28\textwidth}    %trim={<left> <lower> <right> <upper>}
\includegraphics[width=\textwidth,trim={150 70 150 50}, clip]{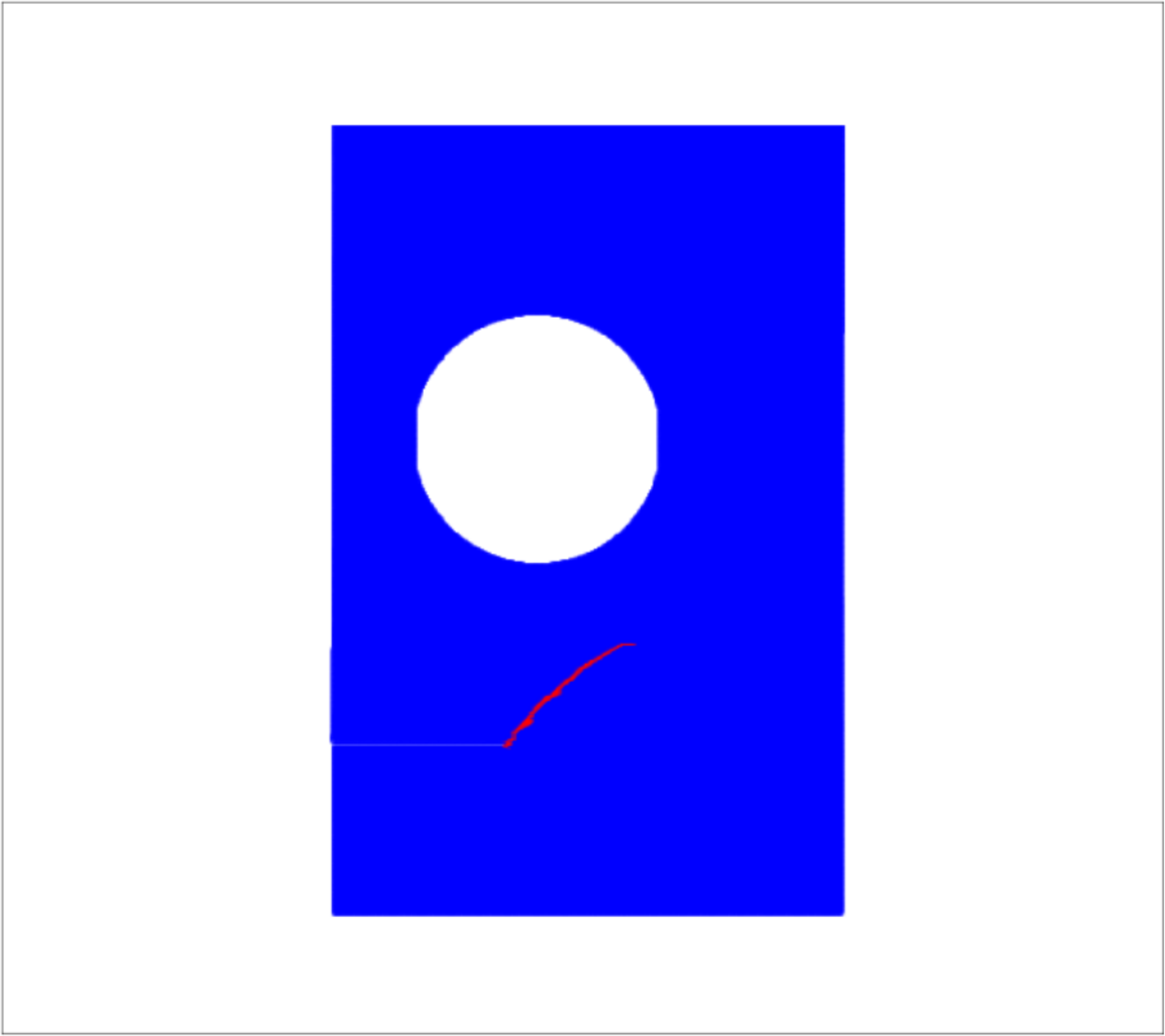}
\caption{Predicted crack path by present framework} 
\end{subfigure}
\begin{subfigure}[t]{0.28\textwidth}    %trim={<left> <lower> <right> <upper>}
\includegraphics[width=0.995\textwidth,trim={150 50 150 50}, clip]{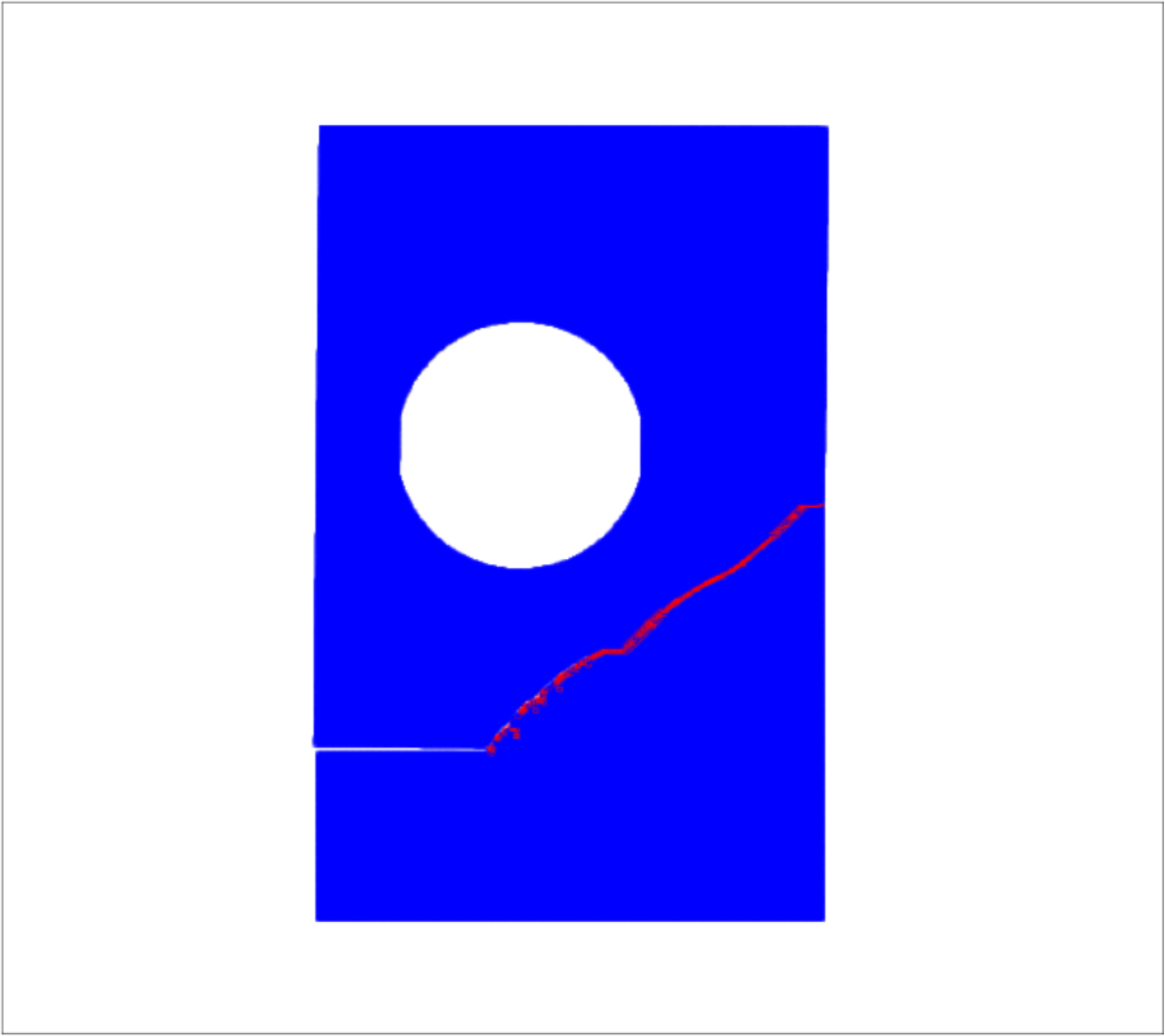}
\end{subfigure}
\caption{Comparision of crack path for notch distance 10 $mm$ from support (case II)}\label{cir_hole_10mm}
\end{figure}

\begin{figure}[hbtp!]
\centering
\begin{subfigure}[t]{0.74\textwidth}    %trim={<left> <lower> <right> <upper>}
\includegraphics[width=\textwidth]{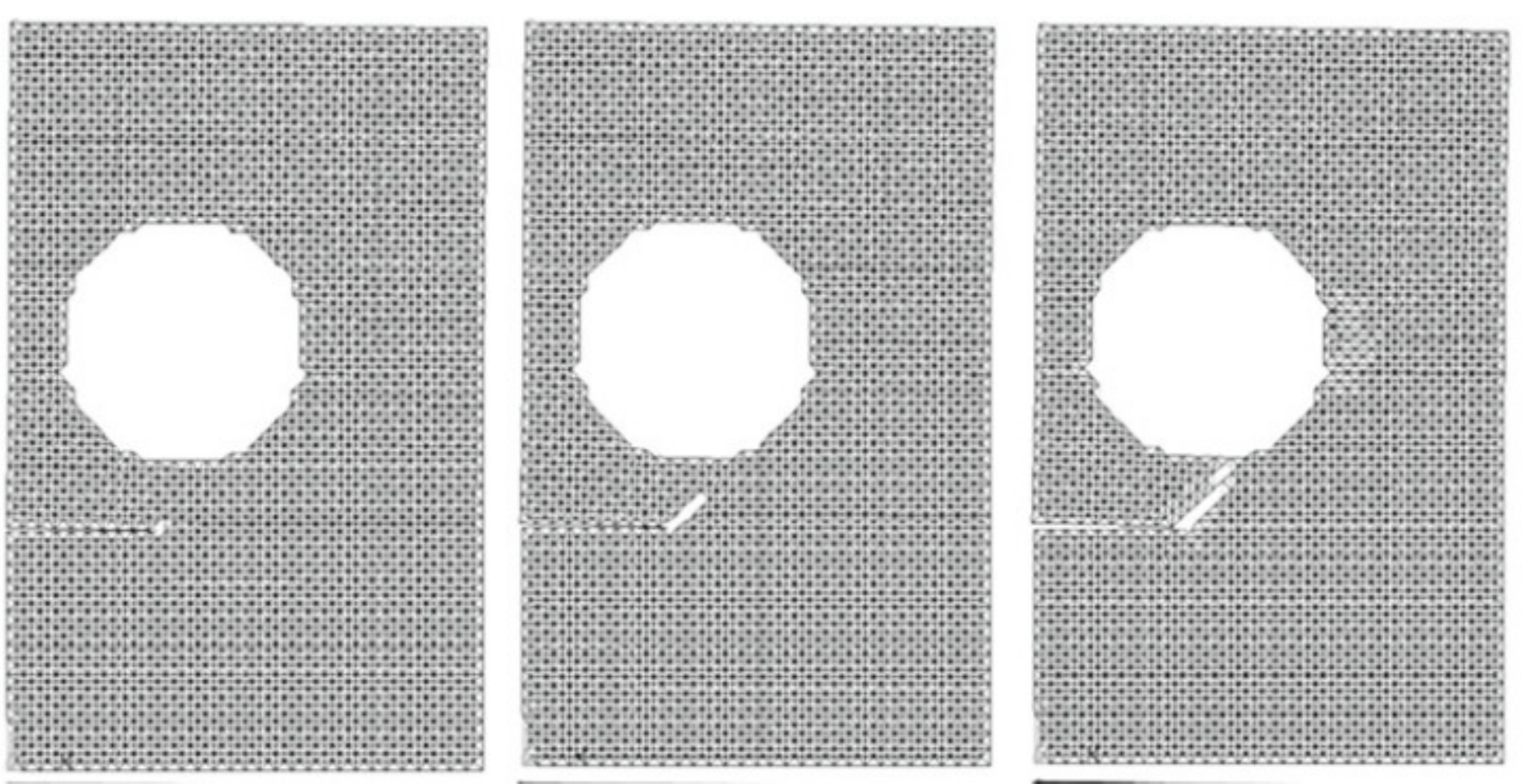}
\caption{Predicted crack path by \cite{kosteski2012crack}} 
\end{subfigure}
\begin{subfigure}[t]{0.82\textwidth}
\includegraphics[width=\textwidth]{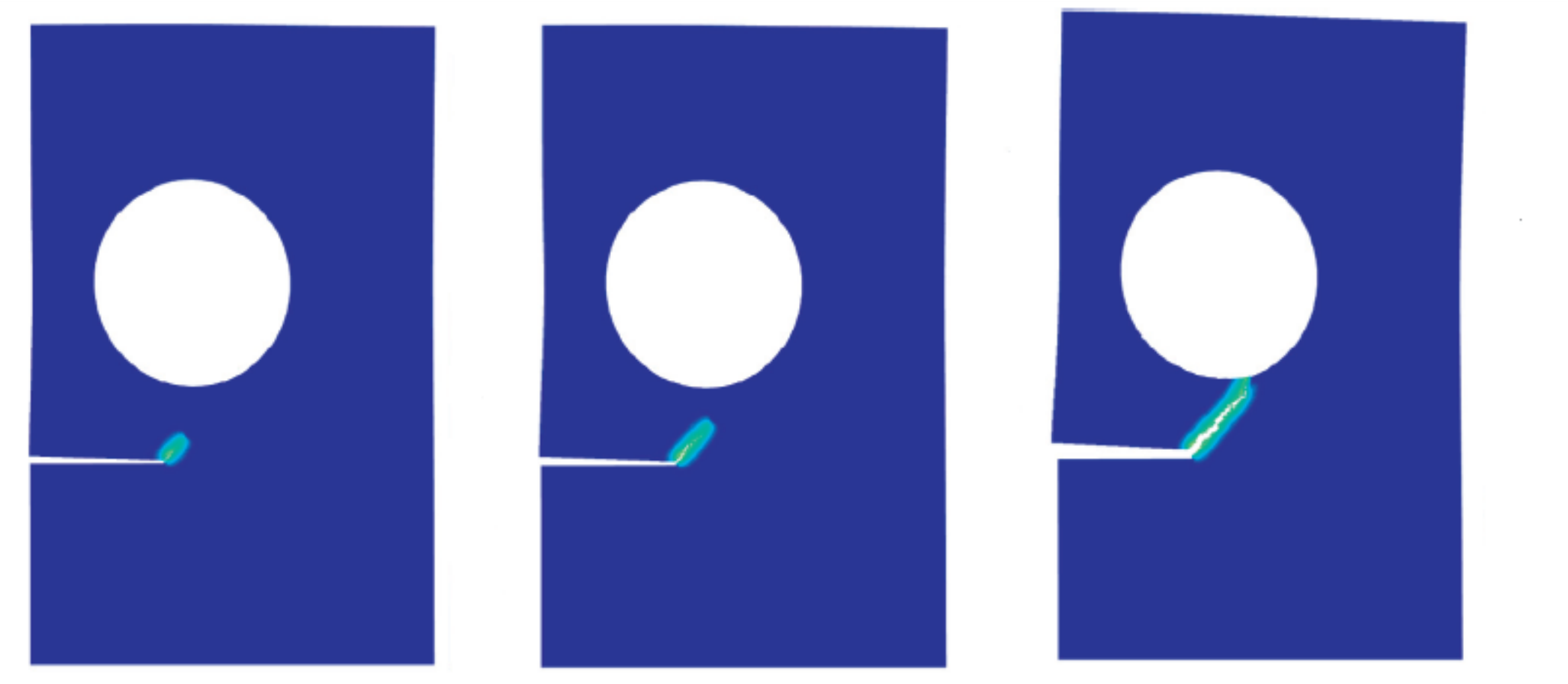}
\caption{Predicted crack path by \cite{ni2017peridynamic}} 
\end{subfigure}
\begin{subfigure}[t]{0.27\textwidth}    %trim={<left> <lower> <right> <upper>}
\includegraphics[width=\textwidth,trim={150 70 150 50}, clip]{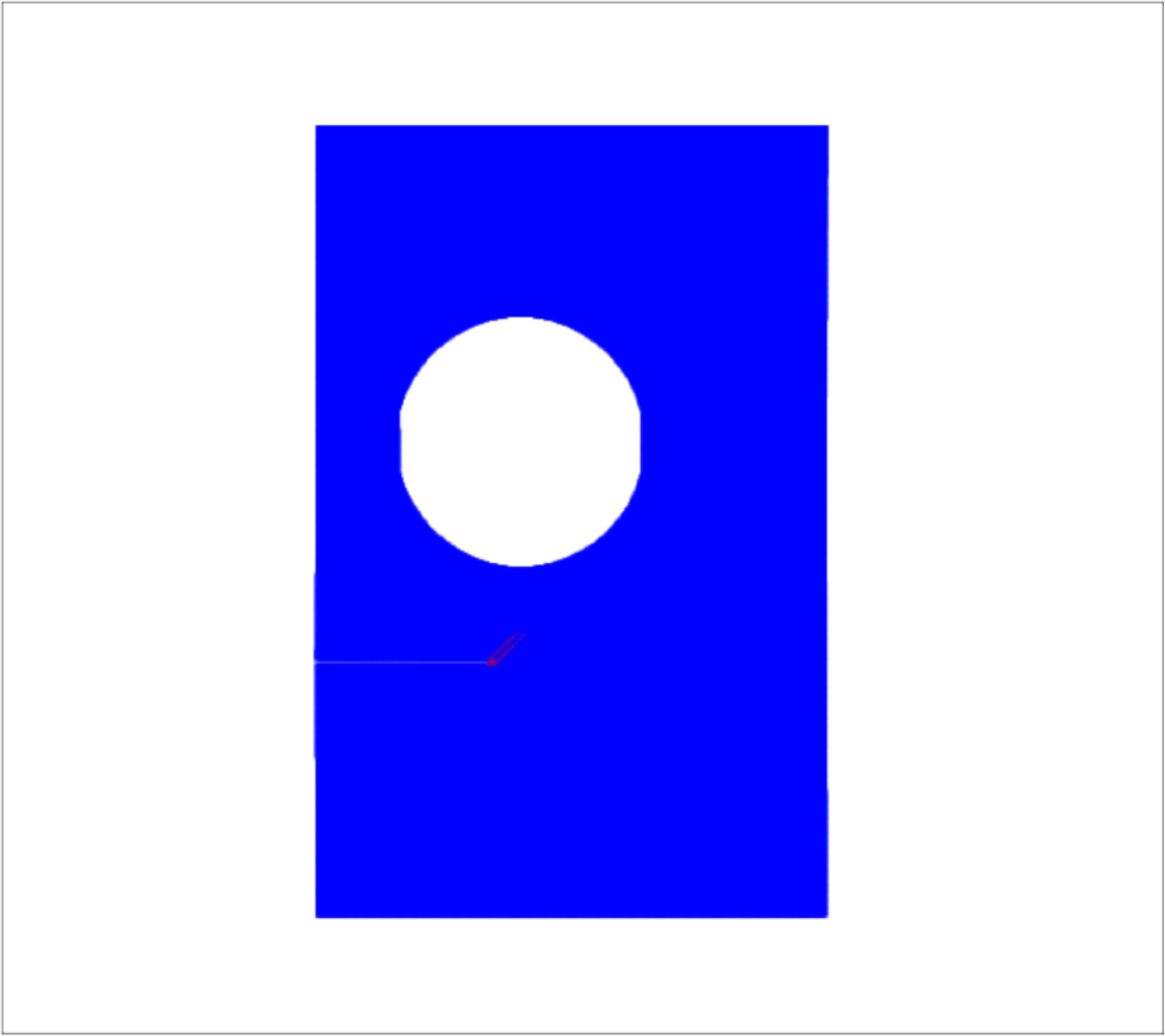}
\end{subfigure}
\begin{subfigure}[t]{0.27\textwidth}    %trim={<left> <lower> <right> <upper>}
\includegraphics[width=\textwidth,trim={150 70 150 50}, clip]{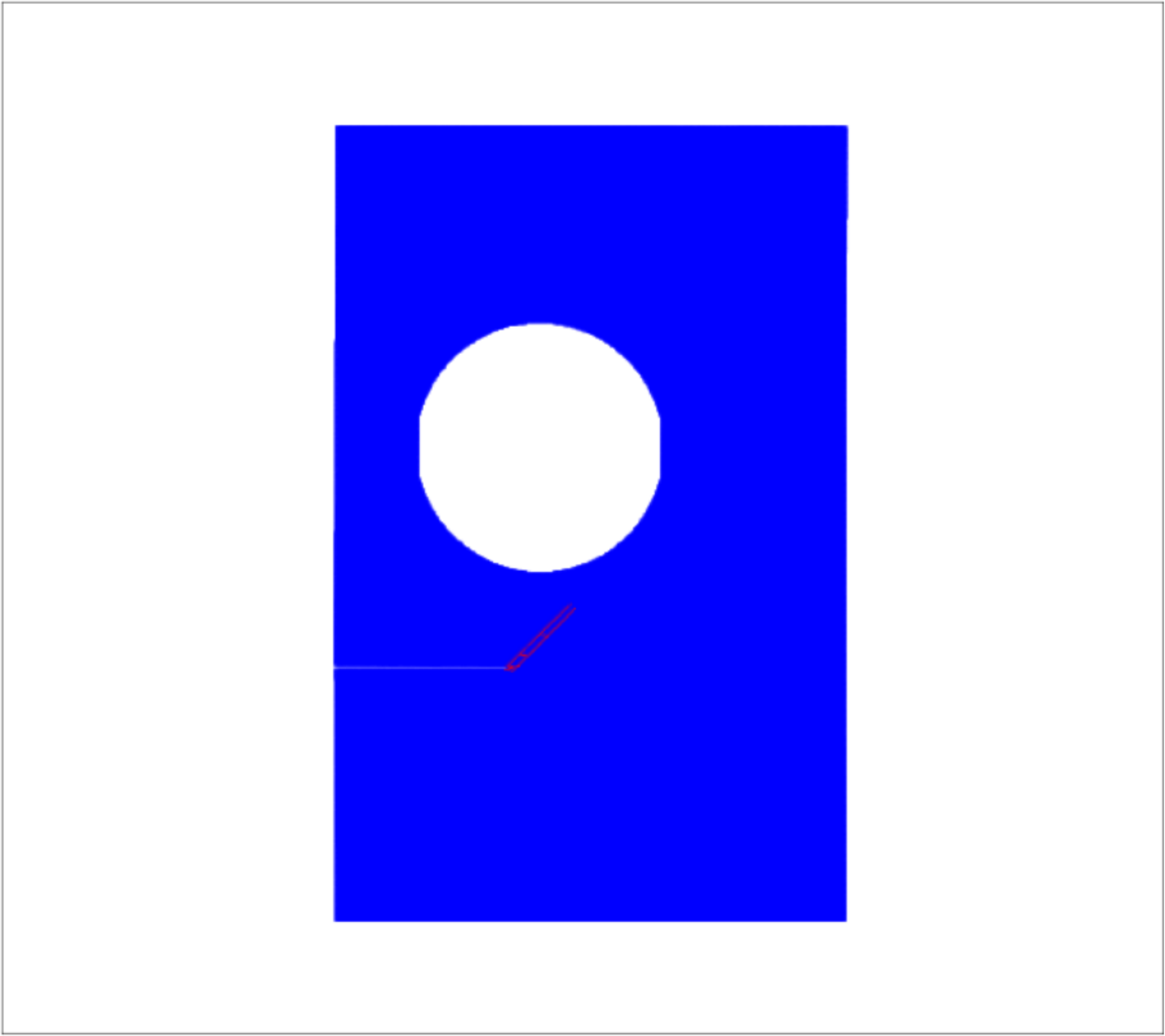}
\caption{Predicted crack path by present framework} 
\end{subfigure}
\begin{subfigure}[t]{0.27\textwidth}    %trim={<left> <lower> <right> <upper>}
\includegraphics[width=0.995\textwidth,trim={150 50 150 50}, clip]{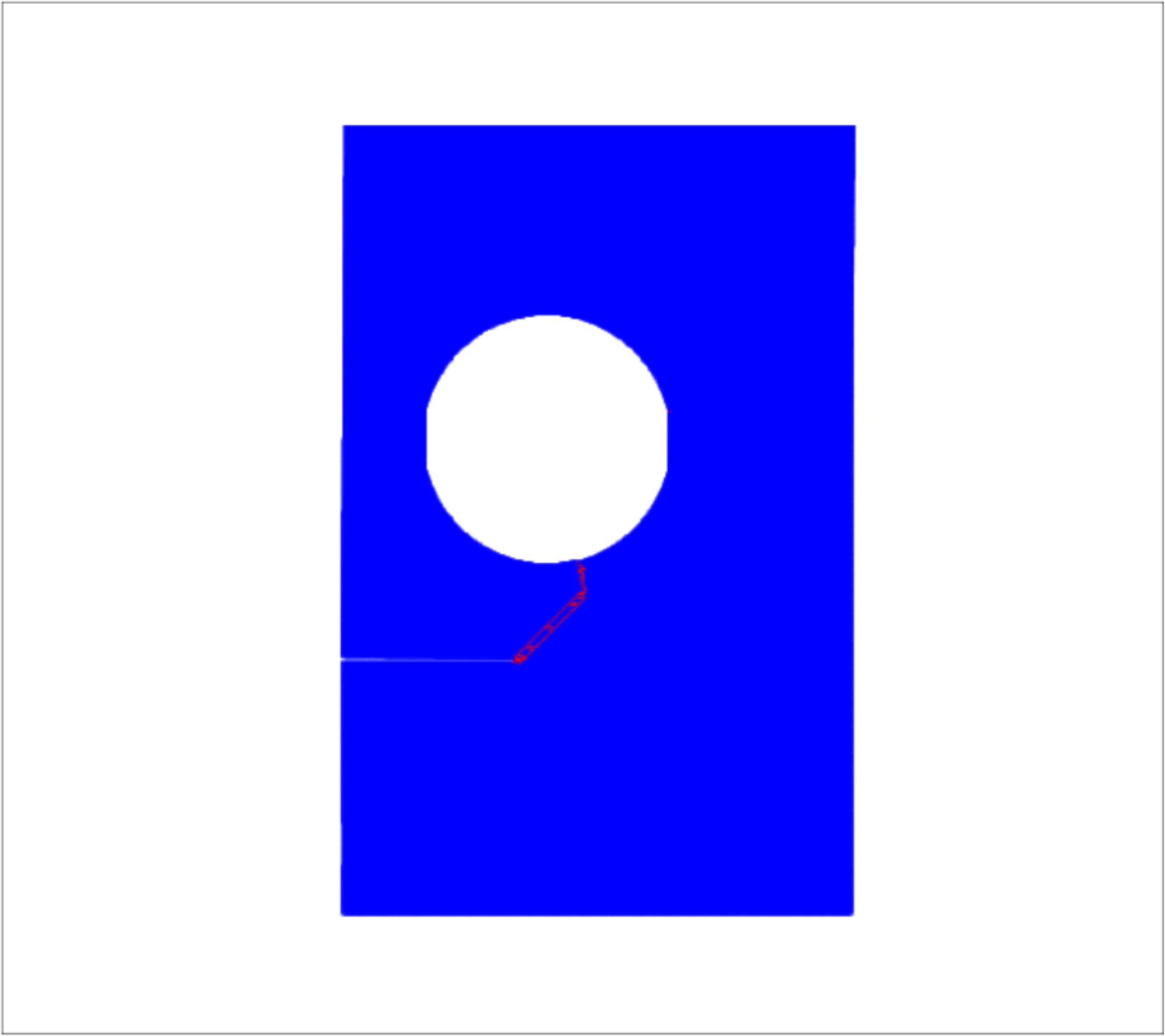}
\end{subfigure}
\caption{Comparision of crack path for notch distance 15 $mm$ from support (case III)}\label{cir_hole_15mm}
\end{figure}

The crack propagation for case II (notch distance 10 mm) and case III (notch distance 15 mm) are given in 
Figure \ref{cir_hole_10mm} and \ref{cir_hole_15mm} respectively. The crack path deviation from the off-centre circular hole depends on the notch location. As the distance from the support increases, the deviation of the crack path from the circular hole decreases. This deviation is largest for the case I (notch distance 5 mm). The crack path interacts with the circular hole for case III. The predicted crack paths for all the three cases are consistent with the observations in \cite{kosteski2012crack, ni2017peridynamic} as shown in the Figure \ref{cir_hole_5mm}, \ref{cir_hole_10mm} and \ref{cir_hole_15mm}.

\subsection{Three dimensional numerical examples}\label{sec-4}
Numerical modelling of arbitrary propagation and interaction of cracks in three dimension (3-D) is a challenging exercise. Crack propagation in 3-D is much more difficult to capture than the same in 2-D, and it poses a real test for a computational framework to demonstrate its efficacy. This motivated the extension of the present computational framework in 3-D. Herein, the \textit{pseudo spring SPH} is explored to model arbitrary crack propagation in the 3-D solid continuum. Towards this end, a few example problems are considered. The first example is the formation of a helical crack in cylindrical chalk subjected to torsional loading at both ends. The Kalthoff-Winkler experiment and the failure of Taylor bullet are also modelled. The predicted fracture surfaces are found to be in good agreement with the experimental results available in the literature. 

\subsubsection{Cylindrical chalk under torsion}
\begin{figure}[hbtp!]
\centering
\begin{subfigure}[t]{0.7\textwidth}    %trim={<left> <lower> <right> <upper>}
\includegraphics[width=\textwidth]{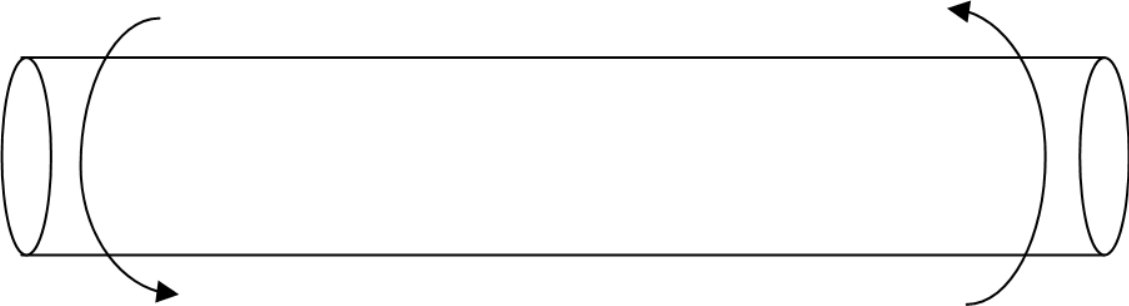}
\end{subfigure}
\begin{subfigure}[t]{0.7\textwidth}
\includegraphics[width=\textwidth]{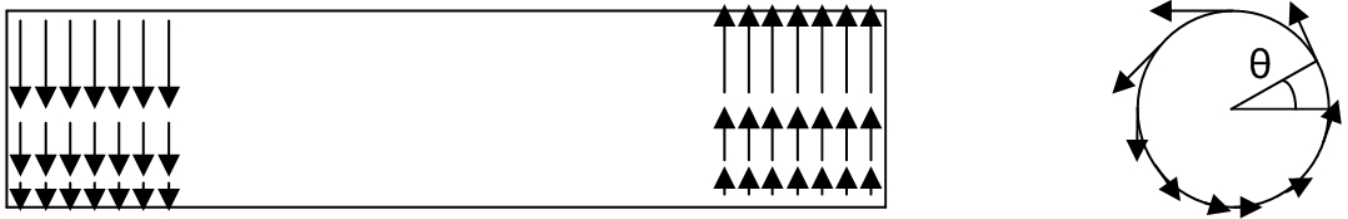} 
\end{subfigure}
\caption{Diagram of chalk under torsion}\label{chalk}
\end{figure}

In this section, the failure of a cylindrical chalk bar under torsional loading is modelled. The diagram of the set-up and loading on the chalk specimen are shown in Figure \ref{chalk}. The specimen has a length of 100 mm and a diameter of 8 mm. The torsional loading is generated by applying a velocity field of $u_{max}$ = 28.3 m/s in the tangential direction to the chalk bar as shown in Figure \ref{chalk}. The distribution of the velocity field is given as  

\begin{equation}   
   u = u_{max} \sin (\frac{\theta + \frac{\pi}{2} }{2})
\end{equation}

The velocity field is applied for a length of 20 mm each at both ends of the specimen. The application of the velocity field is such that it generates a twisting behaviour in the chalk. This has the similar effect of the traction field applied in the tangential direction of the chalk creating torsional loading \citep{bordas2008three, rabczuk2010three}. The material properties and other computational parameters are given in Table \ref{chalk_tab}.

\begin{table}[hbtp!]
\centering
\caption{Parameters for the chalk under torsional loading}\label{chalk_tab}
\begin{tabular}{ccccccccc}
\hline
                           & \multicolumn{4}{c}{Material Proerties}                 & \multicolumn{2}{c}{Discretization} & \multicolumn{2}{c}{Artificial Viscosity}                \\
\multirow{2}{*}{Parameter} & $\rho$     & $E$ & \multirow{2}{*}{$\nu$} & $\sigma^p_{max}$ & $\Delta p$          & $h$          & \multirow{2}{*}{$\beta_1$} & \multirow{2}{*}{$\beta_2$} \\
                           & ($kg/m^3$) & (GPa) &                        &   (MPa)  & (mm)                & (mm)         &                            &                            \\ 
                           \hline
Value                      & 1150       & 2 & 0.18                    &  15     & 0.4                & 0.52         & 1.0                        & 1.0 \\  
\hline                    
\end{tabular}
\end{table}

\begin{figure}[hbtp!]
\centering
\begin{subfigure}[t]{0.49\textwidth}    %trim={<left> <lower> <right> <upper>}
\includegraphics[width=\textwidth]{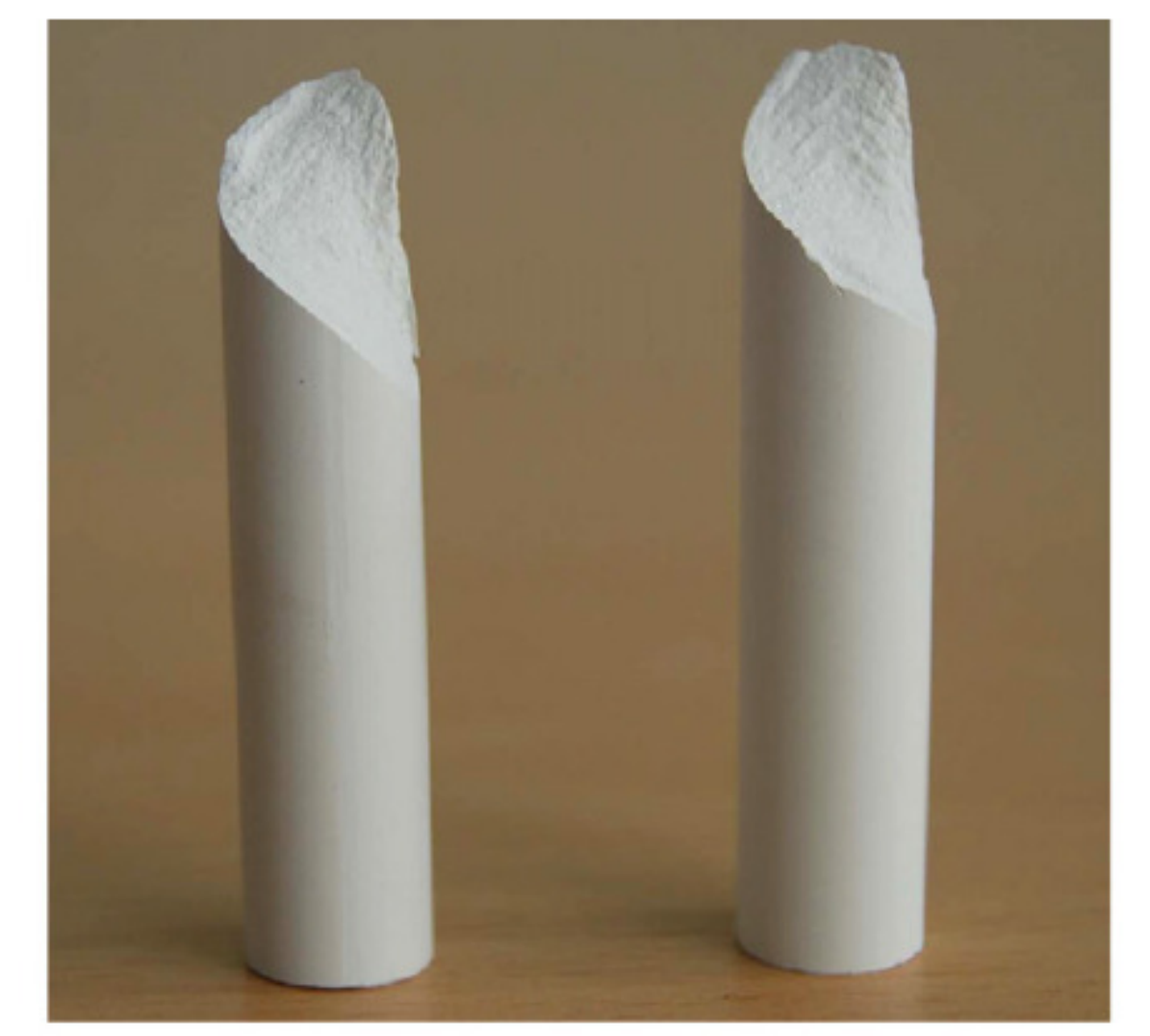}
\caption{Crack surface by \cite{bordas2008three}} 
\end{subfigure}
\begin{subfigure}[t]{0.4\textwidth}    %trim={<left> <lower> <right> <upper>}
\includegraphics[width=\textwidth]{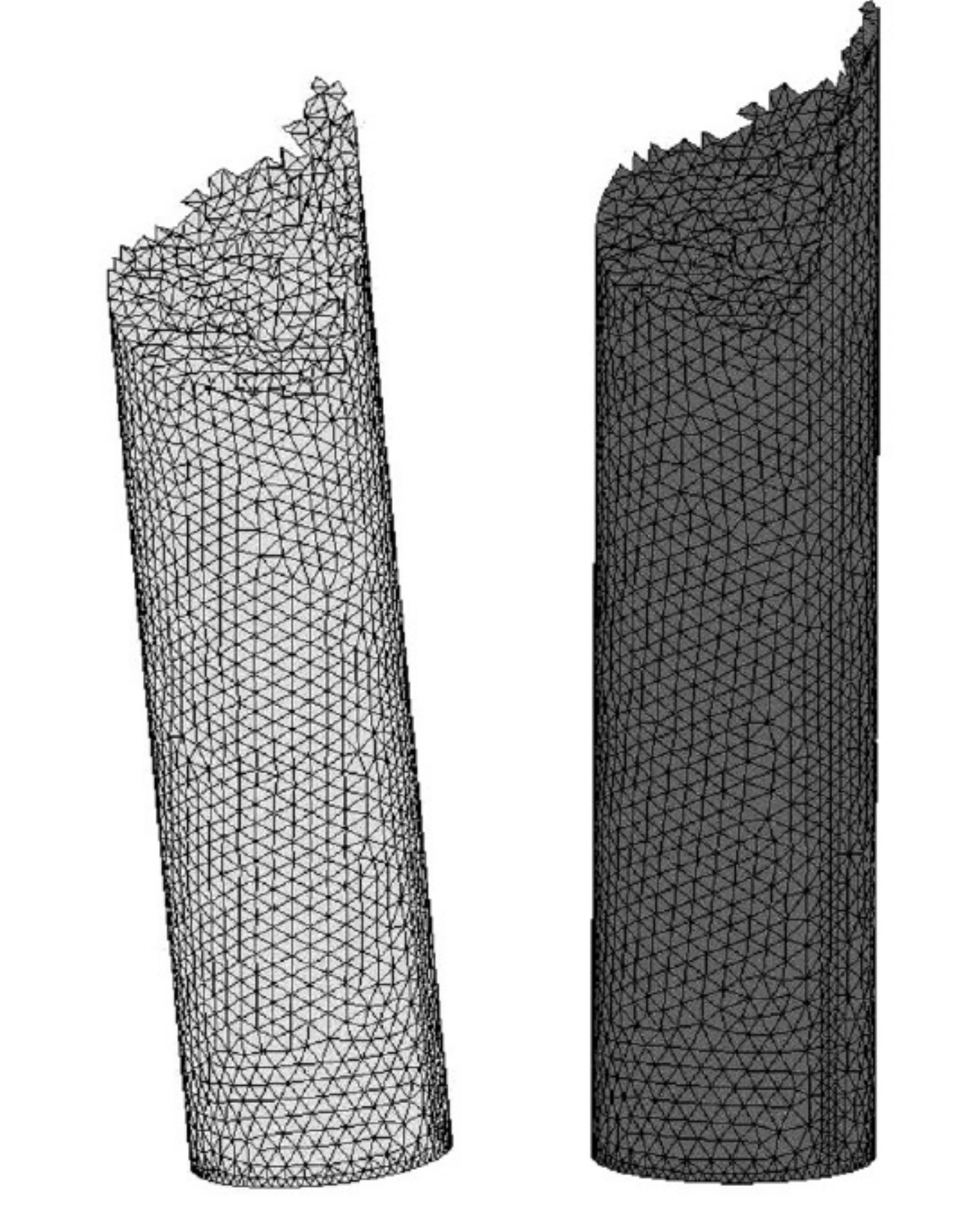}
\caption{Predicted crack surface by \cite{bordas2008three}} 
\end{subfigure}
\begin{subfigure}[t]{0.2\textwidth}    %trim={<left> <lower> <right> <upper>}
\includegraphics[width=\textwidth,trim={150 25 310 25}, clip]{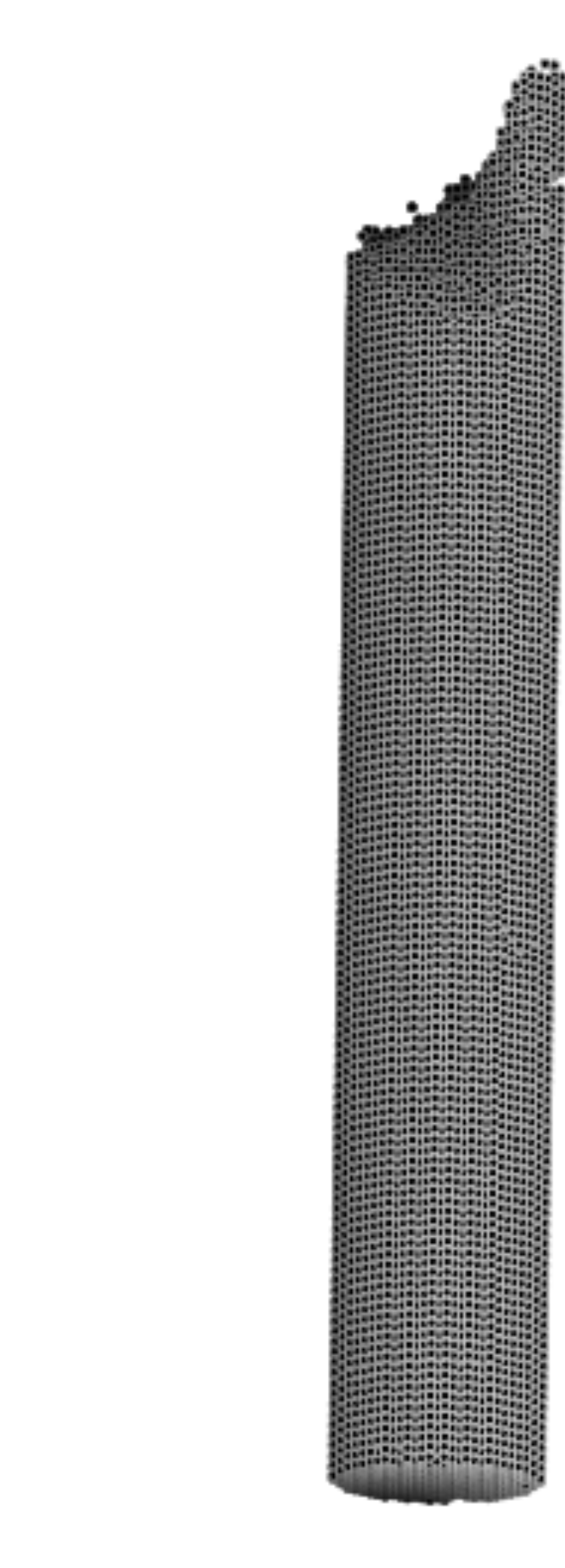}
\caption{Present prediction}
\end{subfigure}
\begin{subfigure}[t]{0.2\textwidth}    %trim={<left> <lower> <right> <upper>}
\includegraphics[width=\textwidth,trim={297 20 172 25}, clip]{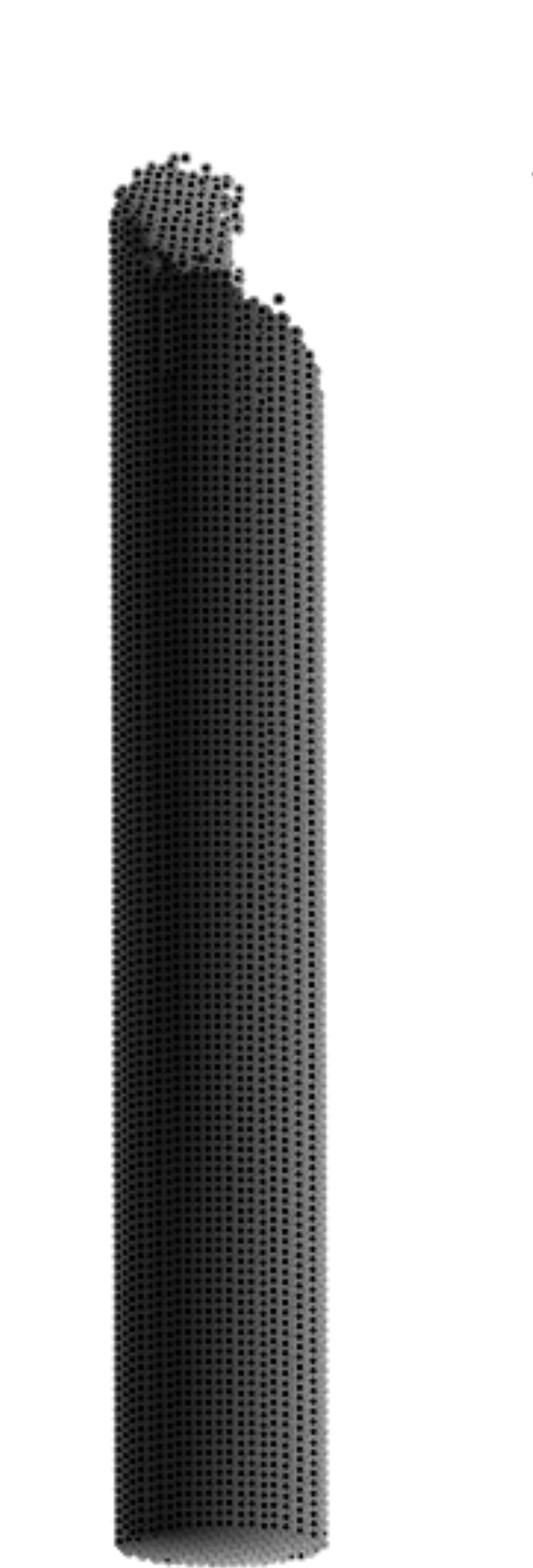}
\caption{Present prediction} 
\end{subfigure}
\caption{A three dimensional chalk under torsion with helical crack surface}\label{chalk_dam}
\end{figure}

The maximum principal stress criterion is used for modelling of the failure. The springs are considered to fail completely when the maximum principal stress exceeds $\sigma^p_{max}$ value. The failure can happen anywhere and results in an arbitrary crack surface as shown in Figure \ref{chalk_dam}. The present fracture surfaces are found to be helical. These surfaces are consistent with the crack surface obtained in \cite{bordas2008three, rabczuk2010three} for the similar problem using the extended element-free Galerkin (XEFG) method. 

\subsubsection{Kalthoff-Winkler experiment}
\begin{figure}[hbtp!]
\centering
\includegraphics[width=0.7\textwidth]{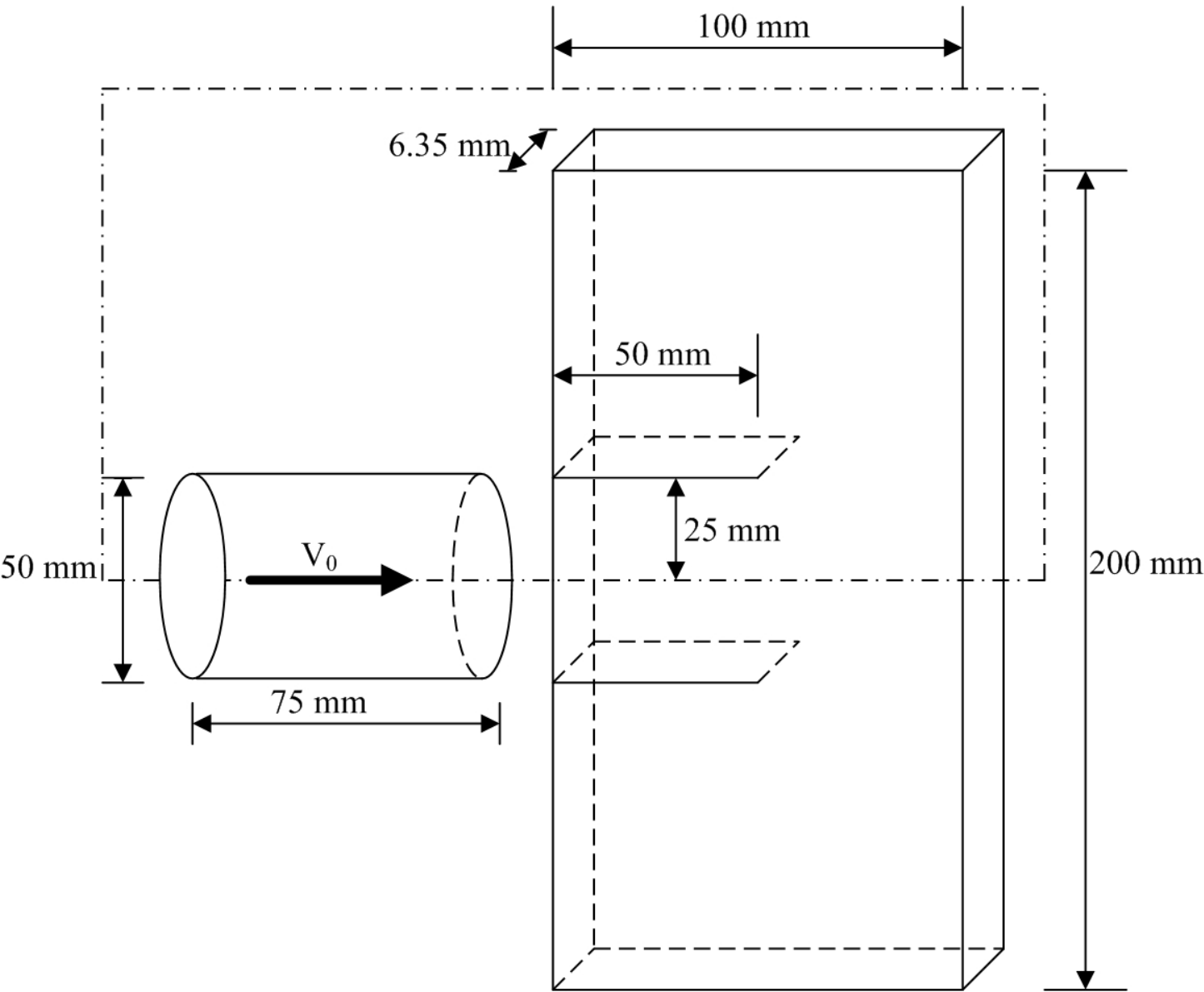}
\caption{Set up for Kalthoff-Winkler experiment}\label{kalthoff}
\end{figure}

As the second example, the Kalthoff-Winkler experiment \citep{kalthoff1988failure} is considered. A cylindrical projectile impacts a plate with two edge notches as shown in Figure \ref{kalthoff}. Numerical simulations of similar problems are performed in \cite{rabczuk2007meshfree, rabczuk2010simple, braun2014new, kosteski2012crack, zhou2016numerical, chakraborty2013pseudo}. At a low-velocity impact, the plate undergoes brittle failure, and crack propagates at an angle 70 $^ \circ $ to the notch direction. The material and SPH parameters are given in Table \ref{kalthoff_tab}. The impact velocity is taken as 16.5 m/s.  

\begin{table}[hbtp!]
\centering
\caption{Parameters for Klathoff-Winkler experiment simulation}\label{kalthoff_tab}
\begin{tabular}{ccccccccc}
\hline
                           & \multicolumn{4}{c}{Material Proerties}                 & \multicolumn{2}{c}{Discretization} & \multicolumn{2}{c}{Artificial Viscosity}                \\
\multirow{2}{*}{Parameter} & $\rho$     & $E$ & \multirow{2}{*}{$\nu$} & $\delta_{tc}$ & $\Delta p$          & $h$          & \multirow{2}{*}{$\beta_1$} & \multirow{2}{*}{$\beta_2$} \\
                           & ($kg/m^3$) & (GPa) &                        &    & (mm)                & (mm)         &                            &                            \\ 
                           \hline
Value                      & 8000       & 190 & 0.3                    &  0.0044     & 0.8                & 1.04         & 1.0                        & 1.0 \\  
\hline                    
\end{tabular}
\end{table}

In the present numerical simulation, half of the projectile-target system is considered due to the symmetry and accordingly a symmetric boundary condition is applied. The Rankine's criteria or critical stretch between particles is used as the failure \citep{chakraborty2013pseudo, braun2014new} criterion. The critical stretch at any time instant is calculated as

\begin{equation}
    \delta_t = \frac{r_{ij} \vert_t - \Delta p}{\Delta p}
\end{equation}

\begin{figure}[hbtp!]
\centering
\begin{subfigure}[t]{0.49\textwidth}    %trim={<left> <lower> <right> <upper>}
\includegraphics[width=\textwidth,trim={10 10 10 10}, clip]{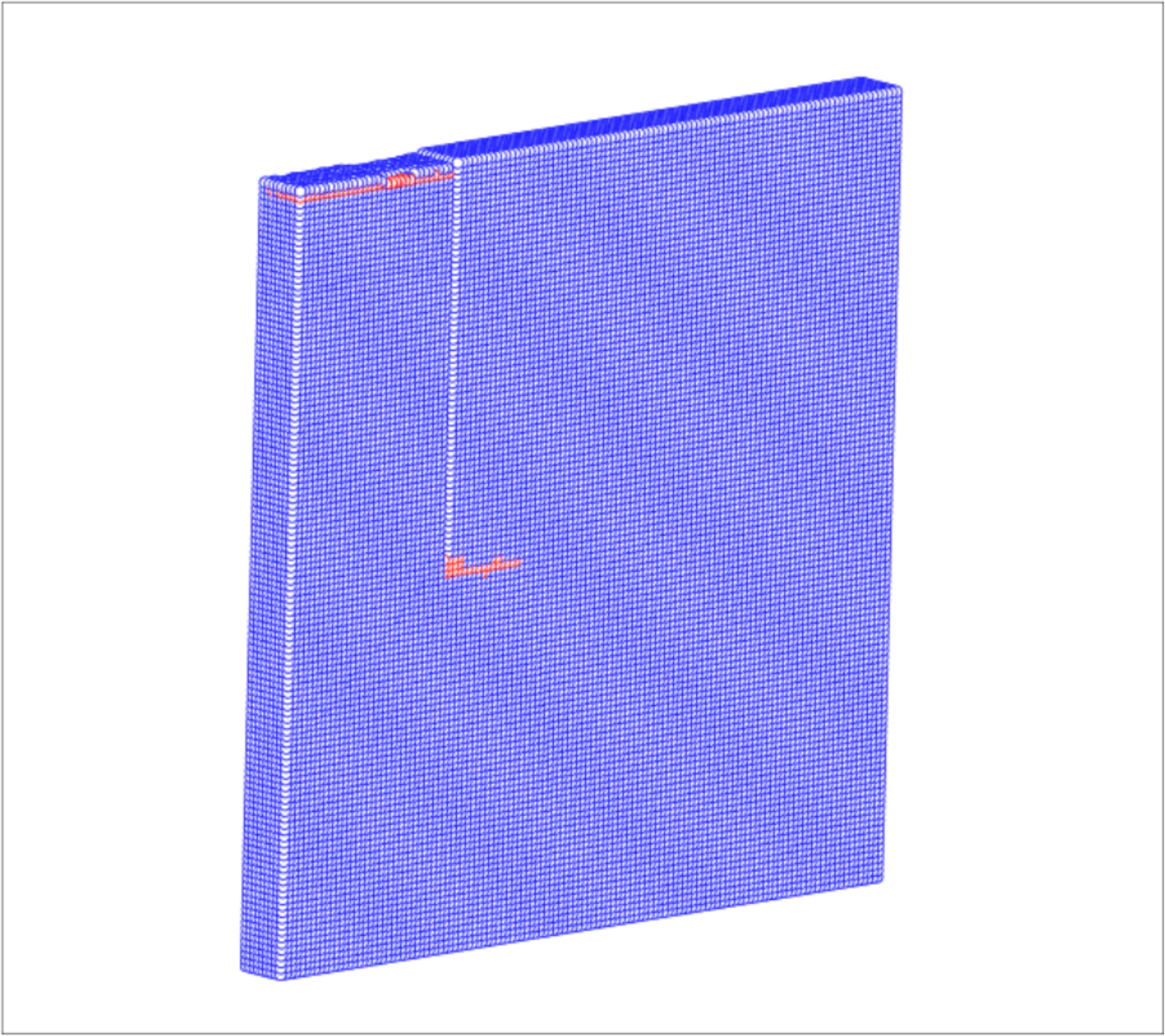}
\caption{} 
\end{subfigure}
\begin{subfigure}[t]{0.49\textwidth}
\includegraphics[width=\textwidth,trim={10 10 10 10}, clip]{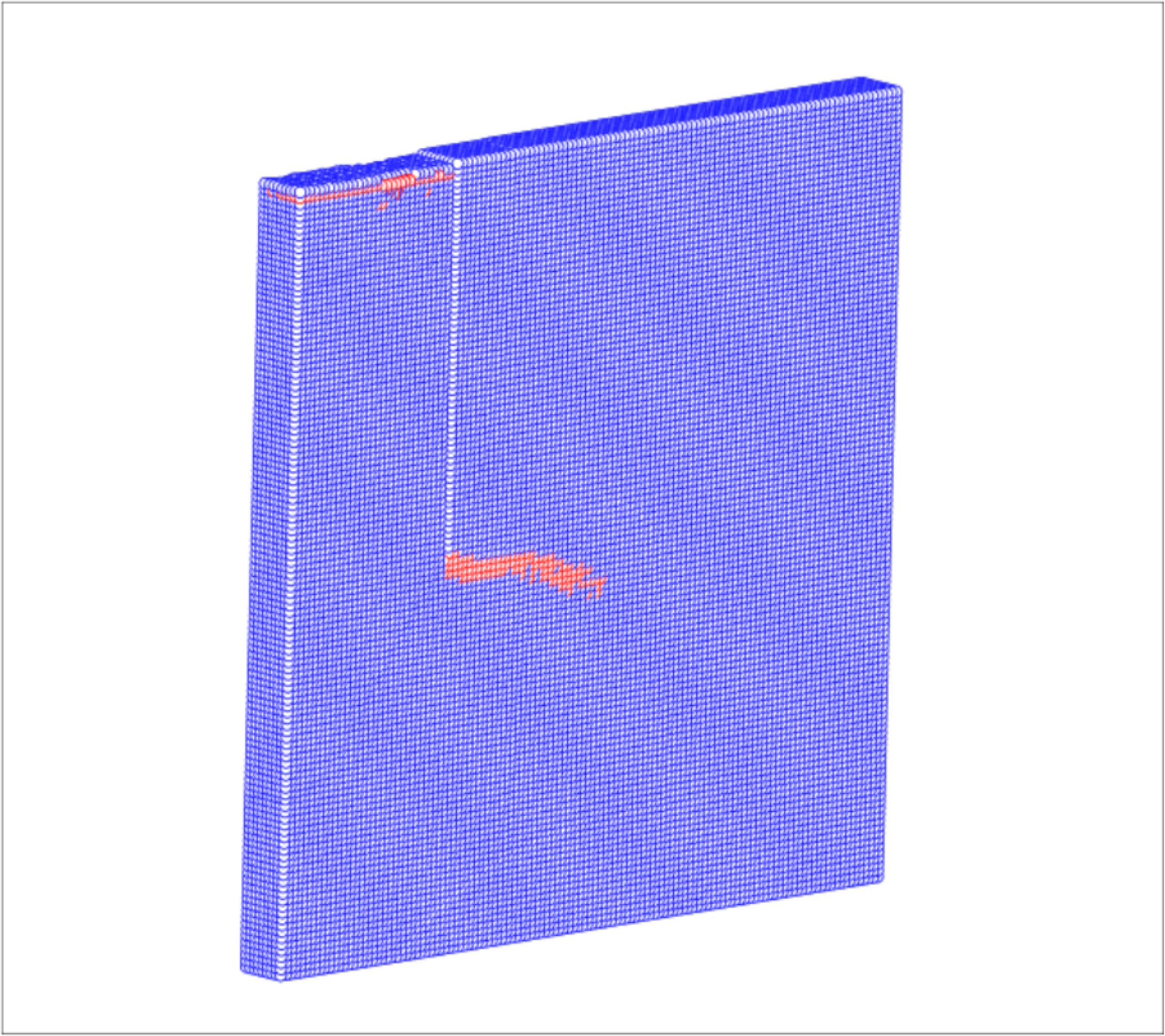}
\caption{} 
\end{subfigure}
\begin{subfigure}[t]{0.49\textwidth}
\includegraphics[width=\textwidth,trim={10 10 10 10}, clip]{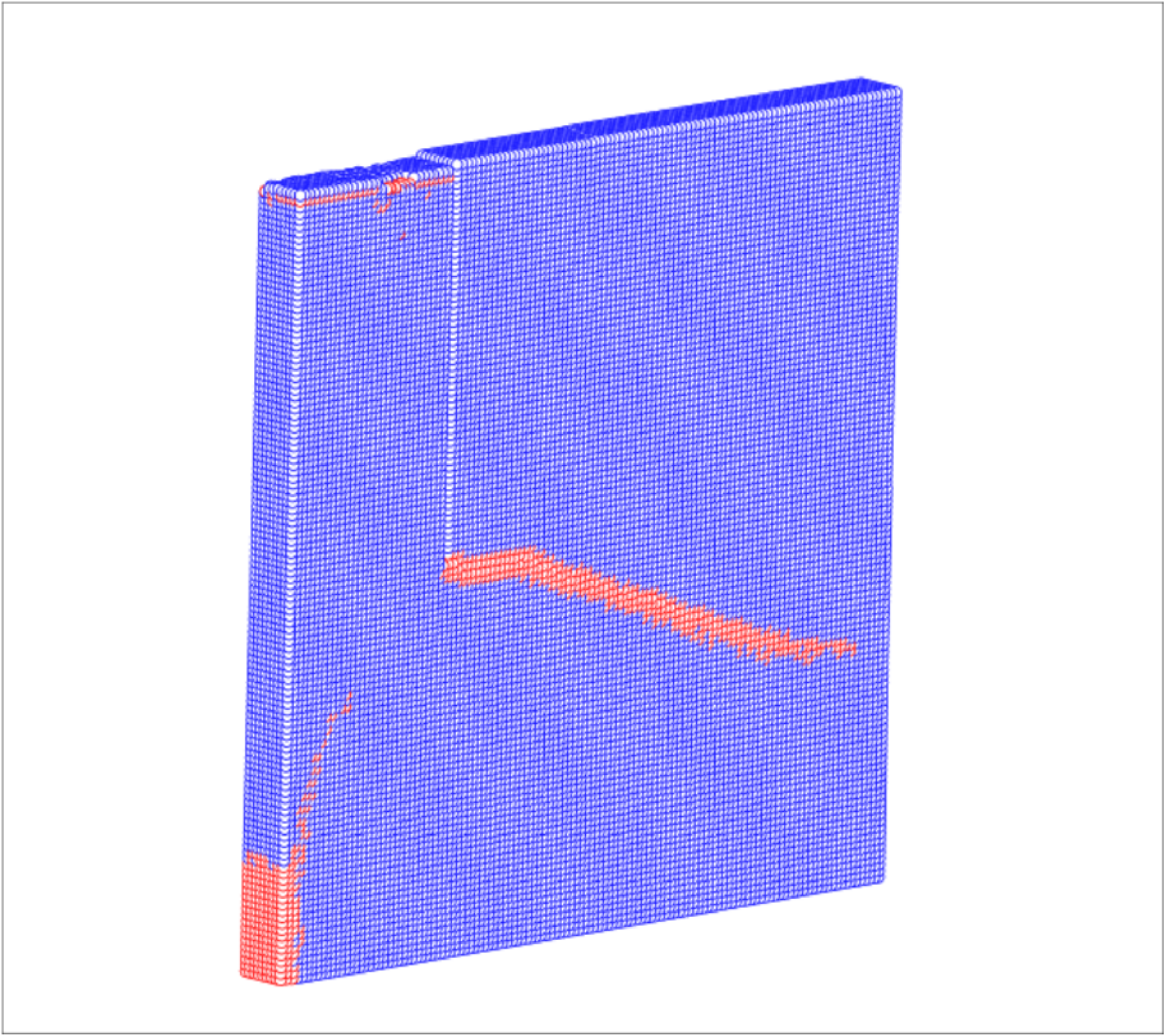}
\caption{} 
\end{subfigure}
\begin{subfigure}[t]{0.49\textwidth}
\includegraphics[width=\textwidth,trim={10 10 10 10}, clip]{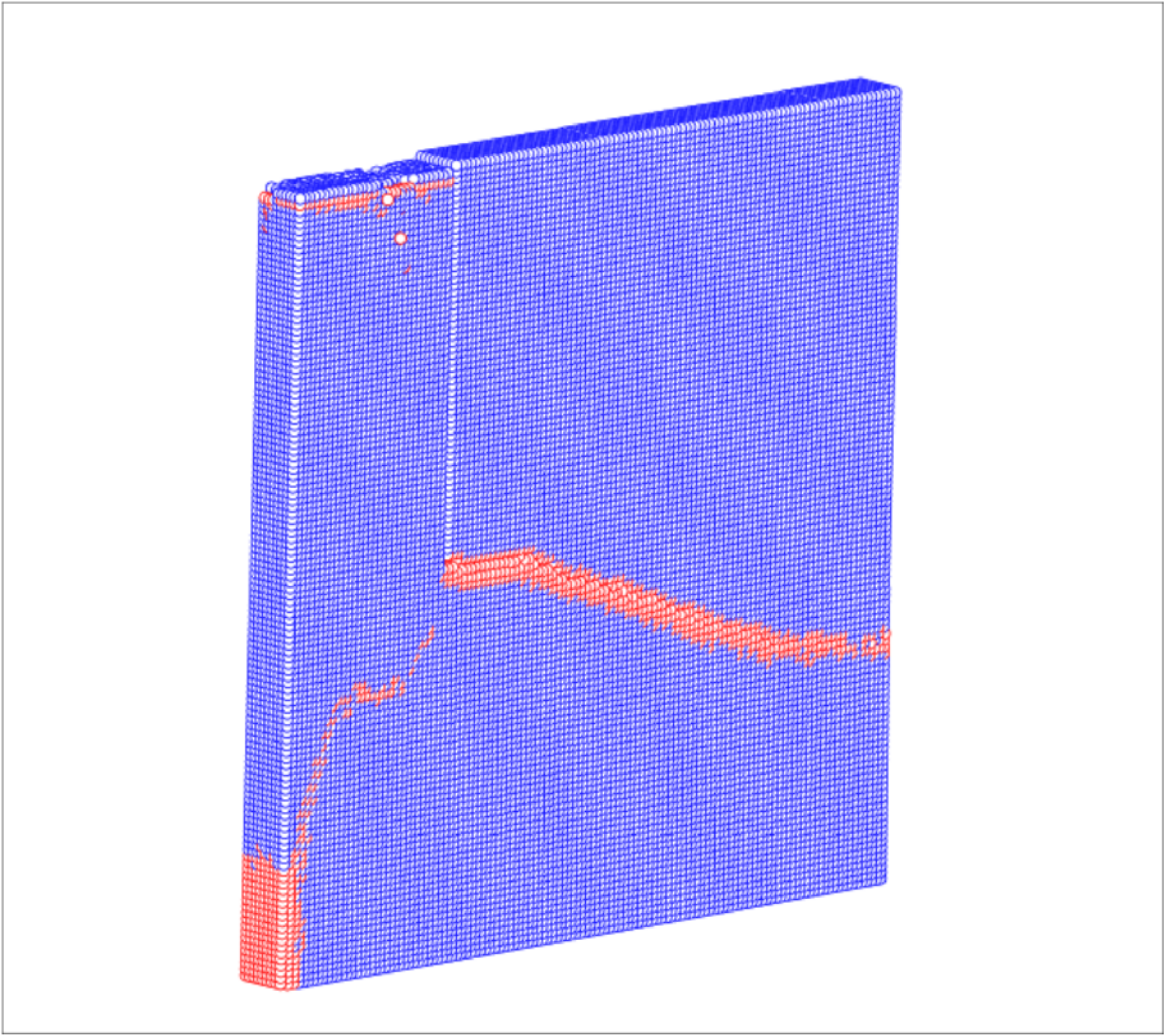}
\caption{} 
\end{subfigure}
\begin{subfigure}[t]{0.6\textwidth}
\includegraphics[width=\textwidth,trim={5 260 5 250}, clip]{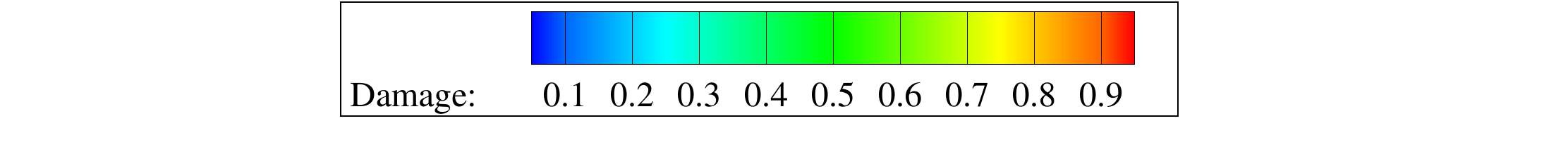} 
\end{subfigure}
\caption{Damage contour in the crack path for Kalthoff-Winkler experiment}\label{kalthoff_crack}
\end{figure}

\begin{figure}[hbtp!]
\centering
\begin{subfigure}[t]{0.49\textwidth}    %trim={<left> <lower> <right> <upper>}
\includegraphics[width=\textwidth,trim={10 10 10 10}, clip]{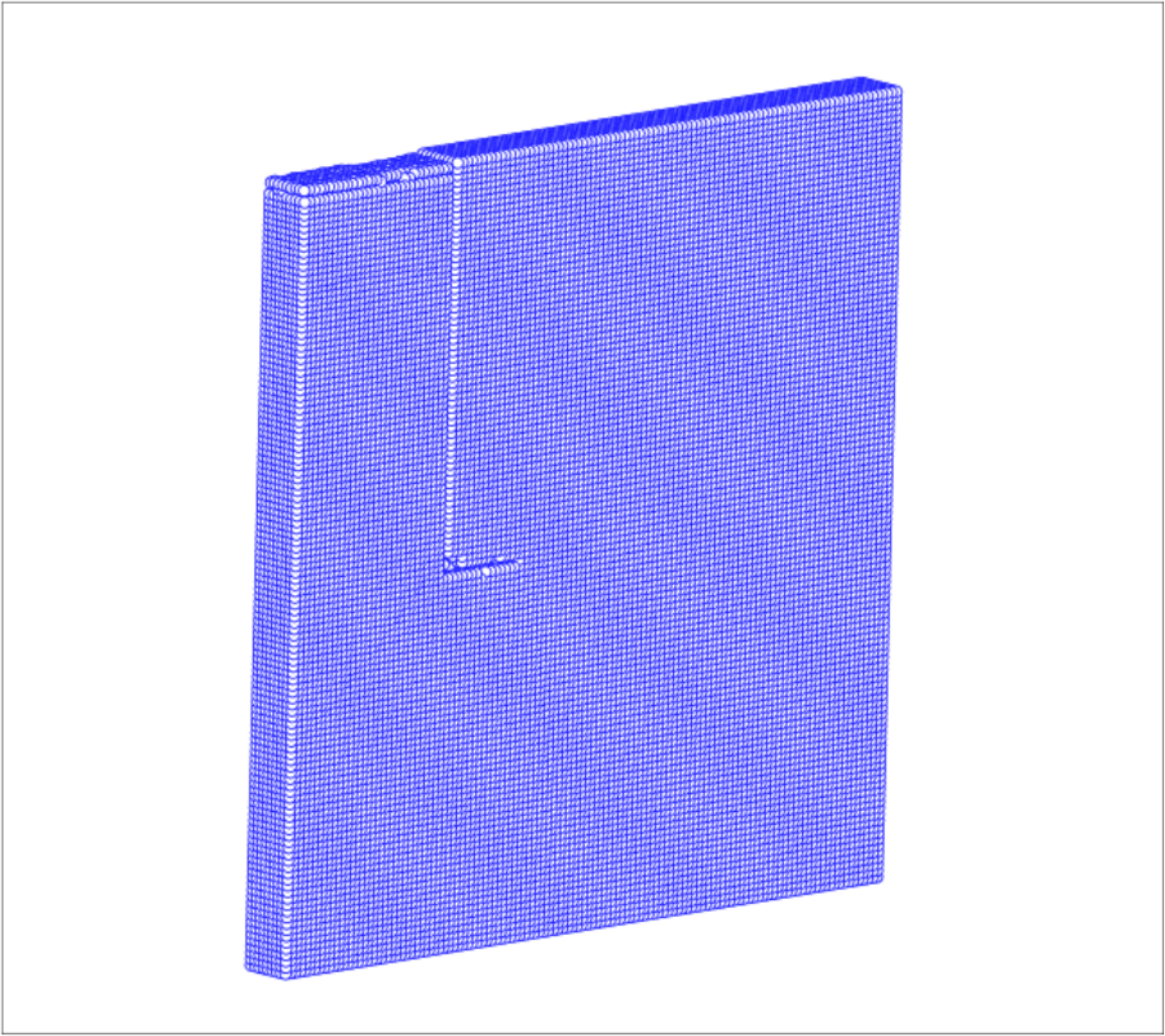}
\caption{} 
\end{subfigure}
\begin{subfigure}[t]{0.49\textwidth}
\includegraphics[width=\textwidth,trim={10 10 10 10}, clip]{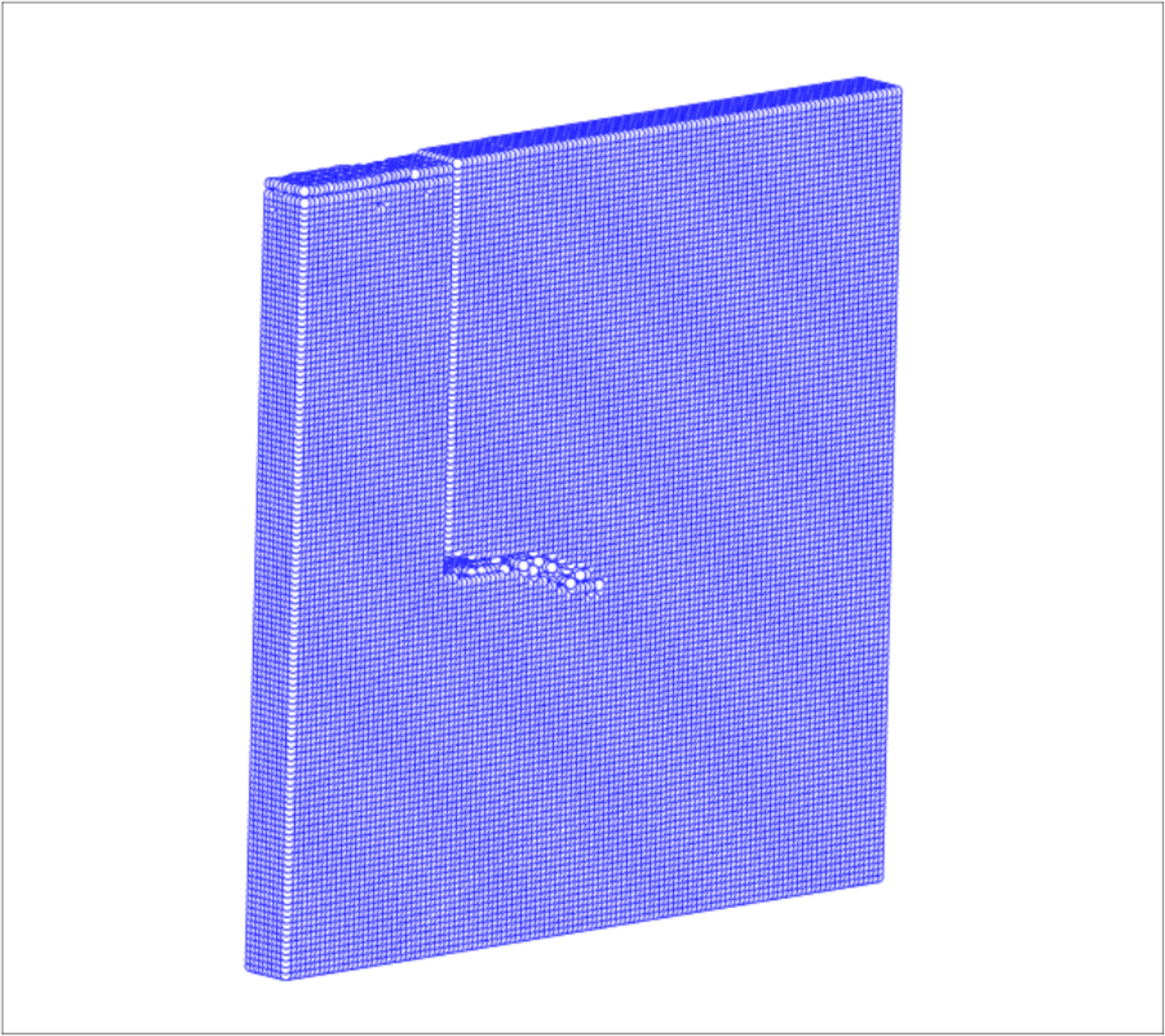}
\caption{} 
\end{subfigure}
\begin{subfigure}[t]{0.49\textwidth}
\includegraphics[width=\textwidth,trim={10 10 10 10}, clip]{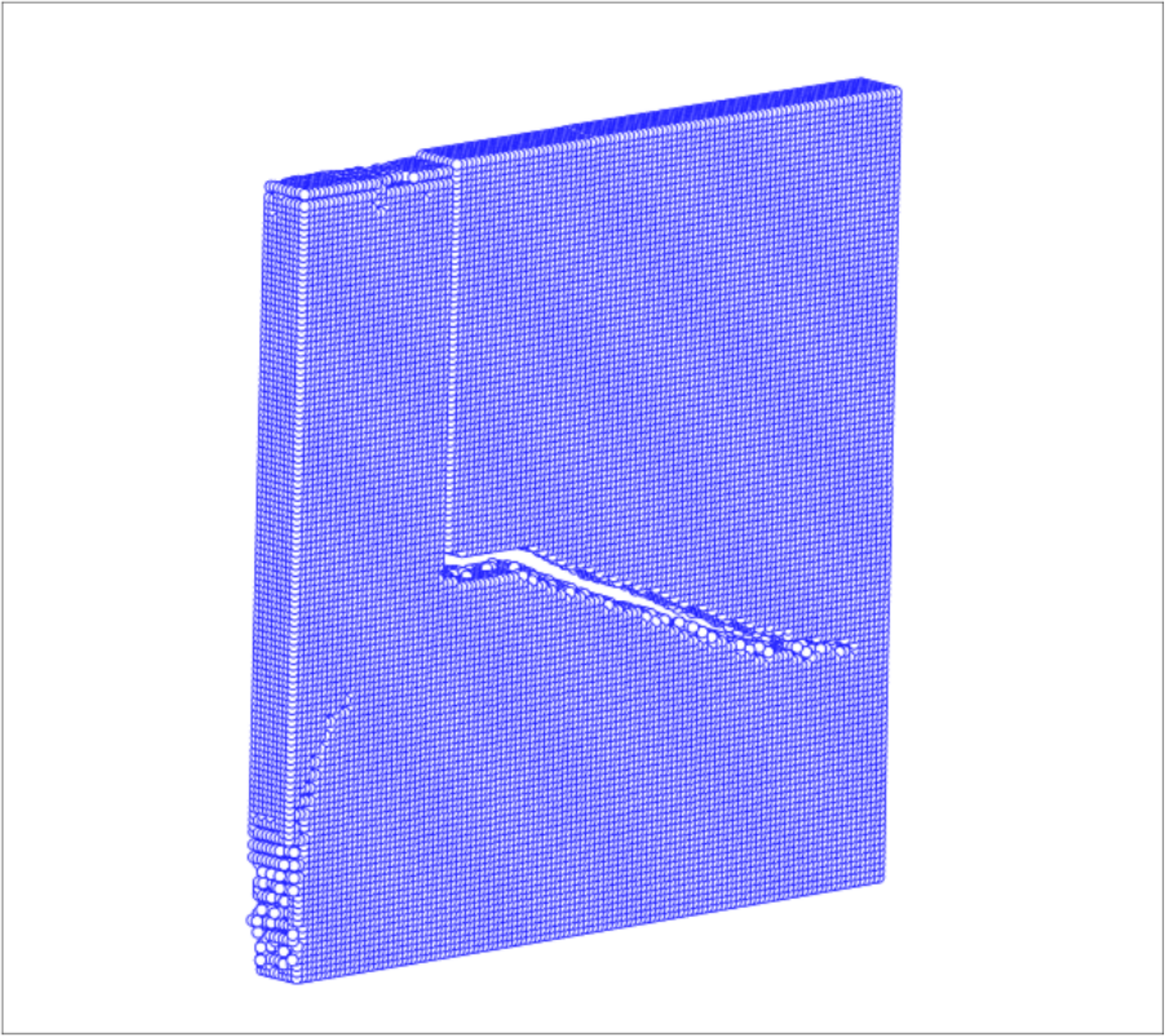}
\caption{} 
\end{subfigure}
\begin{subfigure}[t]{0.49\textwidth}
\includegraphics[width=\textwidth,trim={10 10 10 10}, clip]{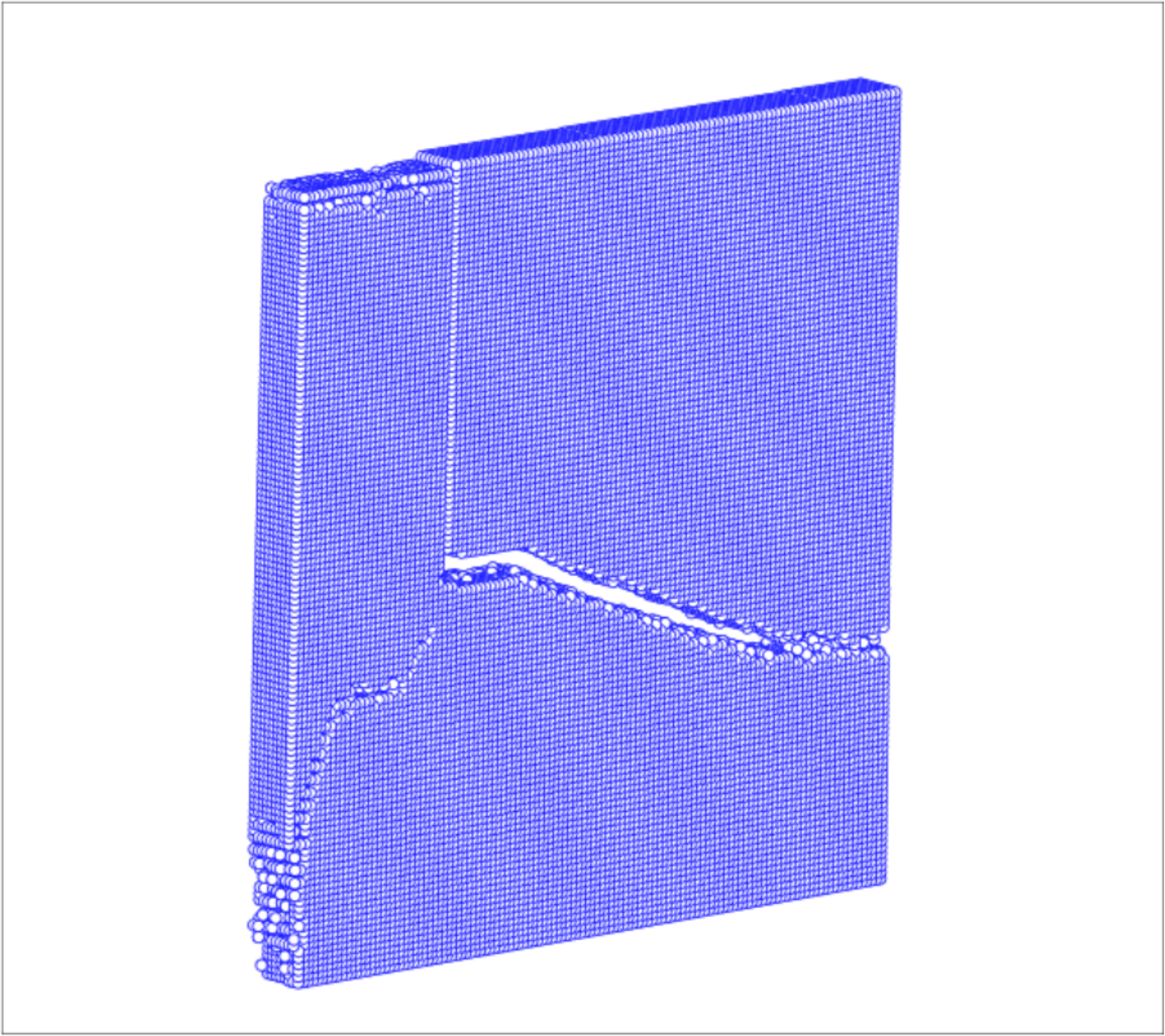}
\caption{} 
\end{subfigure}
\caption{Crack propagation for Kalthoff-Winkler experiment}\label{kalthoff_crack2}
\end{figure}

When the calculated stretch ($\delta_t$) at any time instant exceeds the critical stretch ($\delta_{tc}$) value, the material is assumed to be fully damaged. The crack pattern using the current framework is shown in Figure \ref{kalthoff_crack} and \ref{kalthoff_crack2}. The crack path from the present simulation is found to be approximately 75$^\circ$ which is close to the observed crack path through experiment. A second crack path is found in the numerical results. This is not found in the experimental observation. However, a similar type of second crack path was reported in \cite{belytschko2003dynamic, zhou2016numerical, kosteski2012crack, dipasquale2014crack}. This crack path originates possibly due to the reflection of the wave from the boundary or due to improper estimation of material parameters.

\subsubsection{Taylor bullet impact}
\begin{figure}[hbtp!]
\centering
\includegraphics[width=0.5\textwidth]{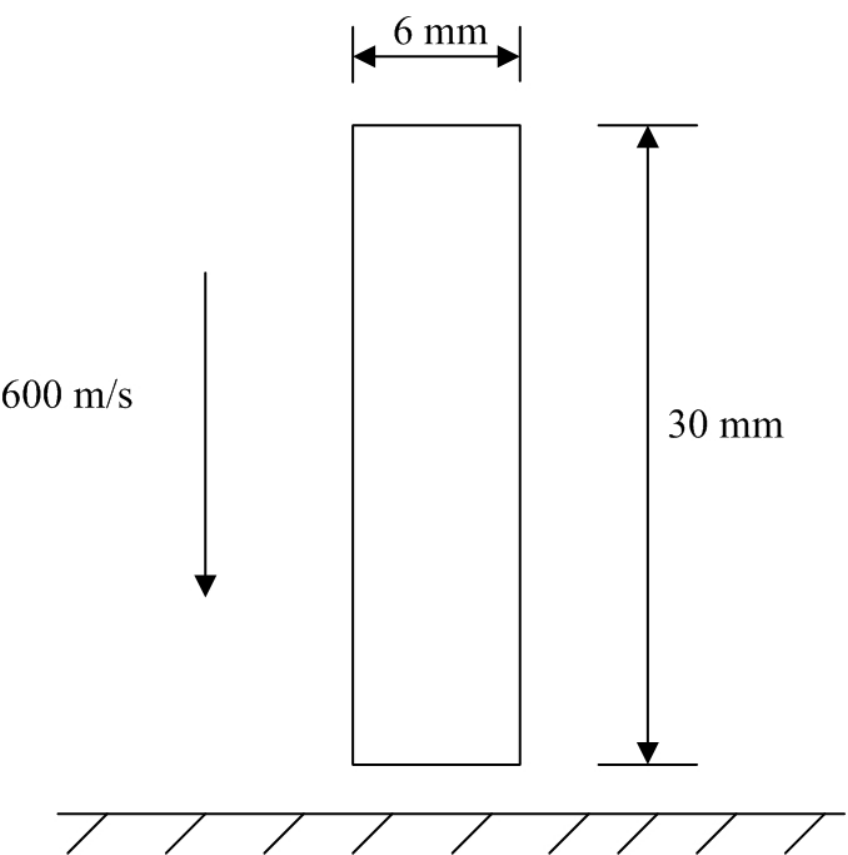}
\caption{Cylindrical projectile impacting rigid surface}\label{tar_con}
\end{figure}

\begin{figure}[hbtp!]
\centering
\includegraphics[width=0.4\textwidth,trim={50 25 50 25}, clip]{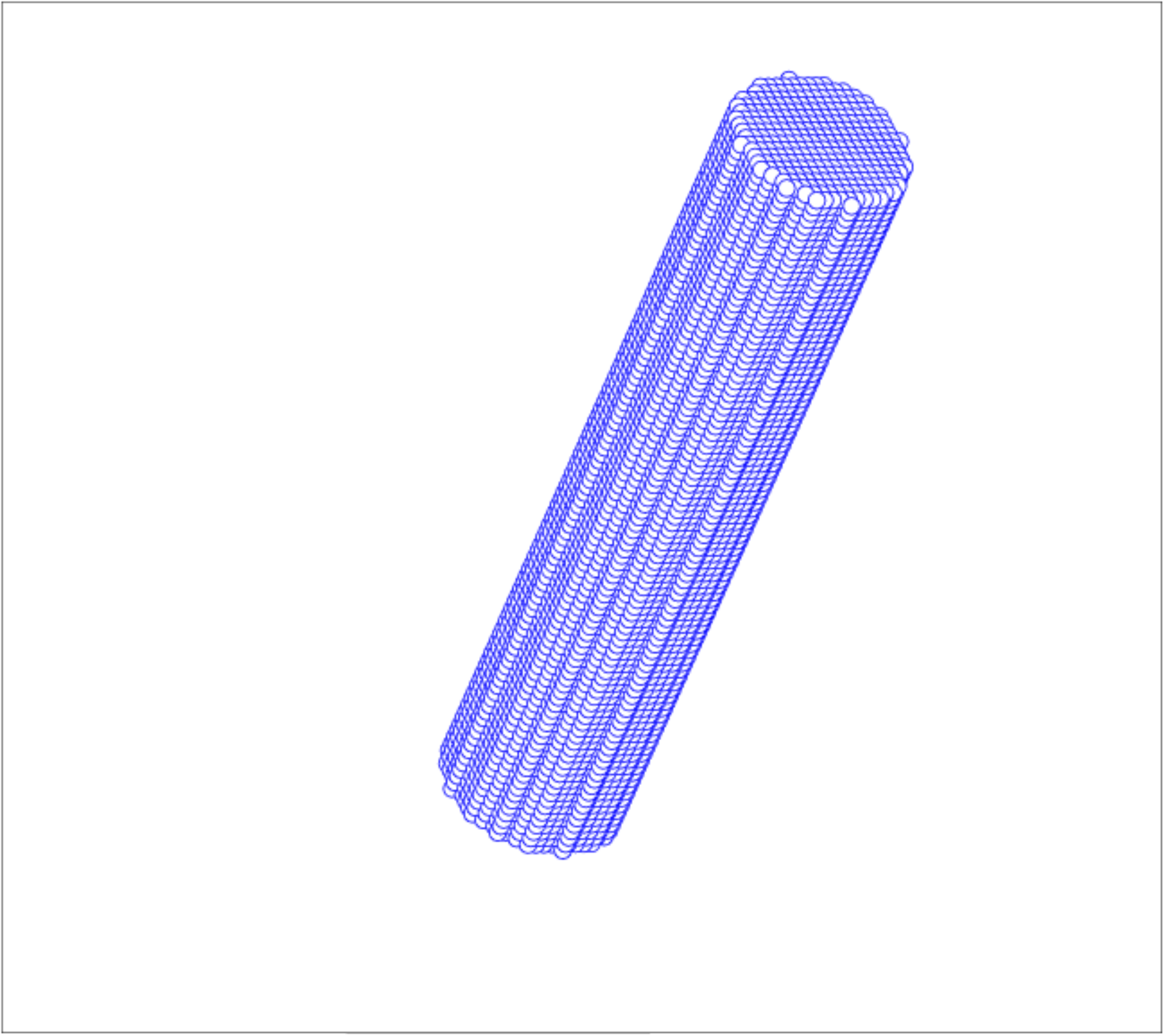}
\caption{Initial particle distribution of the projectile}\label{tar_in}
\end{figure}

\begin{figure}[hbtp!]
\centering
\begin{subfigure}[t]{0.25\textwidth}    %trim={<left> <lower> <right> <upper>}
\includegraphics[width=\textwidth]{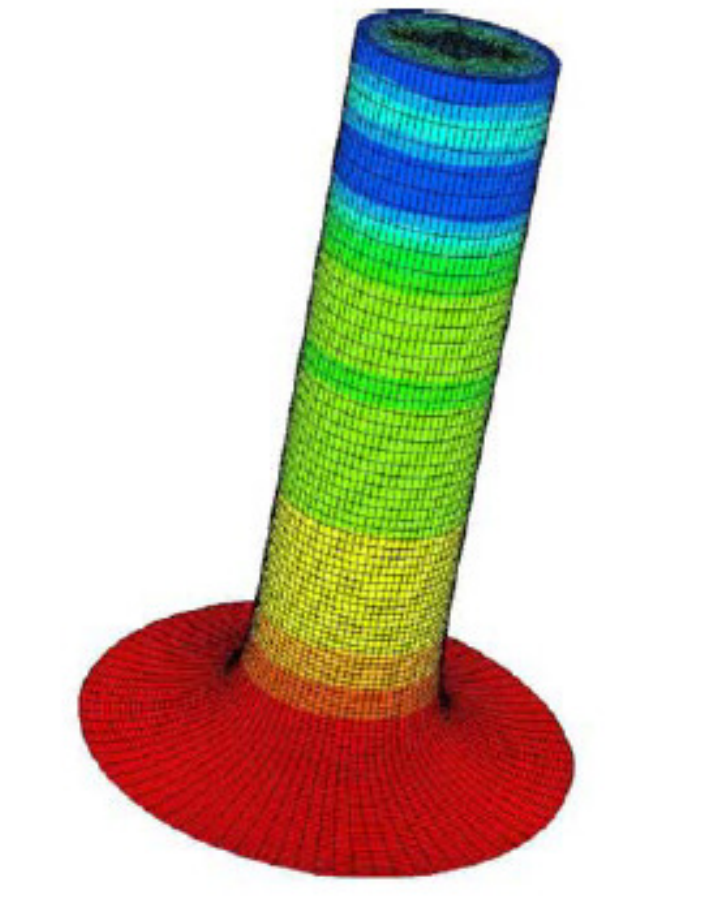}
\caption{Prediction by \cite{teng2005numericalt}} 
\end{subfigure}
\begin{subfigure}[t]{0.25\textwidth}    %trim={<left> <lower> <right> <upper>}
\includegraphics[width=\textwidth]{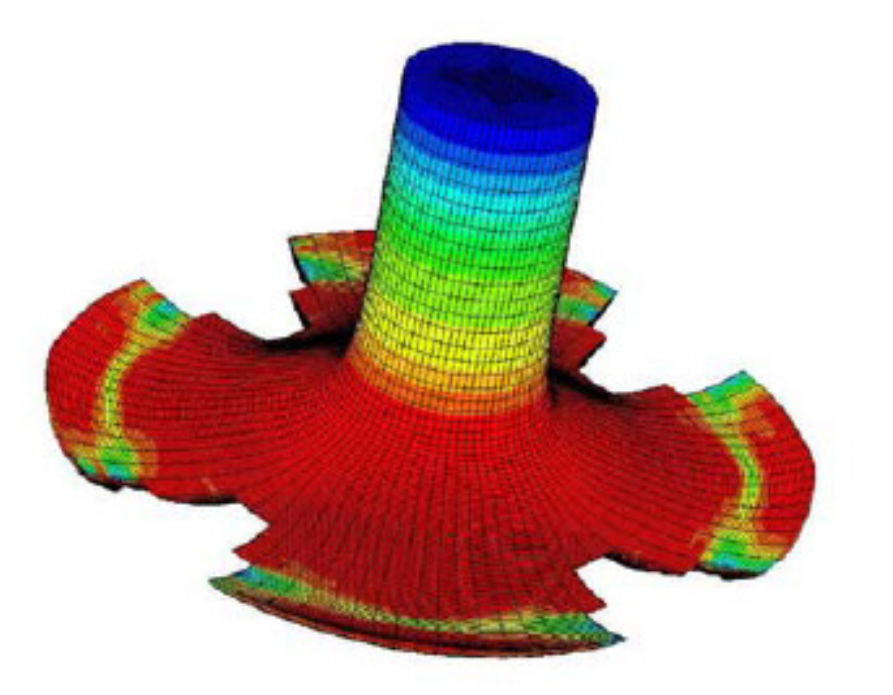}
\caption{Prediction by \cite{teng2005numericalt}} 
\end{subfigure}
\begin{subfigure}[t]{0.45\textwidth}    %trim={<left> <lower> <right> <upper>}
\includegraphics[width=0.8\textwidth,trim={50 25 50 25}, clip]{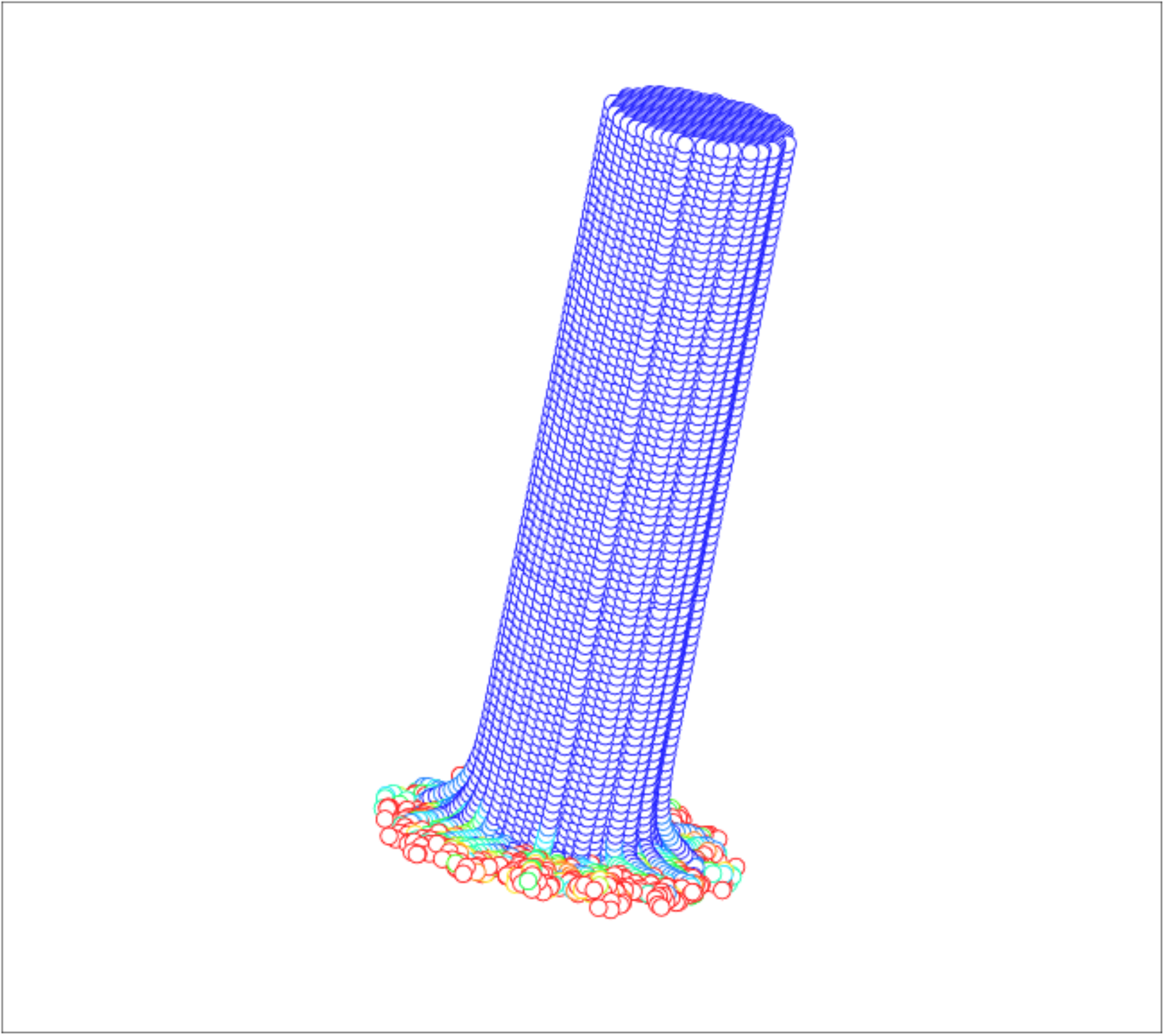}
\caption{Present prediction}
\end{subfigure}
\begin{subfigure}[t]{0.45\textwidth}    %trim={<left> <lower> <right> <upper>}
\includegraphics[width=0.8\textwidth,trim={50 25 50 25}, clip]{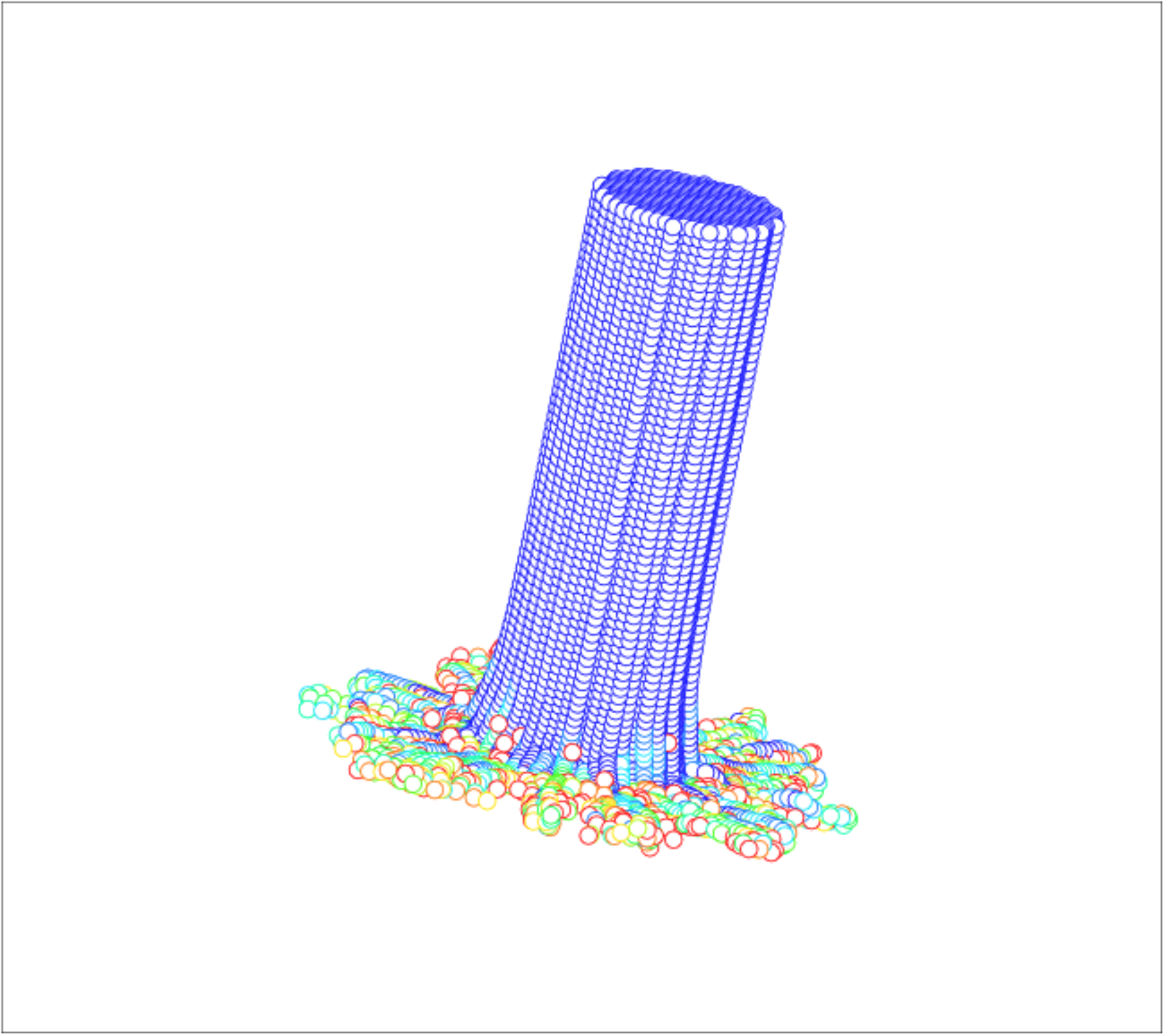}
\caption{Present prediction} 
\end{subfigure}
\begin{subfigure}[t]{0.45\textwidth}    %trim={<left> <lower> <right> <upper>}
\includegraphics[width=0.8\textwidth,trim={50 25 50 25}, clip]{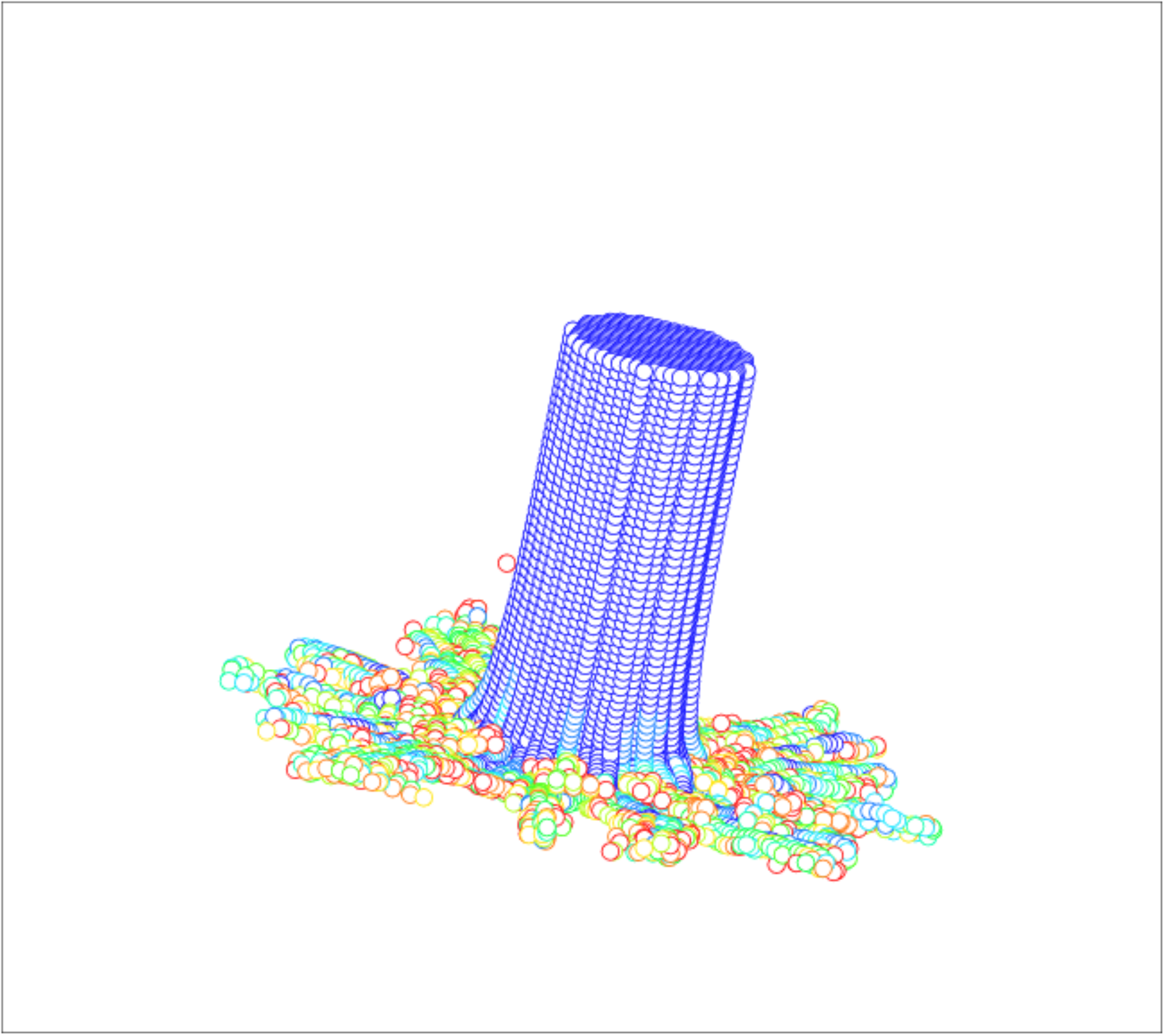}
\caption{Present prediction}
\end{subfigure}
\begin{subfigure}[t]{0.45\textwidth}    %trim={<left> <lower> <right> <upper>}
\includegraphics[width=0.8\textwidth,trim={50 25 50 25}, clip]{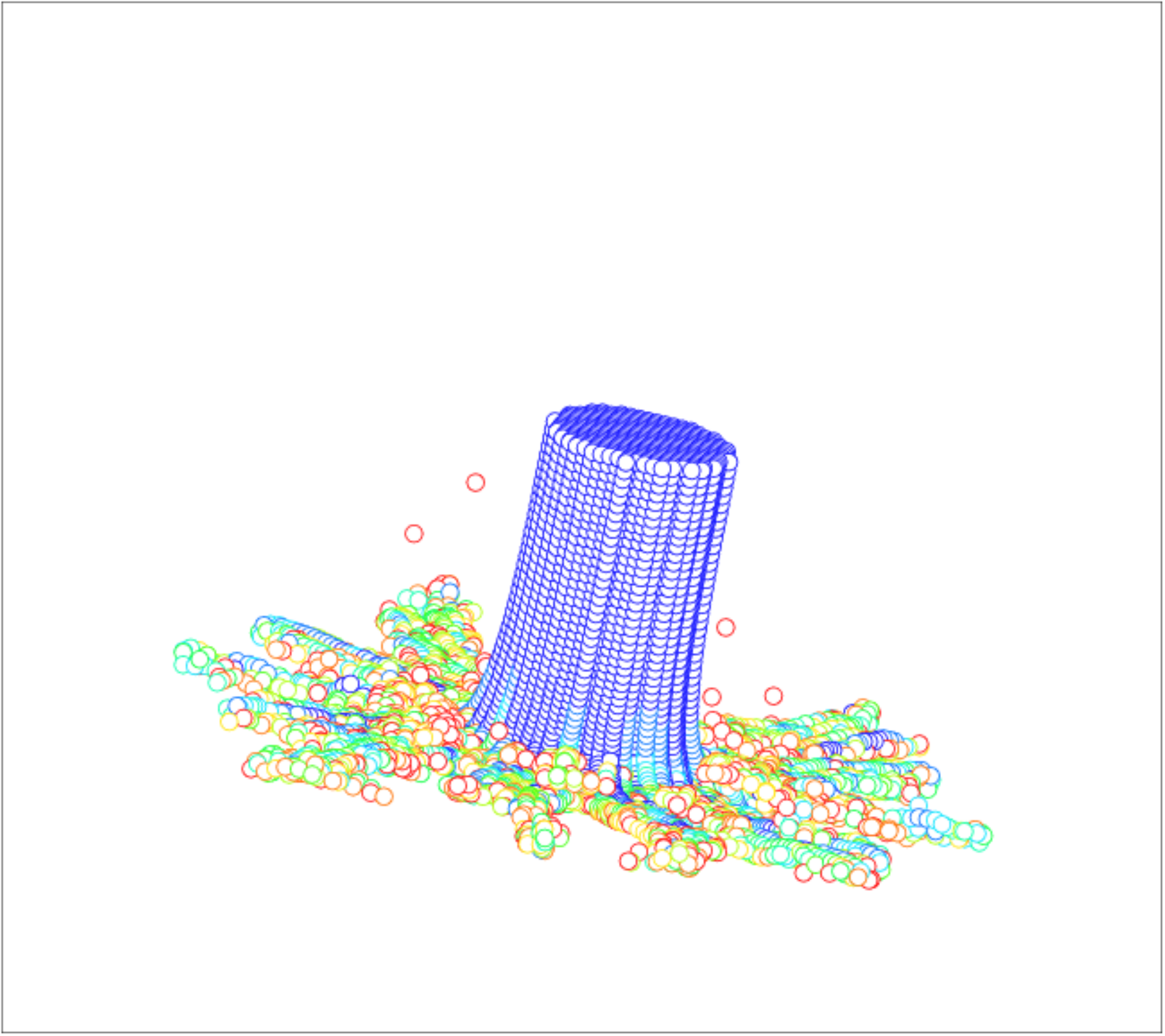}
\caption{Present prediction} 
\end{subfigure}
\begin{subfigure}[t]{0.45\textwidth}    %trim={<left> <lower> <right> <upper>}
\includegraphics[width=0.8\textwidth,trim={50 25 50 25}, clip]{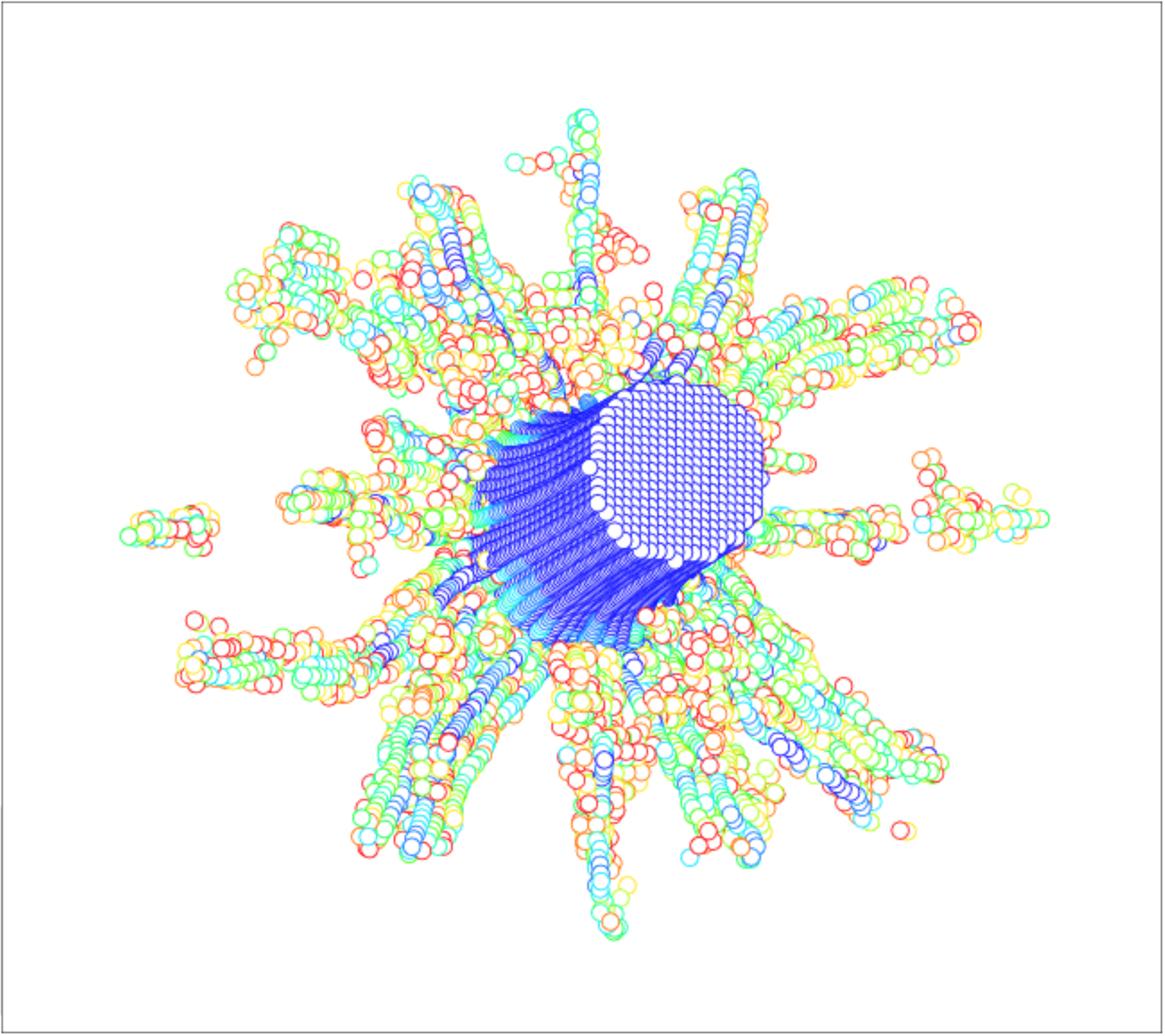}
\caption{Present prediction}
\end{subfigure}
\caption{Petalling fracture mode for cylindrical projectile}\label{tay_pet}
\end{figure}

A flat-ended cylindrical projectile of length 30 mm and diameter 6 mm is moving towards a rigid wall as shown in Figure \ref{tar_con}. The failure of the projectile is modelled in this section. The bullet is made of Weldox 460E steel. The Johnson-Cook material and damage models are used. The material and damage parameters may be found in \cite{islam2017computational}. The SPH parameters are given in Table \ref{table_ta}. Initial impact velocity of the projectile is 600 m/s, and initial particle distribution is shown in Figure \ref{tar_in}. The petalling fracture mode for the projectile is simulated in \cite{teng2005numericalt}. The present framework also predicts petalling fracture mode as shown in Figure \ref{tay_pet}.  

\begin{table}[h!]
\caption{Computational parameters used for Taylor bullet impact}\label{table_ta}

\centering

\begin{tabular}{cccc}

\hline

Inter-Particle                 & Smoothing         & Time-step          & Artificial Viscosity                                \\

Spacing (p)        & Length (h)         & ($\Delta$t)        & Parameters ($\beta_1, \beta_2$)                \\ \hline

0.3 mm                        & 0.6 mm                & $5 \times 10^{-9}$s        & (1.0,1.0)                                                \\ \hline

\end{tabular}

\end{table}

\section{Closure}\label{sec-5}
Numerical modelling of crack initiation, propagation and branching is always a challenging task. Towards this, several attempts are made, but not all are efficient. The element-based methods can simulate the arbitrary cracks but have their own set of disadvantages namely intensive computation for local enrichment, re-meshing, element deletion etc. The mesh-free methods also require enrichments or particle splitting. 

In this work, the SPH along with the `pseudo-spring' analogy is used to model arbitrary crack propagation. The `pseudo-spring' analogy provides a way to model discontinuity in solid continua without any enrichment, particle splitting or crack path tracking. The particles are connected by springs which do not provide any extra stiffness to the system but reduce the level of interaction between particles based on the damage state of the springs. The interaction of particles is only with the immediate neighbours. If a spring fails completely, the interaction between connecting particles ceases. This implies the passing of crack paths through the spring and generation of a new surface. This framework also preserves the simplicity of SPH.

The model is used to study two and three-dimensional problems. First, A pre-notched plate under tensile loading is studied. Initially, the crack propagates in a straight line and then splits into two branches. Both the crack branches propagate until the end of the plate. The crack speed and its path are in good agreement with the experimental and other numerical results from the literature. It is observed that the crack propagates in a straight line without any branches for low-intensity load. However, the crack splits into branches for higher load intensity. If loading intensity is further increased, multiple branches and sub-branches can be observed. Next, a notched plate with a circular hole under tensile load is considered. The position of the notch is changed and the crack path propagation is studied. The propagation of crack path and their interaction with the circular hole are found to be in good agreement with the results from the literature. The computational framework is also used to model arbitrary crack paths and surfaces in three dimensions. The fracture of a chalk bar under torsion is considered. The numerically obtained crack surface is similar to the experimental and numerical results. The Kalthoff-Winkler experiment and the Taylor impact test are also simulated for brittle fracture and the crack paths are found to be comparable to the experimental and numerical results. The `pseudo-spring' SPH is found to be very efficient for the present simulations of crack initiation, propagation and branching without any special treatment. In three dimensional problem, the arbitrary crack surface can also be traced without much difficulty. The methods seem to be very good in tacking discontinuity and easy to implement. It is anticipated that the pseudo-spring analogy has the potential to constitute an efficient computational tool for modelling more complex fracture problems in both 2D and 3D continua.

\section*{Acknowledgement}
The authors would like to acknowledge the Defence Research and Development Organization (DRDO), India for their support of this work. 

\clearpage
\bibliographystyle{elsarticle-num}


%\begin{thebibliography}{00}
%
%%% \bibitem[Author(year)]{label}
%%% Text of bibliographic item
%\end{thebibliography}
\end{document}